\newif\iffigs\figstrue
\documentclass[10pt,a4paper]{article}
\usepackage{latexsym,amssymb,lscape,graphics,setspace}
\usepackage{amsmath}
\usepackage{graphicx}        
\usepackage{longtable}
\usepackage{multirow}
\usepackage{booktabs}
\usepackage{multirow}
\usepackage{color}
\usepackage{slashed,epsfig}
\usepackage{amsfonts}
\usepackage{cite}
\usepackage{verbatim}
\usepackage{colortbl}
\usepackage[table]{xcolor}
\usepackage{hyperref}       
\usepackage{array}
\usepackage{fancyhdr}
\usepackage{tikz}
\usepackage{multicol}
\usepackage{lscape}
\graphicspath{{images/}}

\newcommand{\mathsym}[1]{{}}

\newtheorem{definizione}{Definition}[section]

\newcommand{\bd}{\begin{definizione}}
\newcommand{\ed}{\end{definizione}}

\def\IC{\relax\,\hbox{$\inbar\kern-.3em{\rm C}$}}
\def\IG{\relax\,\hbox{$\inbar\kern-.3em{\rm G}$}}
\def\IB{\relax{\rm I\kern-.18em B}}
\def\ID{\relax{\rm I\kern-.18em D}}
\def\IL{\relax{\rm I\kern-.18em L}}
\def\IF{\relax{\rm I\kern-.18em F}}
\def\IH{\relax{\rm I\kern-.18em H}}
\def\II{\relax{\rm I\kern-.17em I}}
\def\IN{\relax{\rm I\kern-.18em N}}
\def\IP{\relax{\rm I\kern-.18em P}}
\def\IQ{\relax\,\hbox{$\inbar\kern-.3em{\rm Q}$}}
\def\bfzero{\relax\,\hbox{$\inbar\kern-.3em{\rm 0}$}}
\def\IK{\relax{\rm I\kern-.18em K}}
\def\IG{\relax\,\hbox{$\inbar\kern-.3em{\rm G}$}}
 \font\cmss=cmss10 \font\cmsss=cmss10 at 7pt
\def\IR{\relax{\rm I\kern-.18em R}}
\def\ZZ{\relax\ifmmode\mathchoice
{\hbox{\cmss Z\kern-.4em Z}}{\hbox{\cmss Z\kern-.4em Z}}
{\lower.9pt\hbox{\cmsss Z\kern-.4em Z}} {\lower1.2pt\hbox{\cmsss
Z\kern-.4em Z}}\else{\cmss Z\kern-.4em Z}\fi}
\def\bfone{\relax{\rm 1\kern-.35em 1}}

\def\inbar{\vrule height1.5ex width.4pt depth0pt}
\def\bfzero{\relax{\rm I\kern-.18em 0}}
\def\bfone{\relax{\rm 1\kern-.35em 1}}

\DeclareFontFamily{U}{rsf}{} \DeclareFontShape{U}{rsf}{m}{n}{
  <5> <6> rsfs5 <7> <8> <9> rsfs7 <10-> rsfs10}{}
\DeclareMathAlphabet\Scr{U}{rsf}{m}{n}

\newcommand{\ft}[2]{{\textstyle\frac{#1}{#2}}}
\def\tilde{\widetilde}

\def\1bar{1\hskip -.275cm -}
\def\2bar{2\hskip -.275cm -}
\def\3bar{3\hskip -.275cm -}

\newsavebox{\uuunit}
\sbox{\uuunit}
{\setlength{\unitlength}{0.825em}
\begin{picture}(0.6,0.7)
\thinlines
\put(0,0){\line(1,0){0.5}}
\put(0.15,0){\line(0,1){0.7}}
\put(0.35,0){\line(0,1){0.8}}
\multiput(0.3,0.8)(-0.04,-0.02){10}{\rule{0.5pt}{0.5pt}}
\end {picture}}

\makeatletter \@addtoreset{equation}{section} \makeatother

\def\bfone{\relax{\rm 1\kern-.35em 1}}

\def\bfone{\relax{\rm 1\kern-.35em 1}}
\font\cmss=cmss10 \font\cmsss=cmss10 at 7pt

\newcommand{\so}{\mathfrak{so}}

\newcommand{\uu}{\mathfrak{u}}

\begin{document}
\begin{titlepage}
\begin{center}
\vskip 0.2cm
{{\large {\sc General Properties of the Thermo-Metric for\\}}
\vskip 0.1cm 
{\large {\sc CV   event manifolds and  \\}}
\vskip 0.1cm 
  {\large {\sc the magnetization combinatorial scheme } ${}^\dagger$} }\\
 \vskip 1cm {\sc Pietro Fr\'e\,$^{a,b}$ \\ Alexander S. Sorin\,$^{c,b}$ 
 and Mario Trigiante\,$^{d, e}$} \vskip 0.5cm
\smallskip
{\sl \small \frenchspacing ${}^a\,$ {\tt Emeritus Professor of}  Dipartimento di Fisica, 
Universit\`a di Torino, Via P. Giuria 1, I-10125 Torino, Italy \\[2pt]
${}^{b}\,${\tt Senior Consultant of } Additati\&Partners Consulting s.r.l, 
Via Filippo Pacini 36, I-51100 Pistoia, Italy \\[2pt]
${}^c\,$ Center for Quantum Science and Technology, Tel-Aviv University, Tel-Aviv 69978, Israel \\[2pt]
${}^d\,$Dipartimento DISAT, Politecnico di Torino,
C.so Duca degli Abruzzi 24, I-10129 Torino, Italy\\[2pt]
${}^e\,$INFN, Sezione di Torino\\[2pt]
E-mail:  {\tt pietro.fre@unito.it, asorin@tauex.tau.ac.il,\\
 mario.trigiante@polito.it, } }
\begin{abstract}
Following previous results recently obtained by us, both on the general relation between Information Geometry and Geometrical Thermodynamics and on the exact calculation of partition functions for extended Souriau Gibbs distributions on Calabi Vesentini manifolds, in this paper we study  the differential geometry of the corresponding thermo-metrics. A general intriguing scheme is discovered and put into evidence. Apart from a small difference, significantly distinguishing the even from the odd dimensional instance of the microscopic Calabi Vesentini manifolds, the complete thermo-space is flat when no constraint is introduced. Freezing the magnetic fields, which can be done according to complicated combinatorials, forces the thermo-system to evolve on curved submanifolds $\mathfrak{M}_{n|reg}$ of the thermo--space that have a structure depending only on the length of the $n-1$ chain of frozen contiguous magnetic fields. The behavior of the holonomic Riemann tensor components for $\mathfrak{M}_{n|reg}$ is codified by a symmetric matrix with peculiar behavior along special symmetrically arranged submanifolds that, we deem, are responsible for the generation of curvature walls and for the categorical partitioning of the thermo space. The embedding of the curved submanifold $\mathfrak{M}_{n|reg}$  into $\mathbb{R}^{2n-1}$  can be traced back to the vanishing of magnetic fields and, in this case, the flat metric on $\mathbb{R}^{2n-1}$ is the  $\mathfrak{a}_{2n-1}$ simple Lie algebra Cartan matrix, or reconstructed directly, forcing on $\mathbb{R}^{2n-1}$ the standard $\delta_{ij}$ metric and introducing an ingenious immersion algorithm. The second approach  reveals the geometric interpretation of $\mathfrak{M}_{n|reg}$
as a generalized translation hypersurface. Furthermore the boundary at infinity $\partial_\infty\mathfrak{M}_{n|reg}$ has 
a hypercube structure whose face central points and vertices appear, numerically, to be the end-points of all geodesics depending only on their angular slope at the start. Notwithstanding the Euclidean signature of $\mathfrak{M}_{n|reg}$, this general feature is reminiscent of the causal structure at infinity of Lorentzian space-times and of Penrose diagrams.   
\end{abstract}
\vfill
\end{center}
\noindent \parbox{175mm}{\hrulefill}
\par
${}^\dagger$ P.G. Fr\'e acknowledges support by the Company \textit{Additati\&Partners 
Consulting s.r.l} during the development of 
the present research.\\
The work of A.S. Sorin was supported in part by the Center for Integration in Science of the Israel Ministry of Aliyah
and Integration.
\\[5pt]
\end{titlepage}
{\small \tableofcontents} \noindent {}
\newpage
\section{Introduction} \label{introibo}
The present paper is the third episode in the development of a conspicuous and strategic branch of a more general research 
plan, started about two and a half year ago and first publicly announced March 2025, with the foundational paper \cite{pgtstheory}. Within the development of that general programme, whose aim is the geometrical refoundation of Artificial Intelligence on the basis of Cartan Neural Networks (\textbf{CaNNs}),   defined in \cite{TSnaviga,naviga}, (June-July 2025), the present paper counts as the eight episode, the other seven being \cite{pgtstheory,TSnaviga,naviga,tassellandum,axialbeltra,geotermico,secondtemperature}
\par
The conceptual pivot of the \textbf{CaNNs} project is  the construction of Neural Networks that are required to be \underline{\textit{both covariant and interpretable}}, thanks to the removal of point-wise activation functions, the necessary non linearity device being provided, instead, by the exponential map from a solvable Lie algebra to its corresponding solvable Lie group. In that context a rather tight logic based on classical theorems and on  mathematical structures that were developed in the context of Supergravity Theory (see in  particular \cite{pgtstheory} and \cite{TSnaviga}) lead to the conclusion that the most appropriate mathematical modelling of Neural Network layers is indeed provided by non-compact symmetric spaces $\mathrm{U/H}$ (where $\mathrm{U}$ is a simple non-compact Lie group and $\mathrm{H}\subset \mathrm{U}$ its maximal compact subgroup\footnote{More generally, as it starts emerging from a new ongoing research project\cite{toinepietromario}, the mathematical modeling of Cartan Neural Network layers can be provided by all Special K\"ahler Homogeneous manifolds, the
non-symmetric among which might possibly be interpreted as Special K\"ahler Symmetric Spaces with a deformed metric partially breaking the full Isometry Group U.}). 
\par 
The branch in the CaNN project of which the present paper counts as the third episode is related with the \textbf{second pillar of Geometrical Deep Learning}, namely \textbf{Probability Theory}, the \textbf{first pillar} being instead \textbf{Differential Geometry}. Indeed probability distributions on the layers of the CaNN can no longer be assumed to be the classical gaussians, in all of their declinations, that are appropriate only to flat Euclidean spaces $\mathbb{E}^n \simeq \mathbb{R}^n$; rather they  should  be reestablished as Gibbs distributions on those non-compact symmetric spaces, or homogeneous special K\"ahler manifolds, that were singled out as
the correct mathematical modeling of such layers. 
\par
The first episode in the development of this branch was the paper \cite{geotermico} (December 2025) that fixed the conceptual environment and the mission, namely the revisitation 
of \textbf{Information Geometry}\cite{raone,cenzone,amarone} and its identification (at least for \textit{Gibbs like distributions}, appropriate to \textit{Markovian stochastic processes}) with \textbf{Geometrical Thermodynamics}, as it was developed in the noble and distinguished tradition established by the sequence of papers  \cite{gianno1,gianno2,lychaginlecture,Kushner_2020,Lychagin_2020,ludaed,ludaed2,
Ruppeiner_2012b,Ruppeiner_2010,Ruppeiner_2012,Ruppeiner_2013,ruppoRdiag}.
\par 
Within such more general mission, the needs of applications to CaNNs brought to the forefront the issue of \textbf{Souriau thermodynamics} \cite{souriaub1,souriaub2,barbarpapad4,
marlentropia,caldobarbaresco,barbaresco2,barbaresco3,marlegibbs,barbarpapad1,barbarpapad2,barbarpapad3,nebbo}.
\par
As we stressed in \cite{geotermico} and in \cite{secondtemperature}, the latter being the immediate predecessor of the present paper, what goes under the name of \textit{Souriau Thermodynamics} is the construction of
\textit{Gibbs probability measures} on $\mathrm{U/H}$, consistent with the isometry group $\mathrm{U}$ (or a subgroup thereof) of such spaces that necessarily need to be K\"ahlerian.\footnote{As we emphasized in \cite{geotermico}, the original conception of Souriau Thermodynamics, as adopted in the quoted literature  \cite{souriaub1,souriaub2,barbarpapad4,
marlentropia,caldobarbaresco,barbaresco2,barbaresco3,marlegibbs,barbarpapad1,barbarpapad2,barbarpapad3,nebbo},
is based on the notion of \textit{coadjoint orbits} and on the  symplectic structure over them, named after Kirillov-Kostant-Souriau. However, since every $\mathrm{U}$ coadjoint orbit is equivalent to a coset manifold $\mathrm{U/H}$, where $\mathrm{H}$ is the normalizer of the Lie algebra element $\mathfrak{g}\in \mathbb{U}$ labeling the orbit, we always prefer to look at Souriau Gibbs-like distributions as defined on coset manifolds, in particular symmetric spaces that, as we said above, must be K\"ahlerian.}.
\par
In \cite{secondtemperature}, while developing suitable mathematical methods, based on the theory of \textit{Abelian Structures in Integrable Systems}, in order \textit{to calculate explicitly and exactly} the  \textit{partition functions} for the Gibbs-like probability distributions on \textbf{Calabi-Vesentini manifolds} (CV)\footnote{CV manifolds constitute  one of the two infinite towers of K\"ahler non-compact symmetric spaces. The other infinite tower is given by the Siegel half-planes (see \cite{secondtemperature}) and has growing non-compact rank which probably makes much more difficult the exact calculation of partition functions.}, we were led \underline{\textit{by the intrinsic mathematical logic of our constructions}} to \textbf{generalize Souriau thermodynamics} by means of the introduction of \textit{new generalized temperatures}, whose inclusion breaks the $\mathrm{U}$-symmetry of the partition function, not randomly, rather in a precise way generated by the very structure of the $\mathrm{U/H}$ manifold via its admissible
\textbf{compact abelian structures}\footnote{It is a general feature of Symmetry in all sound Physical Theories that the way it can be spontaneously broken is implicitly defined by the very structure of the Symmetric Theory.}. 
\par
In \cite{secondtemperature} we also observed the unexpected emergence of an intriguing and inspiring phenomenon, somewhat reminiscent of spontaneous magnetization in the physics of ferromagnetic systems. This led to a preliminary interpretation of the additional new temperatures that break the $\mathrm{U}$-symmetry of the partition function as \textbf{generalized magnetic fields}.
\par Referring the reader to \cite{secondtemperature} for all details on  the differential geometry of CV manifolds, on 
the theory of compact abelian structures and on the calculational details of the partition functions, in the present paper, after a brief summary of the necessary ingredients developed in \cite{secondtemperature}, we focus on the general properties of the \textbf{Thermo-Metric} streaming from the exact partition functions calculated in \cite{secondtemperature} and on their in depth analysis. Such an analysis reveals, once again, quite unexpected and peculiar properties and brings to the forefront a general combinatorial scheme of reductions that is summarized in the conclusive section \ref{zakliuchenie}, together with the perspectives of its use in Machine Learning both supervised and unsupervised with reinforcement.
\section{Structure of the Riemannian Thermo-Metric on the Thermodynamic Contact Manifold $\mho$}
In order to present our new results we begin by summarizing 
the results obtained in \cite{secondtemperature} concerning the structure of the Thermo-Metric for general Gibbs-like distributions and its specialization to the case of generalized Souriau-like Gibbs distributions on non-compact symmetric spaces $\mathrm{U/H}$ of K\"ahlerian type. 
\subsection{The general case}
Let $\Omega$ be a space of events and let $\mho$ be the corresponding  space of thermodynamic variables, namely of the parameters labeling  a family of probability distributions  $p(\mathbf{q},\boldsymbol{\lambda})$ defined on $\Omega$.  We assume that each probability distribution is  properly normalized:
\begin{equation}\label{probanorma}
  \int_\Omega \, p(\mathbf{q},\boldsymbol{\lambda}) \mathrm{d}[\mathbf{q}] \, = \, 1
\end{equation}
where $\mathrm{d}[\mathbf{q}]$ denotes the integration measure 
(see for instance the comprehensive review \cite{fioresi2023deep}). According with a general consensus in the Statistical/Machine Learning scientific community, if the underlying stochastic processes at the basis of the considered parameterized probability distribution  $p(\mathbf{q},\boldsymbol{\lambda})$ are of \textbf{Markovian type}, then the latter is of Gibbs type namely:
\begin{equation}\label{ciacolone}
  p(\mathbf{q},\boldsymbol{\lambda}) \, = \, \frac{\exp\left[-\boldsymbol{\lambda}\cdot \mathbf{X}(\mathbf{q})\right]}{Z(\boldsymbol{\lambda})} \quad ; \quad 
  Z(\boldsymbol{\lambda}) \equiv \int_\Omega \,\exp\left[-\boldsymbol{\lambda}\cdot \mathbf{X}(\mathbf{q})\right] \, \mathrm{d}[\mathbf{q}]
\end{equation}
where $\mathbf{X}(\mathbf{q})\, = \, X^{i,\dots,n}(\mathbf{q})$ is a suitable collection, depending on the addressed problem, of $n$ \textbf{observable functions} defined over the manifold of events $\Omega$ and the normalization denominator $Z(\boldsymbol{\lambda})$ is named \textbf{the partition function}, while
\begin{equation}\label{calottapolare}
  \mathcal{H}^{sto}(\boldsymbol{\lambda})\, = \, - \, \log\left[Z(\boldsymbol{\lambda}) \right]
\end{equation}
is named the \textbf{stochastic hamiltonian}. The minus sign in front of the logarithm in eq.(\ref{calottapolare}) is essential since the derivatives of $\mathcal{H}^{sto}(\boldsymbol{\lambda})$ with respect to $\lambda_i$ must be the mean  value $x_i$ of the observable function $X_i(q)$ on the microscopic space, which has a unique definition and an absolute sign.
\par
Following the conceptual scheme introduced by  Jaynes in the late fifties of the XXth century \cite{gianno1,gianno2}, further
developed by Lychagin and Roop in the late nineties of the same century and in the first decade of the next one \cite{lychaginlecture,ludaed,ludaed2,Lychagin_2020,Kushner_2020}, also sided by inspiring contributions from Ruppeiner and collaborators \cite{Ruppeiner_2010,Ruppeiner_2012,Ruppeiner_2012b,Ruppeiner_2013,ruppoRdiag,Ruppeiner_2020},
the \textbf{thermodynamic manifold} $\mho$  is always an odd-dimensional \textbf{contact manifold} and its \textit{iso-entropic leaves}, transverse to the Reeb field, are always \textbf{symplectic manifolds} endowed, as it was shown in \cite{secondtemperature}, with a \textbf{K\"ahler metric} that is unique apart from a global scaling factor in front. The sign of such scaling factor must be chosen in such a way that the signature of the metric on the contact-manifold be Euclidean and not Lorentzian. The relevance of such choice of sign was overlooked in the original version of \cite{secondtemperature} and could be discovered only within the context of the present paper that analyses the behavior of thermo--metrics. In the frame of \cite{secondtemperature} such sign was irrelevant and for this reason it was overlooked. We are going to introduce the 
correct choice also in \cite{secondtemperature}.    
\par 
Indeed, as also shown in \cite{secondtemperature}, the \textit{macroscopic thermodynamic manifold $\mho\left[\Omega,\mathbf{X}\right]$}, singled out by the choice of a microscopic manifold $\Omega=\mathcal{M}^{micro}_{m}$ of dimension $m$ and by a choice of $n$ observable functions $\mathbf{X}(q)$ defined over $\mathcal{M}^{micro}_{m}$ is always a $(2n+1)$-dimensional contact manifold $\mho \, = \,\mathcal{M}^{macro}_{2n+1}$  with contact form:
\begin{equation}\label{alphamacro}
  \underbrace{\alpha}_{\text{defined on $\mho\left[\Omega,\mathbf{X}\right]$}} \, = \, \mathrm{d}\mathcal{I} \, + \, \sum_{i=1}^n \lambda_i \, \mathrm{d}x^i
\end{equation}
 Any admissible thermodynamic state can be  identified with a point in the following  $n$-dimensional submanifold  of the contact manifold:
\begin{equation}
\label{crocolo}
    \mathcal{L}_n \, = \, \left\{ \mathcal{I} =
    \mathcal{I}\left(\pmb{\lambda},\mathbf{x}\right)\, , \, x_i \, = \,
    \frac{\partial}{\partial \lambda^i}
    \mathcal{H}\left(\pmb{\lambda}\right)\right\} \, \subset \,\mathcal{M}_{2n+1}
\end{equation}
\par
Given the original contact variety $\mathcal{M}^{2n+1}$
with the contact $1$-form given in eq.(\ref{alphamacro})
the Reeb vector field (see appendix A of \cite{geotermico} for its definition) is
\begin{equation}\label{Reebtermo}
    \mathbf{R}\, = \,\frac{\partial}{\partial \mathcal{I}}
\end{equation}
namely the entropy gradient. 
\par
On the other hand, from the general discussion in appendix A of \cite{geotermico}
we know that every $2n$-dimensional submanifold
of a contact manifold $\mathcal{M}^{2n+1}$  that is
transverse to the Reeb vector field  is a symplectic
manifold $\mathcal{S}^{2n}$ whose symplectic  $2$-form is the restriction
to $\mathcal{S}^{2n}$ of the exterior differential of the contact 1-form:
\begin{equation}\label{rovereto}
    \omega \, = \,
   \mathrm{d}\alpha \mid_{\mathcal{S}^{2n}}
\end{equation}
Hence applying these general notions to the case at hand, we see that the  symplectic variety transverse to Reeb vector
(\ref{Reebtermo}) is given by the following projection map:
\begin{equation}\label{trasversus}
    \pi_{TR} \, : \, \mho \, = \,\mathcal{M}^{2n+1} \, \rightarrow \,
    \mathcal{S}^{2n} \subset \mho \quad ; \quad
    \pi_{TR}(\mathcal{I},\pmb{\lambda},\mathbf{x})\, = \, (\pmb{\lambda},\mathbf{x})
\end{equation}
and the  symplectic $2$-form  is as follows
\begin{equation}\label{adige}
    \omega \, = \, \sum_{i=1}^{n} \mathrm{d}\lambda^i \wedge dx_i \, \quad
    ;\quad \mathrm{d}\alpha \, = \, \pi^\star_{TR}(\omega)
\end{equation}
\par
In \cite{secondtemperature} it was given a definite answer to the following question:
\textit{ since the submanifolds $\mathcal{S}^{2n}$, that are transverse to the Reeb field, are symplectic with a symplectic $2$-form already in Darboux coordinates provided by eq.(\ref{adige}), 
what is the most general  metric on $\mho[\Omega,\mathbf{X}]$ such that its restriction to $\mathcal{S}^{2n}$ admits the 2-form of eq.(\ref{adige}) as the K\"ahler $2$-form of a K\"ahler metric on the same submanifold?}
It was shown that, due to the very simple structure of the projection map $\pi_{TR}$, the  required metric  is as follows\footnote{In eq.(\ref{contattametrica}) the signature of
the K\"ahler metric $ds^2_{\mathcal{S}^{2n}}$ must be $\{+,+,\dots,+\}$ in order for the full metric on the contact manifold to be Euclidean. Considered per se the factor in front of the K\"ahler metric $ds^2_{\mathcal{S}^{2n}}$ is irrelevant,
but introducing it in eq.(\ref{contattametrica}) fixes its sign.} :
\begin{equation}\label{contattametrica}
  ds^2_{\mho}\, = \, \mathrm{d}{\mathcal{I}}^2 \, + \, ds^2_{\mathcal{S}^{2n}}
\end{equation}
where $ds^2_{\mathcal{S}^{2n}}$ is a K\"ahler metric on $\mathcal{S}^{2n}$ and its K\"ahler form $\boldsymbol{\mathcal{K}} $  is equal to $\omega$ as given in eq.(\ref{adige}).
\par
\subsubsection{The symplectic formalism for the Thermo K\"ahler metric}
\label{certosino}
The symplectic formalism for K\"ahler metrics comes into play when the K\"ahler metric has a non-trivial dependence only on one half of the $2n$ coordinates.  In  the thermodynamic case the $n$ abelian isometries form a non-compact abelian group, made of translations.  
Indeed the K\"ahler potential depends only on the imaginary part $v_i$ of the complex variables whose real parts are the mean values $x_i$ of the functions $\mathbf{X}(\mathbf{q})$, and introducing the momenta variables as:
\begin{equation}\label{coreuta}
  \lambda^i \, \equiv \, \frac{\partial \mathcal{K}(\mathbf{v})}{\partial v_i}
\end{equation}
one defines the symplectic real potential by means of the Legendre transform:
\begin{equation}\label{legendroG}
  G(\boldsymbol{\lambda}) \, = \sum_{i=1}^n\, v_i \, \lambda^i \, - \, \mathcal{K}(\mathbf{v})
\end{equation}
which makes sense as long as one is able to invert the relation (\ref{coreuta}) and write the $v_i$ as  functions of the $\lambda^j$. Viceversa if, given the symplectic potential $G(\boldsymbol{\lambda})$, one can retrieve the K\"ahler potential by inverse Legendre transform:
\begin{equation}\label{legendroK}
  \mathcal{K}(\mathbf{v}) \, = \, \sum_{i=1}^n\, v_i \, \lambda^i \, - \, G(\boldsymbol{\lambda})
\end{equation}
where now we have:
\begin{equation}\label{cetra}
  v_i \, \equiv \, \frac{\partial {G}(\boldsymbol{\lambda})}{\partial \lambda^i}
\end{equation}
A very important property is that,  utilizing the real symplectic potential $G(\boldsymbol{\lambda})$, we can always write the full K\"ahler metric in terms of the Hessian:
\begin{equation}\label{tantopercantar}
  \mathrm{H}_{ij}\left(\boldsymbol{\lambda}\right) \, = \, \frac{\partial^2 G(\boldsymbol{\lambda})}{\partial \lambda^i \partial\lambda^j}
\end{equation}
We have\footnote{Since the Hessian of the stochastic Hamiltonian is negattive definite, the minus sign in front of the metric in eq.(\ref{isoentropic}), which per sé is irrelevant, is fixed by the relation with the contact manifold metric (\ref{contattametrica} as we already stressed. This is the minus sign that was overlooked in \cite{secondtemperature}.}:
\begin{eqnarray}
\label{isoentropic}
ds^2_{\mathcal{S}^{2n}} \, = \, - \, \ft 12 \left(\sum_{i,j} \, \mathrm{H}_{ij}\left(\boldsymbol{\lambda}\right) \,\mathrm{d}\lambda^i\,\mathrm{d}\lambda^j\, + \,
\sum_{i,j} \, \mathrm{H}^{-1|ij}\left(\boldsymbol{\lambda}\right) \,\mathrm{d}x_i\,\mathrm{d}x_j\,\right)
\end{eqnarray}
In matrix notation putting all the $2n$ coordinates in a single array $(\lambda^i,x_j)$ enumerated by indices $A,B,\dots$ we have that the metric and the complex structure tensor are respectively given by: 
\begin{eqnarray}
\label{metricandCC}
g_{AB} & = & \ft 12 \, \left( \begin{array}{c|c}
\mathrm{H}_{ij} & 0 \\
\hline
0 & \mathrm{H}^{-1}_{ij}\\
\end{array}\right)\quad ; \quad J^A_{\phantom{A}B} \, = \, \left( \begin{array}{c|c}
0&\mathrm{H}^{-1|ij}  \\
\hline
- \,\mathrm{H}_{ij} & 0\\
\end{array}\right)
\end{eqnarray}
satisfying the obligatory relations:
\begin{equation}\label{tafferuglio}
  J^2 \, = \, - \, \mathrm{Id} \quad ; \quad J^T\cdot g \cdot J \, = \, g
\end{equation}
the first telling us that $J$ is indeed a complex structure, the second telling us that the metric is hermitian with respect to it. The components of the K\"ahler 2-form are given, according with their definition by\footnote{In the following syntetic formula we utilize the convention that the capital indices $A,B$ run on the values $\null^i$ and $\null_i$ for the coordinates $x_i$ and the momenta $\lambda^i$, respectively.}
\begin{eqnarray}
\label{calloforma}
K_{AB} & = &  \left(g \cdot J\right)_{AB} \, = \, \ft 12 \, \left( \begin{array}{c|c}
0&\mathbf{1}_{ij}  \\
\hline
- \,\mathbf{1}_{ij} & 0\\
\end{array}\right)
\end{eqnarray}
which implies that the K\"ahler $2$-form is indeed the symplectic form $\omega$ of eq. (\ref{adige}):
\begin{equation}\label{callo2forma}
  \boldsymbol{\mathcal{K}} \, = \,   \sum_{i} \mathrm{d}\lambda^i \wedge \mathrm{d}x_i  
\end{equation}
Eq.s (\ref{isoentropic}-\ref{metricandCC}) generally apply to all K\"ahler manifolds for which the K\"ahler 2-form, can be put into the Darboux form (\ref{callo2forma}). 
\par
In view of that, the Riemannian metric on the \textbf{thermodynamic contact manifold $\mho\left[\Omega,\mathbf{X}\right]$}, is the following one:
\begin{equation}\label{contactmetric}
  ds^2_{\mho} \, = \, \mathrm{d}{\mathcal{I}}^2 \, - \, \ft 12 \left(\sum_{i,j} \, \mathcal{H}_{ij}\left(\boldsymbol{\lambda}\right) \,\mathrm{d}\lambda^i\,\mathrm{d}\lambda^j\, + \,
\sum_{i,j} \, \mathcal{H}^{-1|ij}\left(\boldsymbol{\lambda}\right) \,\mathrm{d}x_i\,\mathrm{d}x_j\,\right)
\end{equation}
where:
\begin{equation}\label{pordenone}
  \mathcal{H}_{ij}(\boldsymbol{\lambda}) \, = \, \frac{\partial^2 \boldsymbol{\mathcal{H}}^{stoch}(\boldsymbol{\lambda})}{\partial \lambda^i \,\partial \lambda^j}
\end{equation}
is the Hessian of the stochastic hamiltonian  that from a geometrical point of view plays the role of symplectic potential, the K\"ahler potential being its Legendre transform. 

\subsection{The Souriau-like case}
\label{Ktopolino}
As it was stressed in the previous papers \cite{geotermico,secondtemperature}, what is named Souriau's group thermodynamics, is an instance of the previously described general construction  based on the following microscopic data:
\begin{equation}\label{crodinotrotato}
  \mho\left[\Omega, \boldsymbol{X} \right] \, = \, \mho\left[\mathrm{U/H}, \boldsymbol{\mathfrak{P}}_{Killing} \right]
\end{equation}
where $\mathrm{U/H}$ is a non-compact symmetric space, $\mathrm{U}$ being a simple non-compact Lie group and $\mathrm{H}\subset \mathrm{U}$ its maximally compact subgroup, equipped with its unique $\mathrm{U}$-invariant Riemannian metric, while $\boldsymbol{\mathfrak{P}}_{\mathfrak{k}}$ denotes the 
$\text{dim}\mathbb{U}$-dimensional vector of moment-maps associated with the $\text{dim}\mathbb{U}$-dimensional vector of 
\textbf{Killing vector fields}. The above phrasing implies that the considered $\mathrm{U/H}$ symmetric space cannot be generic, rather it should be such that the Killing vector fields are hamiltonian with respect to a symplectic structure compatible with the given metric. 
\par
Summarizing the results of the previous papers \cite{geotermico,secondtemperature}  there is a unique way of meeting such requirements: \textit{the symmetric space $\mathrm{U/H}$ equipped with its $\mathrm{U}$-invariant metric must be a K\"ahler manifold} and the symplectic $2$-form compatible with the metric is the K\"ahler $2$-form:
\begin{equation}\label{carlone}
  \omega \, = \, \boldsymbol{\mathcal{K}}_{\mathrm{U/H}}
\end{equation}
The necessary and sufficient condition for a symmetric space $\mathrm{U/H}$ of the considered type to be a K\"ahler manifold is formulated in terms of the maximal compact subalgebra $\mathbb{H}\subset \mathbb{U}$ as follows:
\begin{equation}\label{kronto}
  \mathbb{H} \, = \, \so_c(2) \oplus \mathbb{H}^\prime \quad ; \quad \left[\so_c(2)\, , \, \mathbb{H}^\prime\right] \, = \, 0
\end{equation}
namely the compact subalgebra should contain a one-dimensional central subalgebra $\so_c(2) \sim \uu(1)$. 
\par
One names  \textbf{generalized temperatures} the parameters $\boldsymbol{\lambda}\,=\,\boldsymbol{\beta}$ that form a vector in the coadjoint representation of the isometry Lie algebra $\mathbb{U}$ and are associated with the functions $\mathbf{X(q)}=\mathfrak{P}_{\mathfrak{k}}(\mathbf{q})$. 
\par
As it was emphasized in \cite{secondtemperature}, the ability to calculate explicitly the partition function integral $Z(\boldsymbol{\beta})$ and to determine the subspace of the Lie Algebra
$\mathbb{T} \subset \mathbb{U}$ whose dual $\mathbb{T}^\star$ contains the available temperature vectors $\boldsymbol{\beta} \in \mathbb{T}^\star$ for which the corresponding integral converges, strictly depends on the theory of \textbf{Abelian Structures}. With the last name one refers to the classification of canonical Darboux coordinate bases on the  microscopic K\"ahler event manifold $\Omega$, such that the microscopic K\"ahler $2$-form takes the same appearance as in classical Hamiltonian Mechanics \textit{i.e.}:
\begin{eqnarray}\label{cosimoprimo}
  \boldsymbol{\mathcal{K}} &\equiv& \mathit{i}\, \frac{\partial^{2} }{\partial z^i\,\partial \bar{z}^{j^\star}}\, \mathcal{K}(z,\bar{z}) \, \mathrm{d}z^i \, \wedge \, \mathrm{d}\bar{z}^{j^\star} \nonumber \\
 \null &=& \sum_{a=1}^{n} \mathrm{d}\mathfrak{p}_a(z,\bar{z}) \,\wedge \, \mathrm{d}\mathfrak{q}^a(z,\bar{z})
\end{eqnarray}
where $z^i$ ($i=1,\dots,n$) is some set of complex coordinates for $\Omega$, $n$ being defined as the complex dimension of the K\"ahler manifold 
\begin{equation}\label{dimensionotto}
  n \, \equiv \, \mathrm{dim}_{\mathbb{C}} \, \Omega \quad ; \quad \mathrm{dim}_{\mathbb{R}} \, \Omega \, = \, 2\, n
\end{equation}
and $\mathcal{K}(z,\bar{z})$ is the K\"ahler potential generating the K\"ahler metric, while $\mathfrak{p}_a(z,\bar{z})$ and $\mathfrak{q}^a(z,\bar{z})$ are \textbf{two sets of real functions} of the complex coordinates $z,\bar{z}$ (or of any other set of $2n$ real coordinates covering the manifold $\Omega$) that are in involution with respect to the Poisson brackets defined by  the K\"ahler metric and its inverse, namely:
\begin{equation}\label{coricidinoA}
  \left\{\mathfrak{p}_a(z,\bar{z})\, , \,\mathfrak{p}_b(z,\bar{z})\right\} \, = \, 0 \quad; \quad
  \left\{\mathfrak{q}^a(z,\bar{z})\, , \,\mathfrak{q}^b(z,\bar{z})\right\} \, = \, 0 \quad; \quad
  \left\{\mathfrak{p}_a(z,\bar{z})\, , \,\mathfrak{q}^b(z,\bar{z})\right\} \, = \, \delta^{b}_a
\end{equation}
By definition, the Poisson bracket is realized as:
\begin{equation}\label{PB1}
  \left\{\boldsymbol{\mathit{f}}(z,\bar{z})\, , \,\boldsymbol{\mathit{g}}(z,\bar{z})\right\} \, \equiv\, 
  \mathit{i}\, g^{ij^\star} \, \left(\frac{\partial \boldsymbol{\mathit{f}} }{\partial z^i}
  \,\frac{\partial \boldsymbol{\mathit{g}}}{\partial \bar{z}^{j^\star}} \, - \,
  \frac{\partial \boldsymbol{\mathit{g}} }{\partial z^i}
  \,\frac{\partial \boldsymbol{\mathit{f}} }{\partial \bar{z}^{j^\star}}\right) 
  \end{equation}
and 
\begin{equation}\label{cartolina}
  g_{ij^\star}\, \equiv  \, \frac{\partial^{2} }{\partial z^i\,\partial \bar{z}^{j^\star}}\, \mathcal{K}(z,\bar{z}) \quad ; \quad  g^{ij^\star} \, = \, 
   \left(g^{-1}\right)^{ij^\star}
\end{equation}
are the K\"ahler metric and its inverse.
\par
From the Hamiltonian viewpoint, a system is said to be {\bf Liouville integrable}, or {\bf exactly solvable},  if one can find $n$ functions $\mathfrak{p}_a(z,\bar{z})$ that satisfy the first of  (\ref{PB1}) and which include the Hamiltonian of the system itself. In this case, $\mathfrak{p}_a(z,\bar{z})$, named \textbf{actions}, represent a complete set of Hamiltonians in involution and, as such, are constants of motion. The definition of a completely canonically integrable system also requires the level manifold of first integrals $\boldsymbol{\mathcal{Q}}_n\subset \Omega$, spanned by the canonically dual coordinates $\mathfrak{q}^a(z,\bar{z})$, to be compact.  By Arnol'd Theorem, $\mathcal{Q}_n$ is then diffeomorphic to an n-torus: $\boldsymbol{\mathcal{Q}}_n\sim \boldsymbol{\mathcal{T}}^n$:
\begin{equation}\label{ntorus}
  \boldsymbol{\mathcal{T}}^{n} \equiv \underbrace{\mathbb{S}^1\times ...\times \mathbb{S}^1}_{n-times}
\end{equation}
However, the existence of n integrals of motion in involution does not require $\boldsymbol{\mathcal{Q}}_n$, spanned by the $\mathfrak{q}^a(z,\bar{z})$ to be compact. With a slight misuse of terminology,  the definition of Liouville integrable systems is extended to these situations. A particularly relevant completely opposite case occurs when $\boldsymbol{\mathcal{Q}}^{n}\simeq \mathbb{R}^n$ and the K\"ahler potential does not depend on the variables  $\mathfrak{q}^a(z,\bar{z})$ that can be identified with the real parts of the complex coordinates in a suitable complex basis: 
\begin{equation}\label{fragerolamo}
  \mathfrak{q}^a(z,\bar{z}) \, = \, \ft 12 \, (z^a+\bar{z}^{a^\star}) \equiv u^a
\end{equation}
It follows that $\mathcal{K}(z,\bar{z})$ depends only on the imaginary parts:
\begin{equation}\label{cortinadilegno}
  \mathcal{K}(z,\bar{z}) \, = \, \mathcal{K}(\boldsymbol{v}) \quad ; \quad v^a \, - \, \ft 12 \, \mathit{i} \, (z^a -\bar{z}^{a^\star}) \quad a=1\,\dots,n  
\end{equation}
and the manifold $\Omega$ admits $n$ translational abelian isometries. 
\par
As shown in \cite{secondtemperature} and recalled above in section \ref{certosino}, this is the case of the K\"ahler metric of the most general macroscopic thermodynamic space $\mho$, the abelian non-compact isometries corresponding to generic shifts in the average values of the observable functions defined over the microscopic event manifold $\Omega$, irrespective of whether the latter is also K\"ahlerian or not and irrespective of what such observable functions are.
\par
Noticeably the same is also true for all \textbf{Homogeneous Special K\"ahler Manifolds} among which we find the \textbf{Non-Compact Symmetric Spaces of K\"ahler type} that we can utilize as mathematical models of the layers in Cartan Neural Networks.
This happens because on the metric--equivalent solvable groups $\mathcal{S}_{\mathfrak{hsk}}$ (see \cite{pgtstheory}), one constructs the relevant K\"ahler metric  introducing a K\"ahler potential of the type described in eq.(\ref{cortinadilegno}) with the specific form:
\begin{equation}\label{cubicone}
  \mathcal{K}(\boldsymbol{v}) \, = \, - \, \log \left(d_{abc} \,v^a\,v^b \,v^c\right)
\end{equation}
where $d_{abc}$ is a constant symmetric tensor with three indices. Indeed the classification of homogeneous special K\"ahler geometries amounts to a classification of the cubic forms defined by the $3$-tensor $d_{abc}$ and endowed with specific necessary properties singled out in the 1980s and early 1990s \cite{deWit:1995tf,toineugenio,SKGaggio3,SKGaggio2,SKGaggio1,specHomgeoA2,specHomgeoA1}.
The translational isometries encoded in this general structure of special K\"ahler manifolds correspond, at the level of their metric--equivalent solvable group $\mathcal{S}_{\mathfrak{hsk}}$ to a \textbf{maximal abelian ideal} $\mathcal{AI}_{max}\subset\mathcal{S}_{\mathfrak{hsk}}$ of dimension equal to one half the dimension of $\mathcal{S}_{\mathfrak{hsk}}$ which is possessed by all such solvable Lie groups. 
\par
In \cite{secondtemperature} we focused on the Calabi-Vesentini manifolds:
\begin{equation}\label{Cvmanigoldi}
  \mathcal{M}_{CV}^{[2,q]}\,  \equiv \, \frac{\mathrm{SO(2,2+q)}}{\mathrm{SO(2)\times SO(2+q)}} 
\end{equation}
each of which, at all time,  must be regarded as one of the two cofactors constituting an item in following infinite series of Special K\"ahler manifolds:
\begin{equation}\label{speckal}
 \mathcal{SK}_{3+q} \, \equiv \, \frac{\mathrm{SL(2,\mathbb{R})}}{\mathrm{SO(2)}}\times
 \frac{\mathrm{SO(2,2+q)}}{\mathrm{SO(2) \times SO(2+q)}}
\end{equation} 
\par
The important point is that in the Darboux coordinate basis, 
following from the construction of the abelian structure, the integral defining the partition function becomes explicitly computable in simple terms. 
\par
The most relevant result of \cite{secondtemperature}  is not limited to the explicit calculation of the partition function with Souriau temperatures in the co-adjoint representation of $\mathbb{U}$: the very architecture of the compact abelian structure (see all details and conceptual issues in \cite{secondtemperature}) suggests that one can include new generalized temperatures  associated with each one of the extra actions. This breaks the full symmetry under $\mathbb{U}$ leading to a sort of \textbf{phase transition pattern similar to that occurring in ferromagnetism}. The derivative of the stochastic Hamiltonian with respect to each of the extra temperatures yields a non-vanishing expectation value to the corresponding action which is a new observable function on the manifold of events. Such non-vanishing expectation value breaks the $\mathrm{U}$-symmetry to a smaller subgroup and it is the statistical analogue of spontaneous magnetization in ferromagnets. 
\par
\subsection{A warning}
However, an important \underline{\textbf{warning}}, should be immediately anticipated. One should not adopt  the \textbf{naive view point} that the \textit{Group of Isometries}
of the \textit{Thermo-Metric}  is the same as $\mathrm{U}$ or one of its subgroups. \textbf{Actually things are much more complicated} and work, as we are going to show below, in a \textbf{counter-intuitive reversed direction}. The case of Gibbs distributions with fully broken $\mathrm{U}$-symmetry corresponds to a Thermo-Metric that has the maximal isometry group; the partial restoration of some of the $\mathrm{U}$-isometries in the Gibbs distribution, is obtained at the price of breaking some of the isometries of the Thermo-Metric and this corresponds to the generation of a progressively more complex Riemann tensor of the latter with highly non trivial sectional curvatures whose maxima, minima or singularities mark the transition walls among different phases. 
\section{Generalized Souriau temperatures and partition functions for CV
manifolds}
\label{sashasecta}
In this section, we briefly review the results of \cite{secondtemperature} for the \textbf{abelian structures} on CV manifolds that include, as starting point, the moment-maps of their compact Cartan subalgebra $\boldsymbol{\mathcal{C}} \subset \mathbb{H} \subset \mathbb{U}$ and we summarize the simple result for the 
exact partition functions.
\subsection{The Souriau Partition Functions for CV manifolds}
A statistical probability distribution defined over a CV space that incorporates the large symmetry $\mathrm{U}$ of its K\"ahler metric, is of the general Gibbs type as defined in eq.(\ref{ciacolone}), where the observable functions $X(q)$ measured on the CV microscopic manifold are chosen to be the moment maps of the Killing vectors generating the $\mathrm{U}$-isometries, as we recalled above.
\par
Hence assuming:
\begin{description}
  \item[A)] that the generalized temperature vector $\boldsymbol{\beta}$ is such that the partition function converges,
  \item[B)] that the generalized temperature vector $\boldsymbol{\beta}$ lies in the $\mathrm{U}$-adjoint orbit\footnote{We do not distinguish among adjoint and coadjoint orbits since the Lie algebras we utilize are simple and the Killing metric is not degenerate (altrough of indefinite signature) and furthermore always standardized to be diagonal.} of the compact subalgebra $\mathbb{H}\subset \mathbb{U}$, 
\end{description}
we can always rotate the temperature vector, by an adjoint transformation to the Cartan subalgebra of the compact subalgebra $\boldsymbol{\mathcal{C}}\subset \mathbb{H}$. Then, after this formal reduction to $\boldsymbol{\mathcal{C}}$:
\begin{equation}\label{reduzia}
 \mathbb{U} \ni\boldsymbol{\beta} \, \stackrel{\text{Adj(g)}}{\Longrightarrow} \boldsymbol{\beta}_c \, \in \, \boldsymbol{\mathcal{C}} \subset \mathbb{H} \subset \mathbb{U}
\end{equation}
we can study the conditions on $\boldsymbol{\beta}_c$ for the integral convergence and the possible $\boldsymbol{\beta} \in \mathbb{U}$ guaranteeing convergence will be those that are in the adjoint orbit of the $\boldsymbol{\beta}_c$ satisfying the found conditions.
The key to obtain the general result later displayed in eq.s(\ref{zolotayaformula}) is the completion of the Abelian Structure programme for which we refer the reader to \cite{secondtemperature}. 
\par
The real dimension of the microscopic CV manifold is always $2n$ where $n=2+q$ but we have to distinguish the two cases $q=\mathrm{odd}$ and $q=\mathrm{even}$ since this decides whether the compact Lie algebra $\so(2+q)$ belongs to the $\boldsymbol{\mathfrak{b}}$-series or to the $\boldsymbol{\mathfrak{d}}$-series.
We introduce therefore the following notations:
\paragraph{Case $\boldsymbol{\mathfrak{b}}$:  $q=2\nu+1$, $\nu\in \mathbb{N}$; $n=3+2\nu$}
\begin{alignat}{8}\label{pagnaccob}
  &\boldsymbol{\lambda}_c^a  &\quad = \quad & \left\{\underbrace{\boldsymbol{\beta}_c^i}_{i=0,1,\dots, \nu+1} \right.\, , \,\left. \underbrace{\boldsymbol{h}^j}_{j=1,\dots,\nu+1}\right\} &\quad ; \quad & (a=1,\dots ,2\nu +3=n) \nonumber \\
  &\boldsymbol{\mathfrak{p}}_a(\boldsymbol{\Upsilon}) &\quad = \quad & \left\{\underbrace{{\mathcal{P}}_{c|i}(\boldsymbol{\Upsilon})}_{i=0,1,\dots, \nu+1} \right.\, , \,\left. \underbrace{\sqrt{\boldsymbol{\mathfrak{C}}^j(\boldsymbol{\Upsilon})}}_{j=1,\dots,
  \nu+1}\right\}&\quad ; \quad & (a=1,\dots ,2\nu +3=n)
\end{alignat}
and
\begin{alignat}{8}\label{pagnacco2b}
  &\boldsymbol{\mathfrak{q}}^a(\boldsymbol{\Upsilon}) \quad &\quad = \quad & \left\{\underbrace{\boldsymbol{\mathfrak{\theta}}^i(\boldsymbol{\Upsilon})}_{i=0,
  1,\dots, \nu+1} \right.\, , \,\left. \underbrace{\boldsymbol{\mathfrak{\psi}}^j(\boldsymbol{\Upsilon})}_{j=1,\dots,\nu+1}\right\}
  &\quad ; \quad & (a=1,\dots ,2\nu +3=n)
\end{alignat}
\paragraph{Case $\boldsymbol{\mathfrak{d}}$:  $q=2s$, $s\in \mathbb{N}$; $n=2+2s$}
\begin{alignat}{8}\label{pagnaccod}
  &\boldsymbol{\lambda}_c^a  &\quad = \quad & \left\{\underbrace{\boldsymbol{\beta}_c^i}_{i=0,1,\dots, s+1} \right.\, , \,\left. \underbrace{\boldsymbol{h}^j}_{j=1,\dots,s}\right\} &\quad ; \quad & (a=1,\dots ,2s +2=n) \nonumber \\
  &\boldsymbol{\mathfrak{p}}_a(\boldsymbol{\Upsilon}) &\quad = \quad & \left\{\underbrace{\mathcal{P}_{c|i}(\boldsymbol{\Upsilon})}_{i=0,1,\dots, s} \right.\, , \,\left. \underbrace{\sqrt{\boldsymbol{\mathfrak{C}}_j(\boldsymbol{\Upsilon})}}_{j=1,\dots,
  s}\right\}&\quad ; \quad & (a=1,\dots ,2s +2=n)
\end{alignat}
and
\begin{alignat}{8}\label{pagnacco2d}
  &\boldsymbol{\mathfrak{q}}^a(\boldsymbol{\Upsilon}) \quad &\quad = \quad & \left\{\underbrace{\boldsymbol{\mathfrak{\theta}}^i(\boldsymbol{\Upsilon})}_{i=0,
  1,\dots, s+1} \right.\, , \,\left. \underbrace{\boldsymbol{\mathfrak{\psi}}^j(\boldsymbol{\Upsilon})}_{j=1,\dots,s}
  \right\}
  &\quad ; \quad & (a=1,\dots ,2s +2=n)
\end{alignat}
where the $\boldsymbol{\mathfrak{p}}_a(\boldsymbol{\Upsilon})$ are the action functions of the solvable coordinates and $\boldsymbol{\mathfrak{q}}^a(\boldsymbol{\Upsilon})$ the corresponding conjugate angle variables that reduce the K\"ahler $2$-form at the Darboux structure of eq. (\ref{cosimoprimo}). 
\par
As it was explained in \cite{secondtemperature}, $\mathcal{P}_i$ with $(i=1,\dots,2+\ft12 q(q+3)$ is our notation for the moment maps of all $\mathbb{H}$-generators. For brevity we denote instead $\mathcal{P}_{c|i}$ ($i=0,1,\dots,\mathfrak{r}+1$) the $2+\mathfrak{r}$ among them that are associated with the Cartan subalgebra of $\mathbb{H}\, = \,\so(2)+\so(2+q)$. Indeed we have set:
\begin{equation}\label{cardiretto}
  \mathfrak{r} \, \equiv \, \text{rank}\left[\so(q)\right] \, = \, \left\{\begin{array}{cc|ccc|c}
   = & \nu & q & = & \text{odd} &\text{Case $\boldsymbol{\mathfrak{b}}$}  \\
   \hline
    = & s & q & = & \text{even} &\text{Case $\boldsymbol{\mathfrak{d}}$}   \\
  \end{array}
  \right.                                                        \end{equation}
\par
The objects denoted by the symbol $\boldsymbol{\mathfrak{C}}_j(\boldsymbol{\Upsilon})$ in eq.s(\ref{pagnaccob}-\ref{pagnacco2d}) are the Casimirs of the 
stabilizer subalgebras of the compact Cartan generators. The rank of the algebra
$\so(2+q)$ is $\mathfrak{r}$ in both cases $q=2\nu+1$ and $q=2s$, so in both cases $\boldsymbol{\mathfrak{b}}$ and $\boldsymbol{\mathfrak{d}}$, the number of Cartan generators of $\mathbb{H}$ is $\mathfrak{r}+1$ and correspondingly there are $\mathfrak{r}+1$ stabilizer subalgebras with $\mathfrak{r}+1$ Casimirs. The question is why in the $\boldsymbol{\mathfrak{b}}$ case the number of utilized Casimirs is $\mathfrak{r}+1$, while in the $\boldsymbol{\mathfrak{d}}$ case, only $\mathfrak{r}$ Casimirs enter the game. The answer is extremely simple. As it
was shown in \cite{secondtemperature}, in the even case the sequential embedding of  subalgebras does not end not with an $\so(3)$ rather with a one-dimensional $\so(2)$. In other words with an algebra spanned by a single generator. Such generator that, by definition, commutes with the $\mathcal{C}_{i}$ Cartan generator is nothing else but the $\mathcal{C}_{\nu+1}$ Cartan generator whose moment map is already included in the list. Therefore the last Casimir action $\sqrt{\mathcal{P}_{{c|\mathfrak{r}+1}}^2}=\mathcal{P}_{{c|\mathfrak{r}+1}}$ is just a repetition and should not be counted twice.
\par
Once the abelian structure is found, the integral defining the partition function can be rewritten as:
\begin{eqnarray}\label{maggiorana}
  Z^{\boldsymbol{\mathfrak{b}},\boldsymbol{\mathfrak{d}}}
  (\boldsymbol{\lambda}) &=&\int_{\mathcal{P}_{\mathfrak{r}}^{\boldsymbol{\mathfrak{b}},\boldsymbol{\mathfrak{d}}}}\,
  \exp\left[-H_{\mathfrak{r}}^{\boldsymbol{\mathfrak{b}},\boldsymbol{\mathfrak{d}}}
  \left(\boldsymbol{\mathfrak{p}},\boldsymbol{\lambda}\right)\right]
  \mathrm{d}^n \boldsymbol{\mathfrak{p}} \times \int_{\boldsymbol{\mathcal{T}}^{n}} \,\mathrm{d}^n \boldsymbol{\mathfrak{q}} 
\end{eqnarray}
where $\boldsymbol{\mathcal{T}}^n$ is the $n$-torus advocated by Arnol'd theorem and spanned by the angles $\boldsymbol{\theta}^i(\boldsymbol{\Upsilon})$ and $\boldsymbol{\psi}^j(\boldsymbol{\Upsilon})$ while the argument of the exponential is the generalized Hamiltonian displayed below
\begin{equation} \label{eq:extended_Hamiltonian}
H_{\mathfrak{r}}^{\boldsymbol{\mathfrak{b}},\boldsymbol{\mathfrak{d}}} \left(\boldsymbol{\lambda}\right) \,=\, 
\left\{
\begin{array}{ccc}
               H_{\mathfrak{r}}^{\boldsymbol{\mathfrak{b}}}\left(\boldsymbol{\mathfrak{p}},
               \boldsymbol{\lambda}\right)
                & = & \sum_{a=1}^{2\nu+3}
\boldsymbol{\lambda}^a \, \boldsymbol{\mathfrak{p}}_a(\boldsymbol{\boldsymbol{\Upsilon}}) \\
                H_{\mathfrak{r}}^{\boldsymbol{\mathfrak{d}}}
                \left(\boldsymbol{\mathfrak{p}},\boldsymbol{\lambda}\right) & = & \sum_{a=1}^{2s+2}
\boldsymbol{\lambda}^a \, \boldsymbol{\mathfrak{p}}_a(\boldsymbol{\boldsymbol{\Upsilon}}) \\
             \end{array}
\right.
\end{equation}
the Liouville integration measure:
\begin{equation}\label{cagnone}
  \mathrm{d}\boldsymbol{\mu} \, = \, \boldsymbol{\mathcal{K}}\wedge \, \dots \wedge \boldsymbol{\mathcal{K}}\, \simeq \, \text{const} \times \mathrm{d}^n \boldsymbol{\mathfrak{p}} \, \times \,\mathrm{d}^n \boldsymbol{\mathfrak{q}} 
\end{equation}
is flat both in the angle and in the action space. 
Before proceeding we need to stress a very important point. The actions $\boldsymbol{\mathfrak{p}}_a$ and the angles $\boldsymbol{\mathfrak{q}}^a$, ($a=1,\dots,n$) can be regarded as a new coordinate frame for the entire manifold, yet one should stress that there is a fundamental difference between the angles $\boldsymbol{\theta}^i$ and the angles $\boldsymbol{\psi}^j$. The first are cyclic coordinates from which the metric coefficient cannot depend since they are the $\mathfrak{r}+1$ parameters of a subgroup $\mathrm{U(1)}^{\mathfrak{r}+1} \subset \mathrm{U}$ of the isometry subgroup.
On the other hand the second set of angles are the parameters of a $\mathrm{U(1)}^{\mathfrak{r}+1}$ or $\mathrm{U(1)}^{\mathfrak{r}}$ group of transformations in the symplectic manifold that are not isometries of its K\"ahler metric. Hence the 
$\boldsymbol{\psi}^j$ are not cyclic variables and the metric coefficients should explicitly depend on them. 
This fundamental property allows the partition functions to be evaluated as an integral over the action polytope $\mathcal{P}_{\mathfrak{r}}^{\boldsymbol{\mathfrak{b}},\boldsymbol{\mathfrak{d}}}$.
The evaluation of the integral (\ref{maggiorana}) is performed through a recursive algorithm that reflects the sub-algebra inclusion chain described in section \ref{cannastorta}.
\par
The partition functions (\ref{maggiorana}) depending on the extended temperature vector $\boldsymbol{\lambda}$ are the normalization denominators of Gibbs distributions of extended Souriau type as it follows:
\begin{eqnarray} \label{extendedGibbs}
\mathbf{G}^{\boldsymbol{\mathfrak{b}}}_\nu\left(\boldsymbol{\lambda},
\boldsymbol{\Upsilon}\right)\, & =&\frac{1}{ Z^{\boldsymbol{\mathfrak{b}}}(\boldsymbol{\lambda})} \times
\exp\left[- \,\beta^0 \mathcal{P}_{c|0}(\boldsymbol{\Upsilon})- \sum_{i=1}^{\nu+1} \beta^i \mathcal{P}_{c|i}(\boldsymbol{\boldsymbol{\Upsilon}}) - \sum_{j=1}^{\nu+1} 
h^j \sqrt{\boldsymbol{\mathfrak{C}}_j(\boldsymbol{\boldsymbol{\Upsilon}})}\right]\nonumber\\
\mathbf{G}^{\boldsymbol{\mathfrak{d}}}_\nu\left(\boldsymbol{\lambda},\boldsymbol{\Upsilon}\right)\, & =&\frac{1}{ Z^{\boldsymbol{\mathfrak{d}}}(\boldsymbol{\lambda})} \times
\exp\left[-  \,\beta^0 \mathcal{P}_{c|0}(\boldsymbol{\Upsilon})- \sum_{i=1}^{s+1} \beta^i \mathcal{P}_{c|i}(\boldsymbol{\boldsymbol{\Upsilon}})- \sum_{j=1}^{s} 
h^j \sqrt{\boldsymbol{\mathfrak{C}}_j(\boldsymbol{\boldsymbol{\Upsilon}})}\right]
\end{eqnarray}
Eq. (\ref{extendedGibbs}) becomes meaningful as soon the additional $\boldsymbol{\mathfrak{C}}_j(\boldsymbol{\Upsilon})$ functions are determined. 
\subsection{The extra actions completing the abelian structure}
\label{cannastorta}
We finally review the determination, obtained in \cite{secondtemperature}, of the additional actions that complete the Abelian Structure which includes the moment maps
of the compact Cartan subalgebra:
\begin{equation}\label{cartolone}
  \mathbb{H} \, = \, \uu(1) \oplus \mathbb{H}^\prime \, = \, \so(2) \oplus \so(2+q)
\end{equation}
As it was shown in \cite{secondtemperature}, by means of an algorithmic construction of a carefully sequentially arranged basis of compact Cartan generators $X_c,\mathcal{C}_{i=1,..,\mathfrak{r}-1}$ (see section 8.2 of \cite{secondtemperature}) one obtains an embedding chain of principal subalgebras and hence of subgroups that are defined by the very choice of the Cartan generators of the compact subalgebra. The two series are the following ones 
\paragraph{The $\boldsymbol{\mathfrak{b}}_\nu$ embedding chain}
\begin{alignat}{9}\label{gianni1}
&\quad\quad\mathbb{H}_1 &\supset &\quad\quad\mathbb{H}_2 &\supset &\quad\quad\mathbb{H}_3&\supset &\dots &\supset &\quad\mathbb{H}_{\nu+1} \nonumber\\
&\,\mathfrak{so}(2\nu+3) &\supset &\,\mathfrak{so}(2\nu+1) &\supset &\,\mathfrak{so}(2\nu-1)&\supset &\dots &\supset &\quad\mathfrak{so}(3) 
\end{alignat}
\paragraph{The $\boldsymbol{\mathfrak{d}}_\nu$ embedding chain}
\begin{alignat}{10}\label{gianni2}
&\quad\quad\mathbb{H}_1 &\supset &\quad\mathbb{H}_2 &\supset &\quad\quad\mathbb{H}_3&\supset &\dots &\supset &\quad\mathbb{H}_{\nu}\,\,&\supset &\quad\mathbb{H}_{\nu+1} \nonumber\\
&\,\mathfrak{so}(2s+2) &\supset &\,\mathfrak{so}(2s) &\supset &\,\mathfrak{so}(2s-2)&\supset &\dots &\supset &\,\mathfrak{so}(4)&\supset &\quad\mathfrak{so}(2) 
\end{alignat}
and they are intrinsically defined, without the use of any explicit matrix index, in any specific coordinate basis by the following intrinsic rule:
\begin{equation}\label{carambolina}
  \mathbb{H}_i\,= \, \mathrm{N}[\mathcal{C}_{i-1},\mathbb{H}_{i-1}] \, - \,\{\mathcal{C}_{i-1}\}
\end{equation}
where $\mathrm{N}[\mathcal{C}_{i-1},\mathbb{H}_{i-1}]$ denotes the normalizer of the Cartan generator $\mathcal{C}_{i-1}$ in the previous subalgebra:
\begin{equation}\label{craniotto}
  \mathrm{N}[\mathcal{C}_{i-1},\mathbb{H}_{i-1}] \, = \,
  \left\{Y\in\mathbb{H}_{i-1} \, \mid \, \left[\mathcal{C}_{i-1}\, , \, Y \right]\, = \, 0  \right\}
\end{equation}
and the subtraction in eq.(\ref{carambolina}) is in the set-theoretic sense. We have already remarked that in the case of the $\boldsymbol{\mathfrak{d}}$-series the last group in the sequence is $\so(2)$ made of a single generator that is the already counted last Cartan generator.
\par
\par
Given the embedding series described above and the complete set of $\mathbb{U}$-generators $J_\Lambda$, normalized as shown in \cite{secondtemperature} one relies on the following formula:
\begin{eqnarray}\label{gelindoelapecora}
  \mathfrak{P} & : & \mathbb{U} \, \longrightarrow \, \mathbb{C}^{\infty}\left(\frac{\mathrm{U}}{\mathrm{H}}\right)\nonumber\\
  \mathfrak{P}_A\left(\boldsymbol{\Upsilon}\right) &=& \ft 12 \, \text{Tr} \left[ X_c\cdot \mathbb{L}^{-1}
  \left(\boldsymbol{\Upsilon}\right)\cdot J_A \cdot \mathbb{L}
  \left(\boldsymbol{\Upsilon}\right)\right]
\end{eqnarray}
 that defines the basic moment-maps of the $\mathbb{U}$ Lie algebra, the extra actions completing the abelian structure are obtained in terms of the Casimirs of the subalgebras $\mathbb{H}_j$, namely:
and one obtains the extra actions completing the abelian structure by setting
\begin{equation}\label{gulini2}
 \sqrt{ \boldsymbol{\mathfrak{C}}_{j}(\boldsymbol{\Upsilon})}  \,\equiv \, \sqrt{\sum_{J_\Lambda\in\mathbb{H}_j} \mathcal{P}^2_\Lambda(\boldsymbol{\Upsilon})}
 \quad ; \quad \quad j=\left\{\begin{array}{cccc}
                                1, & \dots , & \nu+1 & \text{case $\boldsymbol{\mathfrak{b}}$} \\
                                1, & \dots , & s & \text{case $\boldsymbol{\mathfrak{d}}$}
                              \end{array}\right.
\end{equation}
Indeed by the very definition of the chosen construction all the introduced actions 
are in involution as they should, namely they commute with each other in the Poisson bracket. 
\par
\subsubsection{Generalized Souriau Partition Functions}
The final outcome of the  integration yields the generalized Souriau partition functions firstly described in \cite{secondtemperature}.
In order to write their explicit form  as functions of  the components of the generalized temperature vector $\boldsymbol{\lambda}$ it is convenient to introduce the following auxiliary variables 
\begin{eqnarray} \label{eq:Omega_def_analysis}
\mu_k :&=& \beta_0 + \sum_{j=k}^{\nu+1} h_j 
\quad k=1, \dots, \nu+1 \quad \quad \,\, \, ; \quad  \text{case $\boldsymbol{\mathfrak{b}}$} \nonumber\\
\hat{\mu}_k :&=& \beta_0 + \sum_{j=k}^{s} h_j \,, \quad \hat{\mu}_0 := \hat{\mu}_1\,,
\quad k=1, \dots, s \quad ; \quad \text{case $\boldsymbol{\mathfrak{d}}$} 
\end{eqnarray}
\paragraph{Partition function for the $\boldsymbol{\mathfrak{b}}$-series.} In this case the partition functions with the extended temperature vector is the following one $\boldsymbol{\lambda}$
\begin{equation} \label{eq:final_Z_odd_sequential}
Z_{\nu}^{\boldsymbol{\mathfrak{b}}}(\boldsymbol{\lambda}) =  \frac{2^{\nu+1} (2\pi)^{2\nu+3} e^{-\beta_0} }{\beta_0\prod_{k=1}^{\nu+1} [\mu_k^2 - \beta_k^2]}
\end{equation}
\par
The largest possible domain of the parameters $\boldsymbol{\lambda}$ for which the partition function $Z_{\nu}^{\boldsymbol{\mathfrak{b}}}(\boldsymbol{\lambda})$ is convergent and positive definite is the intersection of $\nu+2$  linear half-spaces:
\begin{equation} \label{eq:D_Theta_Odd}
D^{\boldsymbol{\mathfrak{b}}}_{\nu}(\boldsymbol{\lambda}) = \left\{ (\boldsymbol{\beta}, \boldsymbol{h}) \in \mathbb{R}^{2\nu+3} \mid \beta_0 > 0, \quad \mu_k > |\beta_k| \text{ for } k=1, \dots, \nu+1 \right\}
\end{equation}
\paragraph{Partition function for the $\boldsymbol{\mathfrak{d}}$-series.}
Integrating on the appropriate  polytopes one obtainsIntegrating on the appropriate  polytopes one obtains\footnote{In the case q=even we use $q=2s$ ($s\in \mathbb{N}$). This allows to distinguish the even and odd case visually.}: 
\begin{equation} \label{eq:final_Z_even_sequential}
    Z_{s}^{\boldsymbol{\mathfrak{d}}}(\boldsymbol{\mathfrak{\lambda}}) = \frac{2^{s+1} (2\pi)^{2(s+1)}  \hat \mu_1 e^{-\beta_0}}{\beta_0 \prod_{k=1}^{s+1} [\hat \mu_{k-1}^2 - \beta_k^2]}
\end{equation}
The largest possible domain $D_{\boldsymbol{\mathfrak{d}}} \subset \mathbb{R}^{2\nu+1}$ of the parameters $\boldsymbol{\mathfrak{d}}$ for which the partition function $Z_{\nu}^{\boldsymbol{\mathfrak{d}}}(\boldsymbol{\mathfrak{\lambda}}) $ is convergent and positive definite is the intersection of the following linear half-spaces:
\begin{equation} \label{eq:D_Theta_Even}
\hspace{-0.2cm} \hat D_{\boldsymbol{\mathfrak{d}}} = \left\{ (\boldsymbol{\beta}, \boldsymbol{h}) \in \mathbb{R}^{2\nu+1} \mid \beta_0 > 0, \quad \hat \mu_1 > \max(|\beta_1|, |\beta_2|), \quad \mu_{k-1} > |\beta_k| \text{ for } k=2, \dots, \nu \right\}
\end{equation}
\subsubsection{The Adjoint $\mathrm{U}$-orbit limit of the compact Cartan subalgebra (Souriau)}
By setting to zero all the extra temperatures, namely the magnetic fields associated with the Casimirs of the principal subalgebras, \textit{i.e.} setting ${h}^i = 0$ in $ Z_{\nu}^{\boldsymbol{\mathfrak{b}},\boldsymbol{\mathfrak{d}}}(\boldsymbol{\mathfrak{\lambda}})$ we find that $\hat{\mu}_k = \beta_0$ or $\mu_{k}=\beta_0$ for all values of $k$ and we find the Souriau partition functions:
{\large
\begin{eqnarray}
\label{zolotayaformula}
 Z^{\mathfrak{b},\mathfrak{d}}\left(\beta_0,\beta_{i=1,\dots,1
 +\nu}\right) &=& 
c^{\mathfrak{b},\mathfrak{d}} \frac{ (8\pi^2)^{\nu+1} e^{-\beta_0}}{\prod_{i=1}^{\nu+1} (\beta_0^2 - \beta_i^2)}  \\
                   \nonumber\\
  c^{\mathfrak{b}} &=& \frac{2\pi}{\beta_0} \nonumber\\
  c^{\mathfrak{d}} &=& 1 \nonumber
\end{eqnarray}
}
\noindent with a small difference in the denominator between the odd-dimensional and even dimensional case of $q$ which, from the Lie algebra theory point of view, corresponds to the distinction between the series $\mathfrak{b}_{\nu}$ and $\mathfrak{d}_{\nu}$. 
\section{The Thermo-Metric of CV manifolds} 
As we anticipated in the introduction, the study of the K\"ahler Thermo-Metric associated with CV event manifolds introduces several surprises, all encoded in the simple, yet far from trivial, explicit form of the partition functions (\ref{eq:final_Z_odd_sequential}),(\ref{eq:final_Z_even_sequential}), the first of which is a substantial structural difference between the odd and the even $q$ cases. Such a far-reaching structural difference 
follows from the apparently innocent $\hat{\mu}_1$ factor appearing in the numerator of eq.(\ref{eq:final_Z_even_sequential}) and in the equality between $\hat{\mu}_0$ and $\hat{\mu}_1$. Because of this we analyse first the odd case which is simpler. Yet, as we are going to see in section \ref{evencrat}, once the three directions associated with the variables $\beta_1,\beta_2,\hat{\mu}_1$ have been separated, the remaining part of the Thermo--Metric for the even case, associated with the
variables, $\beta_{i}$ ($i>2$), $\hat{\mu}_j$ ($j>1$) is formally identical with the corresponding part ($\beta_{i}$ ($i>2$), ${\mu}_j$ ($j>2$)) of the Thermo--Metric of the odd case upon the identification $\hat{\mu}_j \simeq \mu_{j+1}$. This implies that the non-trivial curvature--deformed submanifolds that arise in the odd case when magnetic fields $h_i$ ($i>1$) are switched off according to various patterns are actually universal and arise identically also in the even case just
\textit{mutatis mutandis}.  
\subsection{The CV Thermo--Metric in the $q=\text{odd}$ case}\label{oddcrat}
The Thermo-Metric is easily obtained by calculating the Hessian matrix of the stochastic Hamiltonian:
\begin{eqnarray}\label{cardinalodd}
ds^2_{\nu} & \equiv &\, - \,\,\mathrm{d}\lambda^i \, \mathrm{d}\lambda^j \, \frac{\partial^2}{\partial\lambda^i \, \partial\lambda^j}\,\mathcal{H}^{sto}_{\nu}(\boldsymbol{\lambda}) \quad ; \quad i,j=1,\dots, 2\nu +3 \nonumber\\
 \mathcal{H}^{sto}_{\nu}(\boldsymbol{\lambda})& \equiv & - \,\log\left[Z_{\nu}^{\boldsymbol{\mathfrak{b}}}(\boldsymbol{\lambda})\right]\nonumber\\
 \boldsymbol{\lambda}& = & \left\{\beta_0,\,\beta_1,\, \dots , \beta_{\nu+1}, \, h^1,\, \dots, h^{\nu+1} \right\}
\end{eqnarray}
where $Z_{\nu}^{\boldsymbol{\mathfrak{b}}}(\boldsymbol{\lambda})$ is defined by eq.(\ref{eq:final_Z_odd_sequential}) and then by rewriting every thing in terms of the variables $\beta_k$ and
$\mu_k$. With trivial algebra one obtains:
\begin{eqnarray}
ds^2_{\nu} &=& \,\frac{\mathrm{d}\beta_0^2}{\beta_0^2} \, + \, \sum_{i=1}^{\nu+1}\,\frac{1}{\left(\beta_i^2-\mu^2_i\right)^2}\,
\left[2\,\left(\beta_i^2+\mu^2_i\right)\,(d\beta_i^2 + d\mu_i^2)\, - \, 8 \beta_i \, \mu_i \,d\beta_i \,d\mu_i\right] \label{distaquadodd}
\end{eqnarray}
If one calculates the Riemann tensor of the above $(2\nu+3)$-dimensional metric one finds that it vanishes identically, so that such a Riemann manifold is flat. Indeed there is a simple coordinate transformation that puts the metric 
(\ref{distaquadodd}) into the form of a standard Euclidean $\mathbb{R}^{2\nu+3}$ metric. Such transformation is the following one:
\begin{equation}\label{roncoscrivia}
  \beta_0 \, = \, \exp[\rho_0] \quad ; \quad \beta_i \, = \, \exp[\rho_i] \, - \, \exp[\rho_{\nu+1+i}]\quad ; \quad \mu_i \, = \, \exp[\rho_i] \, + \, \exp[\rho_{\nu+1+i}] \quad : \quad i=1,\dots,\nu+1
\end{equation}
Indeed upon use of the transformation (\ref{roncoscrivia}) in eq.(\ref{distaquadodd}), we obtain:
\begin{equation}\label{crullone}
  ds^2_{\nu}\, = \,  \mathrm{d}\rho_0^2\, + \, \sum_{a=1}^{2\nu+2}\mathrm{d}\rho_a^2
\end{equation}
At the same time the magnetizations, namely the average values of the square roots of the Casimirs, have the following form:
\begin{eqnarray}\label{brontine}
  \langle \sqrt{ \boldsymbol{\mathfrak{C}}_{k}(\boldsymbol{\Upsilon})} \rangle & = &\frac{\partial \mathcal{H}^{sto}_{\nu}}{\partial h^k} \, = \,\sum _{i=1}^k \frac{2 \mu _i}{\mu _i^2-\beta _i^2}\nonumber\\
  &=&\ft 12 \,\sum _{i=1}^k \left(\exp[- \rho_i] -\exp[-\rho_{\nu+1+i}]\right)
\end{eqnarray}
As one sees from its transformation into (\ref{crullone}), the flat Thermo-Metric (\ref{distaquadodd}) of the macroscopic manifold $\mho$ has a very large group of isometries, namely the Euclidean group $\mathbb{E}^{2\nu+3}=\mathrm{ISO}(2\nu+3)$, while the generic configuration of mean values of the actions:
\begin{equation}\label{meanvalori}
  x_a \, \equiv \, \langle \boldsymbol{\mathfrak{p}}_a\left(\boldsymbol{\Upsilon}\right)\rangle
  \,= \, \frac{\partial \mathcal{H}^{sto}_{\nu}}{\partial\lambda^a} 
\end{equation}
breaks completely the $\mathrm{U}$-isometry of the microscopic
manifold $\Omega$. 
\subsubsection{Spontaneous symmetry breaking}
On the other hand, as proven in \cite{geotermico,secondtemperature} if we set all the magnetic fields to zero $h^i\rightarrow 0$, the partition function becomes $\mathrm{U}$-invariant and the average values of the actions should fall into a  co-adjoint $\mathrm{U}$-orbit. This happens if we take first the Souriau limit of eq.(\ref{zolotayaformula}) and then we take the derivatives with respect to the $\boldsymbol{\beta}$-temperatures. Yet if we first compute the magnetizations (\ref{brontine}) and then we perform the limit
$h^i\rightarrow 0$, things go differently. Indeed we find:
\begin{equation}\label{carnevaleviareggio}
  \langle \sqrt{ \boldsymbol{\mathfrak{C}}_{k}(\boldsymbol{\Upsilon})} \rangle_{\boldsymbol{h}=0} \, = \, \sum _{i=1}^k \frac{2 \beta_0}{\beta_0^2-\beta _i^2} \, \neq \, 0
\end{equation}
and the $(2\nu+3)$-vector of mean action values:
\begin{eqnarray}\label{freshka}
 && x^{[0]}_a \, = \, \langle \boldsymbol{\mathfrak{p}}_a\left(\boldsymbol{\Upsilon}\right)\rangle_{\boldsymbol{h}=0}
  \, = \,\nonumber\\
  && \left\{ \underbrace{1+\frac{1}{\beta_0}+2\sum_{i=1}^{\nu+1}\frac{\beta_0}{\beta_0^2-\beta_i^2},\,
  \frac{2\,\beta_1}{\beta_1^2-\beta_0^2},\, \dots,\,\frac{2\,\beta_{\nu+1}}{\beta_{\nu+1}^2-\beta_0^2}}_{\text{mean values of $\mathcal{P}_{c|i}$}},\, \mid\,\underbrace{\frac{2\,\beta_0}{\beta_0^2-\beta_1^2},\,\dots,\sum _{i=1}^k \frac{2 \beta_0}{\beta_0^2-\beta _i^2},\, \dots, \,\sum _{i=1}^{\nu+1} \frac{2 \beta_0}{\beta_0^2-\beta _i^2}}_{\text{spontaneous magnetizations}}  \right\} \nonumber\\
\end{eqnarray}
does not fall into any $\mathrm{U}$-coadjoint orbit, although its first half does it and coincides with the calculation of the
mean values of the compact Cartan moment maps $\mathcal{P}_{c|i}(\boldsymbol{\Upsilon})$ starting from the $\mathrm{U}$-invariant partition functions of the Souriau limit (\ref{zolotayaformula}). Indeed in the Souriau limit the Gibbs distribution can be written in the following completely U-invariant form:
\begin{alignat}{4}\label{cerusico}
  & G(\boldsymbol{\beta},\boldsymbol{\Upsilon}) &\quad = \quad &\quad  \frac{\exp\left[-\boldsymbol{\beta}\cdot\boldsymbol{\mathfrak{P}}
  \left(\boldsymbol{\Upsilon} \right)\right]}{Z(\boldsymbol{\beta})}\nonumber\\
  & Z(\boldsymbol{\beta}) &\quad = \quad &\quad \int_{\mathrm{U/H}}
  \, \exp\left[-\boldsymbol{\beta}\cdot\boldsymbol{\mathfrak{P}}
  \left(\boldsymbol{\Upsilon} \right)\right] \, \boldsymbol{\mathrm{d}\mu}\left(\boldsymbol{\Upsilon}\right)
\end{alignat}
where $\boldsymbol{\mathfrak{P}}_A\left(\boldsymbol{\Upsilon} \right)$ are the
Killing vector moment maps defined below:
\begin{eqnarray}\label{gelindoelapecora}
  \mathfrak{P} & : & \mathbb{U} \, \longrightarrow \, \mathbb{C}^{\infty}\left(\frac{\mathrm{U}}{\mathrm{H}}\right)\nonumber\\
  \mathfrak{P}_A\left(\boldsymbol{\Upsilon}\right) &=& \ft 12 \, \text{Tr} \left[ X_c\cdot \mathbb{L}^{-1}
  \left(\boldsymbol{\Upsilon}\right)\cdot J_A \cdot \mathbb{L}
  \left(\boldsymbol{\Upsilon}\right)\right]
\end{eqnarray}
(see\cite{secondtemperature}) that provide 
a Poissonian realization of the full $\mathbb{U}$-Lie algebra:
\begin{equation}\label{quirito}
  \left \{\boldsymbol{\mathfrak{P}}_A\left(\boldsymbol{\Upsilon} \right) \, , \,
  \boldsymbol{\mathfrak{P}}_B\left(\boldsymbol{\Upsilon} \right) \right\} \, = \, 
  \mathit{f}_{AB}^{\phantom{AB}C} \, \boldsymbol{\mathfrak{P}}_C\left(\boldsymbol{\Upsilon} \right)
\end{equation}
The derivative with respect to $\beta_A$ ($A=1,\dots, \text{dim}\mathbb{U})$ of minus the logarithm of the U-invariant partition function $Z(\boldsymbol{\beta})$ displayed in eq.(\ref{cerusico}) provides the average value of the moment maps $\boldsymbol{\mathfrak{P}}_A\left(\boldsymbol{\Upsilon} \right) $:
\begin{equation}\label{geminato}
  x_A \, = \, \langle \mathfrak{P}_A\left(\boldsymbol{\Upsilon}\right) \rangle \, = \,  \frac{1}{Z(\boldsymbol{\beta})} \times \int_{\mathrm{U/H}}
  \, \mathfrak{P}_A\left(\boldsymbol{\Upsilon}\right)\, \exp\left[-\boldsymbol{\beta}\cdot\boldsymbol{\mathfrak{P}}
  \left(\boldsymbol{\Upsilon} \right)\right] \, \boldsymbol{\mathrm{d}\mu}\left(\boldsymbol{\Upsilon}\right)
\end{equation}
Considering the action of a generic U element $\mathfrak{g}\in \mathrm{U}$ in the adjoint representation $\text{Adj}[\mathfrak{g}]_A^{\phantom{A}B}$ we have:
\begin{eqnarray}
\label{crantello}
  \text{Adj}[\mathfrak{g}]_A^{\phantom{A}B}\, x_B  &=& \frac{1}{Z(\boldsymbol{\beta})} \times \int_{\mathrm{U/H}}
  \,  \text{Adj}[\mathfrak{g}]_A^{\phantom{A}B} \mathfrak{P}_B\left(\boldsymbol{\Upsilon}\right)\, \exp\left[-\boldsymbol{\beta}\cdot\boldsymbol{\mathfrak{P}}
  \left(\boldsymbol{\Upsilon} \right)\right] \, \boldsymbol{\mathrm{d}\mu}\left(\boldsymbol{\Upsilon}\right) 
\end{eqnarray}
Consider now the element $\mathfrak{g}_0 \in \mathrm{U}$ such that:
\begin{equation}\label{forbillone}
  \boldsymbol{\beta} \, = \, \boldsymbol{\beta}_0 \cdot \text{Adj}\left[\mathfrak{g}_0\right]
\end{equation}
where $\boldsymbol{\beta}_0 \in \boldsymbol{\mathcal{C}} \subset 
\mathbb{H} \subset \mathbb{U}$ lies in the compact Cartan subalgebra, all the other component vanishing. The existence of 
$\boldsymbol{\beta}_0 \in \boldsymbol{\mathcal{C}}$ and $\mathfrak{g}_0$ follows from the fundamental hypothesis that $\boldsymbol{\beta}$ should be in the adjoint orbit of the compact Cartan subalgebra. Utilizing the $\mathrm{U}$-invariance of the integration measure we get:
\begin{eqnarray}
\label{cartonato}
  x_A &=& \frac{1}{Z(\boldsymbol{\beta})} \times \int_{\mathrm{U/H}}
  \,  \mathfrak{P}_A\left(\boldsymbol{\Upsilon}\right)\, \exp\left[-\boldsymbol{\beta}_0\cdot\text{Adj}\left[\mathfrak{g}_0\right]\boldsymbol{\mathfrak{P}}
  \left(\boldsymbol{\Upsilon} \right)\right] \, \boldsymbol{\mathrm{d}\mu}\left(\boldsymbol{\Upsilon}\right)\nonumber  \\
  &=& \frac{1}{Z(\boldsymbol{\beta}_0)} \times \int_{\mathrm{U/H}}
  \,  \mathfrak{P}_A\left(\boldsymbol{\Upsilon}\right)\, \exp\left[-\boldsymbol{\beta}_0\cdot\boldsymbol{\mathfrak{P}}
  \left(\mathfrak{g}_0[\boldsymbol{\Upsilon}] \right)\right] \, \boldsymbol{\mathrm{d}\mu}\left(\mathfrak{g}_0[\boldsymbol{\Upsilon}]\right) \nonumber\\
   &=& \frac{1}{Z(\boldsymbol{\beta}_0)} \times \int_{\mathrm{U/H}}
  \,  \mathfrak{P}_A\left(\boldsymbol{\mathfrak{g}}_0^{-1}[\boldsymbol{\Upsilon}^\prime]\right)\, \exp\left[-\boldsymbol{\beta}_0\cdot\boldsymbol{\mathfrak{P}}
  \left(\boldsymbol{\Upsilon}^\prime \right)\right] \, \boldsymbol{\mathrm{d}\mu}\left(\boldsymbol{\Upsilon}^\prime\right) \nonumber\\
  &=&\text{Adj}\left[\mathfrak{g}_0^{-1}\right] \underbrace{ \frac{1}{Z(\boldsymbol{\beta}_0)} \times \int_{\mathrm{U/H}}
  \,  \boldsymbol{\mathcal{P}}_c\left({\Upsilon}^\prime\right)\, \exp\left[-\boldsymbol{\beta}_0\cdot\boldsymbol{\mathfrak{P}}
  \left(\boldsymbol{\Upsilon}^\prime \right)\right] \, \boldsymbol{\mathrm{d}\mu}\left(\boldsymbol{\Upsilon}^\prime\right)}_{\text{mean values of the compact moment maps}}
  \end{eqnarray}
  Hence in the Souriau limit the mean values of all the moment maps in the adjoint representation of $\mathbb{U}$ are obtained as the adjoint orbit of the mean values of the moment maps in the compact Cartan subalgebra, namely the first half of the $x^{[0]}$ vector in eq.(\ref{freshka}). The same cannot be said of the second half of the same vector, \textit{i.e.} the mean values of the square roots of the Casimir functions. The latter do no transform linearly under the action of full group $\mathrm{U}$ and accordingly we do not have a linear adjoint transformation of the associated magnetic fields.
\section{Curved submanifolds of the Thermo Manifold and the
magnetic combinatorial scheme}
Setting $r$ of the magnetic fields $h_i$ to zero corresponds, from the geometric point of view, to singling out in the Thermo--manifold a submanifold $\mathcal{M}_{r}\subset \mho$ of codimension $r$. The pull-back to such $\mathcal{M}_{r}$ of the flat $\mho$-metric produces a Riemannian metric that typically is no longer flat. It is interesting to analyse the resulting Riemannian space, its geometry and geodesics. The magnetic fields can be set to zero one by one or in pairs or in triples and so on. This allows us to define a complex sequential combinatorials. We begin by considering the vanishing of a single $h_j$. 
\subsection{Vanishing of one magnetic field and the universal 
three-space $\mathfrak{B}_{3|reg}$}
\label{unovanni}
If we switch off the j-th magnetic field $h_j\rightarrow 0$ 
we find that $\mu_j = \mu_{j+1} = w$, where $w$ is a new name for the identified coordinate. Let us also name $s_1=\beta_j$ and
$s_2=\beta_{j+1}$. Introducing these names and the identification in the metric $ds^2_{\nu}$ as written in eq.(\ref{distaquadodd}) we find:
\begin{eqnarray}\label{bacherozzo}
  ds^2_{\nu} & = & \,\frac{\mathrm{d}\beta_0^2}{\beta_0^2} \, + \, \sum_{i=1}^{j-1}\frac{1}{\left(\beta_i^2-\mu^2_i\right)^2}\,
\left[2\,\left(\beta_i^2+\mu^2_i\right)\,(d\beta_i^2 + d\mu_i^2)\, - \, 8 \beta_i \, \mu_i \,d\beta_i \,d\mu_i\right] \nonumber\\
  &&\, + \, \sum_{i=j+1}^{\nu+1}\frac{1}{\left(\beta_i^2-\mu^2_i\right)^2}\,
\left[2\,\left(\beta_i^2+\mu^2_i\right)\,(d\beta_i^2 + d\mu_i^2)\, - \, 8 \beta_i \, \mu_i \,d\beta_i \,d\mu_i\right]\nonumber\\
  &&\, + \, ds^2_{\mathfrak{B}_3}(s_1,s_2,w)
\end{eqnarray}
where
\begin{eqnarray}\label{olunitre}
 ds^2_{\mathfrak{B}_3}(s_1,s_2,w) &=&\mathrm{d}w^2
   \left(\frac{1}{(s_1+w)^2}+\frac{1}{(s_1-w)^2}+\frac{1}{(s_2-w)^
   2}+\frac{1}{(s_2+w)^2}\right)\nonumber\\
   &&- \frac{8 \,s_1\, w \,\mathrm{d}s_1\, \mathrm{d}w}{\left(s_1^2-w^2\right)^2}+\mathrm{d}s_1^2
   \left(\frac{1}{(s_1+w)^2}+\frac{1}{(s_1-w)^2}\right)-\frac{8\,
   s_2\, w \,\mathrm{d}s_2 \, \mathrm{d}w}{\left(s_2^2-w^2\right)^2}+\mathrm{d}s_2^2
   \left(\frac{1}{(s_2+w)^2}+\frac{1}{(s_2-w)^2}\right)\nonumber\\
\end{eqnarray}
Applying the transformation (\ref{roncoscrivia}) to $ds^2_{\nu}-ds^2_{uni3}$ we find a $(2\nu-1)$-dimensional flat metric spanned by all the previous $\rho$ coordinates except
$\rho_{j},\rho_{j+1},\rho_{\nu+1+j},\rho_{\nu+2+j}$. Let us rename these latter (in the same order) as:
\begin{equation}\label{frullatodipesca}
 \left\{\rho_{j},\,\rho_{j+1},\, \rho_{\nu+1+j},\,\rho_{\nu+2+j}\right\} \, \longrightarrow \, \left\{\xi_1, \, \xi_2\, \, \xi_3, \, \xi_4\right\}
\end{equation}
These changes of names are performed in order to show the universality of the mechanism and of the geometry that arises.
The four coordinates $\xi_i$ span a flat $\mathbb{E}_4 \simeq\mathbb{R}^4$ manifold with metric:
\begin{equation}\label{crocchetta}
  ds_{\mathbb{E}_4}^2 \, =\,\sum_{i=1}^4 d\xi_i^2
\end{equation}
This standard $\mathbb{E}_4$ flat  space is complementary to the residual $\mathbb{R}^{2\nu-1}$ flat space spanned  by the other $\rho$.s, and it is reduced, by fixing to zero the magnetic field  $h_j$, to a three-dimensional hypersurface 
$\mathfrak{B}_{3|reg}$ whose metric, generated by the pull-back of the immersion map:
\begin{equation}\label{immersione4}
  \iota_{[4]} \quad : \quad \mathfrak{B}_{3|reg} \, \longrightarrow \, \mathbb{E}_4
\end{equation}
is the one displayed in eq.(\ref{olunitre}):
\begin{equation}\label{iota4star}
   \iota_{[4]}^\star\left[ds_{\mathbb{E}_4}^2\right] \, = \,  ds^2_{\mathfrak{B}_3}(s_1,s_2,w)
\end{equation}
Hence, whenever we switch off one magnetic field, irrespective of its choice, we always find that the original flat Riemaniann manifold $\mho_{2\nu+3}\simeq \mathbb{R}^{2\nu +3}$ contracts to the following direct product 
\begin{eqnarray}\label{direprudo}
  \mho_{2\nu+3}& \mapsto &\mathbb{R}^{2\nu-1} \times \mathfrak{B}_{3|reg} \nonumber\\
   \mathfrak{B}_{3|reg}& \subset & \mathbb{E}_4
\end{eqnarray}
where the metric of the universal three-dimensional submanifold 
$\mathfrak{B}_{3|reg}$ is $ds^2_{\mathfrak{B}_3}(s_1,s_2,w)$ as displayed in eq.(\ref{olunitre}). 
\begin{figure}[htb]
\begin{center}
\includegraphics[width=13cm]{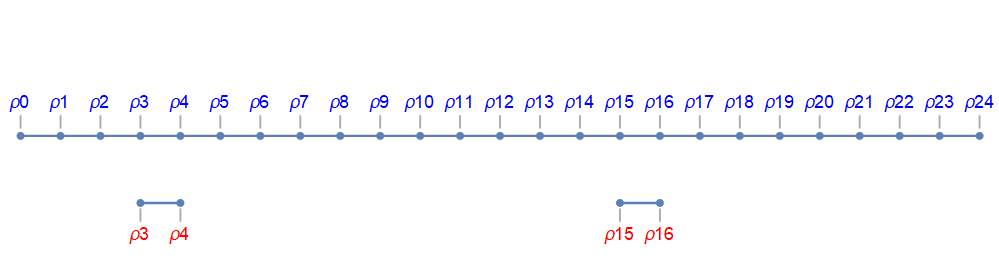}\\
\includegraphics[width=13cm]{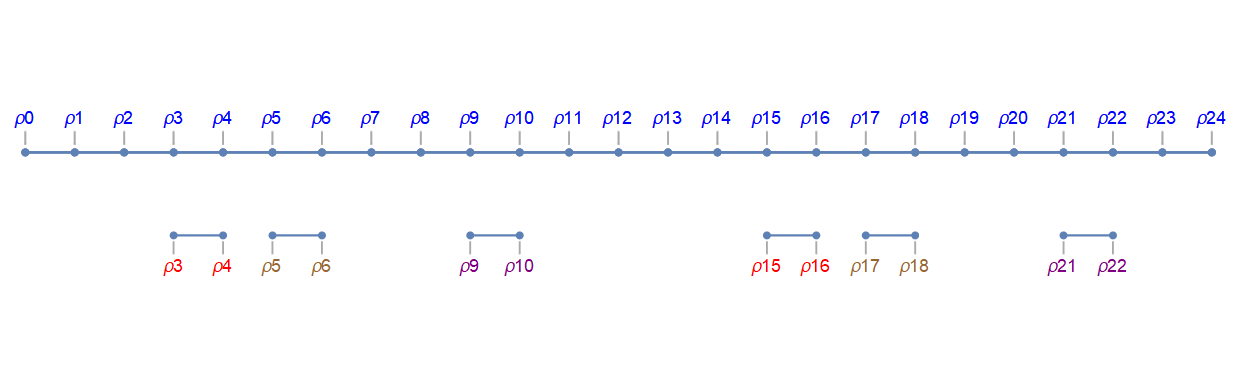}
\caption{\label{sequenze1} In this figure, choosing the case $\nu=11$ which yields a flat Thermo-Manifold $\mho_{25}$ with 25 dimensions, we display the sequence of the $25$ coordinates $\rho_0,\rho_1,\dots\,\rho_{24}$, illustrating what happens by
switching off just one magnetic field or a discontinuous sequence of magnetic fields. In the first picture we show the effect of setting $h_3=0$. The four $\rho$ coordinates marked in red span an $\mathbb{R}^4$ that is separated off from the remaining $21$-dimensional flat space and makes one copy of the universal $\mathfrak{B}_{3|reg}$ manifold, whose geometry we study in section \ref{uni32model}. In the second figure, with reference to the  same ambient space $\mho^{25}$, we show the effect of setting to zero $h_3,h_5,h_5$ the three groups of four coordinates, respectively marked in red, in brown and in purple
contract each to one independent copy  of the 3-dimensional surface $\mathfrak{B}_{3|reg}$ }
\end{center}
\end{figure}
\par
We refer the reader to Fig.\ref{sequenze1} in order to grasp  a conceptual view of the combinatorials of manifolds $\mathfrak{B}_{3|reg}$ that can arise setting  a discontinuous  sequence of magnetic fields to zero.
\par
Such combinatorials, however, becomes much more complicated if we consider also the vanishing of uninterrupted sequences of magnetic fields. Indeed, as we are going to see in section \ref{uninmodel}, every time we set simultaneously to zero an uninterrupted sequence of $(n-1)$ magnetic fields:
\begin{equation}\label{cornicione}
  h_j \, = \, h_{j+1} \, = \, \dots \, = \,h_{j+n-2}\, = \,0
\end{equation}
we split from the flat manifold $\mho_{2\nu +3}$ a flat subspace
$\mathbb{E}_{2n} \simeq \mathbb{R}^{2n}$ spanned by $2n$ among the $2\nu+3$ coordinates $\rho$ that, analogously to eq.(\ref{frullatodipesca}), we can rename $\left\{\xi_1,\dots,\xi_{2n}\right\}$. By effect of the $(n-1)$ conditions (\ref{cornicione}) the flat space $\mathbb{E}_{2n}$ 
with metric:
\begin{equation}\label{crocchetta2n}
  ds_{\mathbb{E}_{2n}}^2 \, \equiv\,\sum_{i=1}^{2n} d\xi_i^2
\end{equation}
contracts to a curved submanifold of dimension $(n+1)$: 
 \begin{equation}\label{cuntrattafoglia}
   \mathbb{E}_{2n} \, \stackrel{\text{$(n-1)$ h.s }\to\, 0}{\Longrightarrow} \, \mathfrak{B}_{n+1|reg}
 \end{equation}
defining an immersion map: 
\begin{equation}\label{immersione2n}
  \iota_{[2n]} \quad : \quad \mathfrak{B}_{n+1|reg} \, \longrightarrow \, \mathbb{E}_{2n}
\end{equation}
whose pull-back, on its turn, defines the Riemannian metric of
the submanifold $\mathfrak{B}_{n+1|reg}$:
\begin{equation}\label{iota2nstar}
 ds^2_{\mathfrak{B}_{n+1}}(s_1,\dots, s_n,w)  \, \equiv \  \iota_{[2n]}^\star\left[ds_{\mathbb{E}_{2n}}^2\right]  
\end{equation} 
The geometries of all the manifolds $\mathfrak{B}_{n+1}$ equipped with the pull-back metric (\ref{iota2nstar})  are very similar and display the same features, the most relevant of which is the following: each  $\mathfrak{B}_{n+1}$
further splits as follows:
\begin{equation}\label{strippato}
  \mathfrak{B}_{n+1}\, \simeq \, \mathbb{R} \, \times \, \mathfrak{M}_{n|reg}
\end{equation}
where $\left(\mathfrak{M}_{n|reg}, g^{[n]}\right)$ is an $n$-dimensional Riemannian space whose metric $g^{[n]}$ (defined by the pull-back of the immersion map in $\mathbb{E}_{2n}$) has a general form with peculiar distinctive properties. Before coming to the discussion of the general case (section \ref{uninmodel}) in the next section \ref{uni32model} we analyse in depth the case $n=2$ that corresponds to what we started with in the present section, namely the vanishing of a single isolated magnetic field. The properties of the $2$-dimensional manifold $\mathfrak{M}_{2|reg}$ are already an inspiring illustration of what happens in the general case.
\subsection{Analysis of $\mathfrak{B}_{3|reg}$ geometry}
\label{uni32model}
\begin{figure}
\begin{center}
\includegraphics[width=9cm]{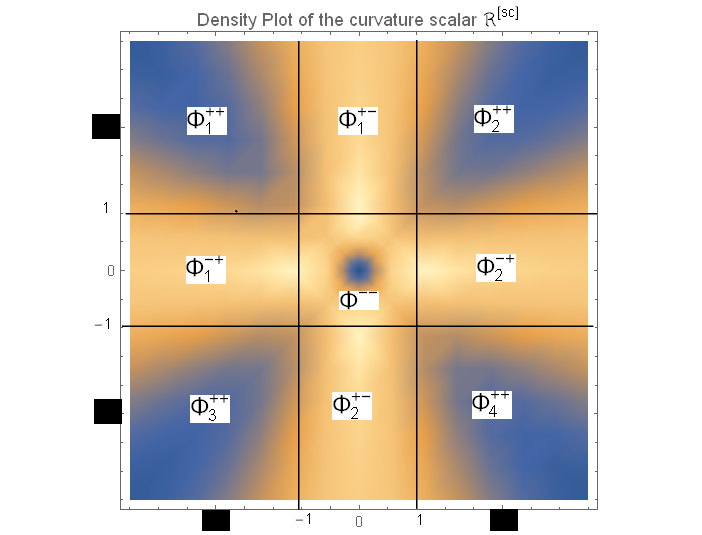}
\includegraphics[width=7cm]{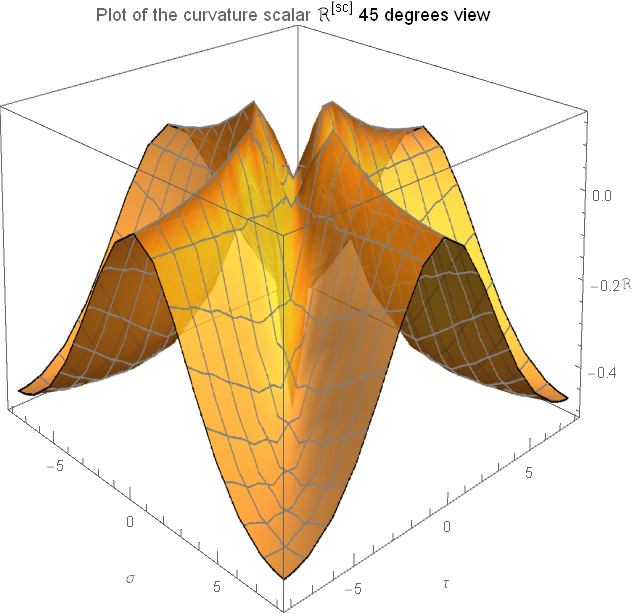}
\includegraphics[width=7cm]{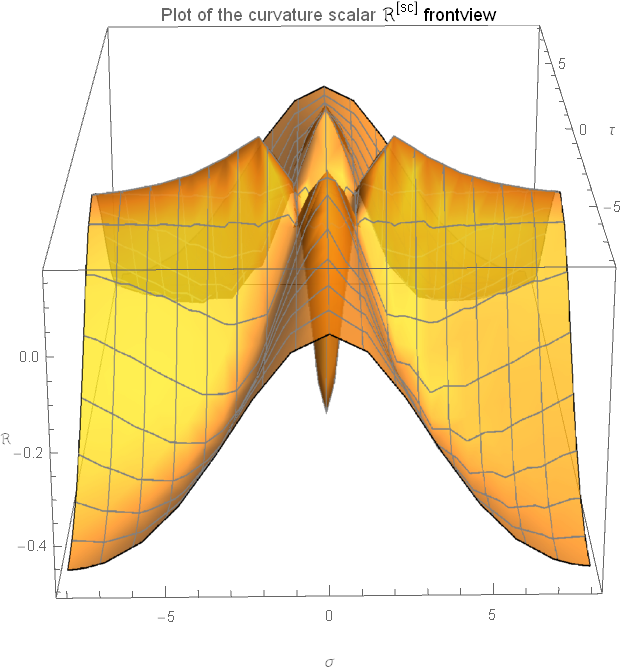}
 \caption{\label{curva2odd} In this picture we illustrate the curvature of the universal 2-dimensional model arising when one puts to zero one of the magnetic fields in the case of the
 $q=odd$ CV Thermodynamics. As we advocate later in section \ref{evencrat}, the same model describes also the curved geometry that arises in the $q=\text{even case}$ when we set to zero any magnetic field except the first $h_1$. On the left we present a density plot. The bluish color corresponds to lower curvature (negative of higher absolute value as the color becomes more deeply blue) while the yellowish color corresponds to higher curvature (less and less negative as the color becomes brighter yellow). On the right and below the curvature scalar is plotted in three dimension. As one sees the curvature is never singular and it is mostly negative and upper bounded. Its behaviour divides the $\sigma_1,\sigma_2$ plane into 5 regions: a central deep valley with four branches that meet at a local deep minimum at the origin. In addition we have four mountain ridges of higher that separate the branches of the central valley from each other. The mountain ridges stretch along the four principal axes $\pm \sigma_1$ and $\pm \tau_2$. In the first picture, displaying the density plot of the curvature, we have marked the lines that partition the manifold into the four macro regions mentioned in eq. (\ref{fourregions}), each of which, with the exception of the central region $\Phi^{-,-}$ is further split into either four or two disconnected components, making up a total of 9 components. In section \ref{geointerpret} we discuss the geometrical interpretation of such regions. It is important to stress that the splitting of the $\mathbb{R}^2$  manifold in the mentioned 9 components is precisely governed by the behaviour of the curvature scalar.}
\end{center}
\end{figure}

In order to analyse the differential geometry of the three-dimensional metric encoded in eq.(\ref{olunitre}), it is convenient to introduce a \textbf{dreibein} description. A convenient choice is provided by the following triplet of $1$-forms:
\begin{eqnarray}\label{dreibein1}
  \boldsymbol{e}^1 & = &\frac{\sqrt{2} \,\mathrm{d}s_1 \, \sqrt{s_1^2+w^2}}{s_1^2-w^2}-\frac{2 \sqrt{2}
   \,s_1 \,w \,\mathrm{d}w}{\left(s_1^2-w^2\right) \sqrt{s_1^2+w^2}}                    \nonumber\\ 
  \boldsymbol{e}^2 & = & \frac{\sqrt{2} \,\mathrm{d}s_2 \, \sqrt{s_2^2+w^2}}{s_2^2-w^2}-\frac{2 \sqrt{2}\,
   s_2 \, w \,\mathrm{d}w}{\left(s_2^2-w^2\right) \sqrt{s_2^2+w^2}}                   \nonumber\\
  \boldsymbol{e}^3 & = & \frac{\sqrt{2} \,\mathrm{d}w \, \sqrt{s_1^2+s_2^2+2 w^2}}{\sqrt{s_1^2+w^2}
   \sqrt{s_2^2+w^2}}                   \nonumber\\
\end{eqnarray}
in such a way that:
\begin{equation}\label{tregambette}
  ds^2_{\mathfrak{B}_3}(s_1,s_2,w) \, = \, \sum_{i=1}^3 \, \boldsymbol{e}^i\times \boldsymbol{e}^i
\end{equation}
Calculating the spin connection $\omega^{ij}$, the curvature 2-form $d\omega +\omega \wedge \omega$ and then the Ricci tensor with flat anholonomic indices, we find that the latter has one zero eigenvalue and two identical non vanishing ones. This suggests to perform an $\mathrm{O(3)}$ local gauge rotation on the dreibein that diagonalizes the Ricci tensor. Such transformation is provided by the following orthogonal $3\times3$-matrix:
\begin{equation}\label{sfratto}
\mathcal{O} \, =\, \left(
\begin{array}{ccc}
 \frac{s_1}{\sqrt{2} \sqrt{s_1^2+w^2}} & \frac{s_2}{\sqrt{2}
   \sqrt{s_2^2+w^2}} & \frac{w}{\sqrt{2}
   \sqrt{\frac{\left(s_1^2+w^2\right)
   \left(s_2^2+w^2\right)}{s_1^2+s_2^2+2 w^2}}} \\
 -\frac{w}{\sqrt{\frac{\left(s_1^2+w^2\right) \left(s_2^2+2
   w^2\right)}{s_1^2+s_2^2+2 w^2}}} & 0 &
   \frac{s_1}{\sqrt{\frac{\left(s_1^2+w^2\right) \left(s_2^2+2
   w^2\right)}{s_2^2+w^2}}} \\
 -\frac{s_1 s_2}{\sqrt{2} \sqrt{\left(s_1^2+w^2\right) \left(s_2^2+2
   w^2\right)}} & \frac{1}{\sqrt{\frac{s_2^2}{s_2^2+2 w^2}+1}} &
   -\frac{s_2 w}{\sqrt{2} \sqrt{\frac{\left(s_1^2+w^2\right)
   \left(s_2^2+w^2\right) \left(s_2^2+2 w^2\right)}{s_1^2+s_2^2+2
   w^2}}} \\
\end{array}
\right)
\end{equation}
Introducing the new dreibein
\begin{equation}\label{nuovodrei}
  \hat{\boldsymbol{e}}^i \, = \, \mathcal{O}^{ij} \, \boldsymbol{e}^j
\end{equation}
which take the following form:
\begin{eqnarray}\label{hatdreibein}
  \hat{\boldsymbol{e}}^1 & = & \frac{w \,\mathrm{d}w \left(s_1^2+s_2^2-2 w^2\right)+s_2\, \mathrm{d}s_2
   \left(w^2-s_1^2\right)+s_1 \,\mathrm{d}s_1
   \left(w^2-s_2^2\right)}{\left(s_1^2-w^2\right)
   \left(w^2-s_2^2\right)}                    \nonumber\\ 
  \hat{\boldsymbol{e}}^2 & = &\frac{(w \,\mathrm{d}s_1-s_1 \, \mathrm{d}w) \sqrt{\frac{2 s_1^2}{s_2^2+2
   w^2}+2}}{w^2-s_1^2}                    \nonumber\\ 
  \hat{\boldsymbol{e}}^3 & = & \frac{-s_2\, w \,\mathrm{d}w \left(-3 s_1^2+s_2^2+2 w^2\right)-\mathrm{d}s_2\, (s_1-w)
   (s_1+w) \left(s_2^2+2 w^2\right)+s_1\, s_2\, \mathrm{d}s_1\, (s_2-w)
   (s_2+w)}{\left(w^2-s_1^2\right) (s_2-w) (s_2+w) \sqrt{s_2^2+2 w^2}}                \nonumber\\ 
\end{eqnarray}
we find that the first dreibein $\hat{\boldsymbol{e}}^1$ is actually an exact $1$-form since it is the differential of a function:
\begin{equation}\label{ciurlamanico}
  \hat{\boldsymbol{e}}^1 \, = \, \mathrm{d}f(s_1,s_2,w) \quad ; \quad f(s_1,s_2,w) \, = \, \frac{1}{2} \log \left(\left(w^2-s_1^2\right)
   \left(w^2-s_2^2\right)\right)
\end{equation}
This suggests that one direction is flat and in order to single out the independent coordinate spanning the residual non-trivial 2-space the appropriate coordinate change is the following:
\begin{equation}\label{ciulatoprimo}
  s_1\,= \, \sigma_1 \,w \quad ; \quad s_2 \,= \, \sigma_2 \,w \quad ; \quad \varrho \, = \, \log \left(w^2 \sqrt{\left(\sigma_1 ^2-1\right) \left(\sigma_2
   ^2-1\right)}\right)
\end{equation}
Expressed in terms of the new coordinates ${\varrho,\sigma_1,\sigma_2}$ the new dreibein $\hat{\boldsymbol{e}}$
become
\begin{eqnarray}\label{hatdreibeinnew}
  \hat{\boldsymbol{e}}^1 & = & \mathrm{d}\varrho                \nonumber\\ 
  \hat{\boldsymbol{e}}^2 & = &  -\frac{\mathrm{d}\sigma_1 \sqrt{\frac{2 \sigma_1 ^2}{\sigma_2
   ^2+2}+2}}{\sigma_1 ^2-1}           \nonumber\\ 
  \hat{\boldsymbol{e}}^3 & = & \frac{\left(\sigma_1 ^2-1\right) \left(\sigma_2 ^2+2\right) \mathrm{d}\sigma_2
   +\sigma_1  \left(\sigma_2 -\sigma_2 ^3\right) \mathrm{d}\sigma_1
   }{\left(\sigma_1 ^2-1\right) \left(\sigma_2 ^2-1\right)
   \sqrt{\sigma_2 ^2+2}}                \nonumber\\ 
\end{eqnarray}
The result (\ref{hatdreibeinnew}) means that the additional flat direction that survives is given by the following combination of original coordinates:
\begin{equation}\label{addapiatta}
  \varrho=\log \left(\mu_j^2 \sqrt{\left[\left(\frac{\beta_j}{\mu_j}\right)^2-1\right] \left[\left(\frac{\beta_{j+1}}{\mu_j}\right)^2-1\right]}\right)
\end{equation}
and  the final outcome is that the three-dimensional manifold 
$\mathfrak{MU}_{3}$ is actually the following tensor product:
\begin{equation}\label{friscella}
  \mathfrak{B}_{3|reg} \, = \, \mathbb{R} \times \mathfrak{M}_{2|reg}
\end{equation}
The residual non trivial $2$-dimensional space is spanned by the coordinates $\sigma_1\,=\,\frac{\beta_j}{\mu_j}$ and $\sigma_2\,=\,\frac{\beta_{j+1}}{\mu_j}$ and geometrically described by the \textbf{zweibein} provided by $\left\{V^1,V^2\right\} =\left\{ \hat{\boldsymbol{e}}^2, \hat{\boldsymbol{e}}^3\right\}$ as given in equation (\ref{hatdreibeinnew}). Calculating the spin-connection one finds:
\begin{equation}\label{connospinno}
  \omega^{12} \, = \, \frac{\sigma_1 ^2 \sigma_2  \left(\sigma_2 ^2-1\right) }{\left(\sigma_2
   ^2+2\right)^{3/2} \left(\sigma_1 ^2+\sigma_2
   ^2+2\right)}\,V^1\, -\, \frac{2 \sigma_1  \left(\sigma_2 ^2-1\right)
   }{\left(\sigma_2 ^2+2\right)^2 \sqrt{\frac{2 \sigma_1
   ^2}{\sigma_2 ^2+2}+2}}\,V^2
\end{equation}
and by exterior derivative one obtains the curvature 2-form:
\begin{equation}\label{shurotto}
 \mathrm{d}\omega^{12} \, \equiv \, \mathfrak{R}^{12}\, = \, \mathcal{K}(\sigma_1,\sigma_2)\, V^1 \wedge V^2 \quad ; \quad \mathcal{K}(\sigma_1,\sigma_2)\, = \, - \, \frac{\left(\sigma_1 ^2-1\right) \left(\sigma_2
   ^2-1\right)}{\left(\sigma_1 ^2+\sigma_2 ^2+2\right)^2}
\end{equation}
Recalling that the flat metric is $\, \eta_{ij} \, = \,  \delta_{ij}$ it follows that the curvature scalar is:
\begin{equation}\label{curvscal}
\mathfrak{R}^{[sc]}\, = \, \mathfrak{R}^{ij}_{ij} \, = \, 2 \, \mathcal{K}(\sigma,\tau)
\end{equation}
As one sees, the curvature scalar for the $2$-manifold spanned by the coordinates $(\sigma_1,\sigma_2)\in \mathbb{R}^2$ equipped with the metric:
\begin{equation}\label{praemium}
  ds^2_{\mathfrak{M}_{2|reg}} \, = \,  (\hat{\boldsymbol{e}}^2)^2 + (\hat{\boldsymbol{e}}^3)^2 
\end{equation}
is nowhere singular on $\mathbb{R}^2$ and it is bounded both from above and from below. Although our original domain of definition of the metric requires $\mu_i>|\beta_i|$, and thus $|\sigma_i|<1$, in what follows we shall extend the metric by regularity, and allow $\sigma_i$ to vary in $\mathbb{R}^2$. The sign of $\mathcal{K}(\sigma_1,\sigma_2)$ depends on the signs of the two factors in the numerator. Indeed one can split $\mathbb{R}^2$ into 4 regions:
\begin{equation}\label{splittone}
  \mathbb{R}^2 \, = \, 
  \Phi^{[+,+]} \cup \Phi^{[+,-]}\cup\Phi^{[-,+]}\cup\Phi^{[-,-]}
\end{equation}
where:
\begin{eqnarray}
\label{fourregions}
  \Phi^{[+,+]} &\equiv & \left\{(\sigma_1,\sigma_2) \in \mathbb{R}^2 \, \mid \, |\sigma_1| \geq 1 \quad \& \quad |\sigma_2|\geq 1 \right \} \nonumber\\
 \Phi^{[-,+]} &\equiv & \left\{(\sigma_1,\sigma_2) \in \mathbb{R}^2 \, \mid \, |\sigma_1| \leq 1 \quad \& \quad |\sigma_2|\geq 1 \right \} \nonumber\\
 \Phi^{[+,-]} &\equiv & \left\{(\sigma_1,\sigma_2) \in \mathbb{R}^2 \, \mid \, |\sigma_1| \geq 1 \quad \& \quad |\sigma_2|\leq 1 \right \} \nonumber\\
\Phi_{sq} \, \equiv \, \Phi^{[-,-]} &\equiv & \left\{(\sigma_1,\sigma_2) \in \mathbb{R}^2 \, \mid \, |\sigma_1| \leq 1 \quad \& \quad |\sigma_2|\leq 1 \right \} 
\end{eqnarray}
and conclude that the curvature is negative on $\Phi^{[+,+]}$ and $\Phi^{[-,-]}$ while it is positive on  $\Phi^{[\pm,\mp]}$.
The behaviour of the curvature is illustrated in Fig. \ref{curva2odd}.
The effect of the curvature can be appreciated by studying the geodesic curves. Yet, before launching into the construction of such geodesics, it is convenient to consider the geometrical interpretation of the Riemannian manifold $\mathfrak{M}_{2|reg}$.
\subsubsection{Geometric interpretation of the $2D$-manifold $\mathfrak{M}_{2|reg}$}
\label{geointerpret}
We have seen that the behavior of the curvature scalar suggests the partition of the manifold $\mathbb{R}^2$ equipped with the metric (\ref{praemium}), that  explicitly reads as follows:
\begin{equation}\label{dsM2odd}
 ds^2_{\mathfrak{M}_{2|reg}} \, = \,  \frac{\left(\sigma_1
   ^2+2\right) \mathrm{d}\sigma_1^2}{\left(\sigma_1
   ^2-1\right)^2}\, +\, \frac{\left(\sigma_2 ^2+2\right) \mathrm{d}\sigma_2^2}{\left(\sigma_2 ^2-1\right)^2}\,-\,\frac{2 \, \sigma_1 \, \sigma_2 \, \mathrm{d}\sigma_1 \, \mathrm{d}\sigma_2}{\left(\sigma_1
   ^2-1\right) \left(\sigma_2 ^2-1\right)} 
\end{equation}
into the 4 macroregions (\ref{fourregions}). Such a partition is much more than a mnemonic rule to distinguish positive and negative support domains of the curvature function, as we show below. Indeed let us recall that Cartan Hadamard manifolds  are defined as Riemannian manifolds that:
\begin{enumerate}
  \item are simply connected,
  \item have everywhere a non-positive sectional curvature.
\end{enumerate}
From this it follows that the full $\mathbb{R}^2$ equipped with the metric (\ref{dsM2odd}) cannot be a Cartan-Hadamard manifold, yet, considering its partition (\ref{fourregions}) with the further subdivision into 9 components displayed in Fig. \ref{curva2odd}, we are able to extract from it $5$ Cartan-Hadamard, geodesically complete manifolds, plus $4$ additional simply connected manifolds that are also geodesically complete but not Cartan-Hadamard. 
\par
Recalling Cartan-Hadamard theorem stating that each $n$-manifold 
fulfilling the two above mentioned defining properties, is diffeomorphic to $\mathbb{R}^n$, we see that the first step in the chosen programmatic direction is that of mapping 
each of the $9$ regions of Fig. \ref{curva2odd} into a full copy of $\mathbb{R}^2$. This can be done by means of the branches of the two functions
$\tanh(x)$, $\coth(x)$ and of their inverses. Indeed by suitable combinations of such ingredients each of the 9 regions can be surjectively mapped into $\mathbb{R}^2$:
\begin{eqnarray}
\label{flogistico}
  \mathit{s}_{+,+|i} &=& \Phi^{+,+}_{i} \, \stackrel{\text{surjective}}{\longrightarrow} \, \mathbb{R}^2 \quad ; \quad i=1,2,3,4 \nonumber \\
\mathit{s}_{+,-|i} &=& \Phi^{+,-}_{i} \, \stackrel{\text{surjective}}{\longrightarrow} \, \mathbb{R}^2 \quad ; \quad i=1,2 \nonumber \\
\mathit{s}_{-,+|i} &=& \Phi^{-,+}_{i} \, \stackrel{\text{surjective}}{\longrightarrow} \, \mathbb{R}^2 \quad ; \quad i=1,2 \nonumber \\
\mathit{s}_{-,-} &=& \Phi^{-,-} \, \stackrel{\text{surjective}}{\longrightarrow} \, \mathbb{R}^2 
\end{eqnarray}
What we discover is that each of the above maps can be completed, by means of the addition of a third simple function of the variables $(\sigma_1,\sigma_2)$, well defined on the corresponding region $\Phi^{\pm,\pm}_i$ and representing the third component $z$, to a map:
\begin{equation}\label{granarolo}
  \mathit{m}^{\pm,\pm}_i \quad : \quad \Phi^{\pm,\pm}_i \, \longrightarrow \, \mathbb{R}^3
\end{equation}
in such a way that the pull-back on $\Phi^{\pm,\pm}_i$ of the standard (negative) flat Euclidean metric of $\mathbb{R}^3$:
\begin{eqnarray}
ds^{2}_{\mathbb{E}_3}=dx^{2}+dy^{2}+dz^{2}
\label{flat_space}
\end{eqnarray}
is the metric (\ref{dsM2odd}):
\begin{equation}\label{montepgliaccio}
  \mathit{m}^\star_{\pm,\pm|i} \left(ds^{2}_{\mathbb{E}^3}\right) \, = \, ds^2_{\mathfrak{M}_{2|reg}}\mid_{\Phi^{\pm,\pm}_i}
\end{equation}
\par
We begin with the case of the connected bounded region $\Phi_{sq}$, which is the physical domain of convergence for the corresponding partition function integral, (i.e.  the orbit of the compact Cartan subalgebra $\mathbb{H}\subset \mathbb{U}$ of the microscopic CV manifold.) 
The entire $\mathbb{R}^2$ plane can be mapped into the square $\Phi_{sq}$ by means of the standard map:
\begin{eqnarray}\label{squashing}
  \mathit{s}_{-,-}^{-1}& : & \mathbb{R}^2 \, \longrightarrow \, \Phi_{sq}  \nonumber\\
  \forall (x,y) \in \mathbb{R}^2 &:& \mathit{s}^{-1}_{-,-}\left[(x,y) \right] \, \equiv \, \left(\tanh\left[\frac{x}{\sqrt{2}}\right]\, , \, \tanh\left[\frac{y}{\sqrt{2}}\right] \right) \, = \,\left(\sigma_1,\sigma_2\right) \, \in \, \Phi_{sq}
\end{eqnarray}
The inverse transformation is: 
\begin{eqnarray}\label{invsquashing}
 \mathit{s}_{sq} \, \equiv \, \mathit{s}_{-,-}& : & \Phi_{sq}  \, \longrightarrow \, \mathbb{R}^2 \nonumber\\
  \forall (\sigma_1,\sigma_2) \in \Phi_{sq} &:& \mathit{s}_{sq}\left[(\sigma_1,\sigma_2) \right] \, \equiv \, \left(\frac{1}{\sqrt{2}}\log\left[\frac{1+\sigma_1}{1-\sigma_1}\right]\, , \, \frac{1}{\sqrt{2}}\log\left[\frac{1+\sigma_2}{1-\sigma_2}\right] \right) \, = \,\left(x,y\right) \, \in \, \mathbb{R}^2
\end{eqnarray}
and it is surjective. 
\par This is just standard and it has nothing to do with the metric (\ref{dsM2odd}). As we anticipated above, the non trivial thing is that the  map $\mathit{s}_{sq}$ in eq.(\ref{invsquashing})
can be extended to a new map $\mathit{m}_{sq}$ from $\Phi_{sq}$ to $\mathbb{R}^3$, namely:
\begin{eqnarray}\label{em:map}
  \mathit{m}_{sq}& : & \Phi_{sq}  \, \longrightarrow \, \mathbb{R}^3 \nonumber\\
  \forall (\sigma,\tau) \in \Phi_{sq} &:& \mathit{m}_{sq}\left[(\sigma,\tau) \right] \, \equiv \, \left(\frac{1}{\sqrt{2}}\log\left[\frac{1+\sigma}{1-\sigma}\right]\, , \, \frac{1}{\sqrt{2}}\log\left[\frac{1+\tau}{1-\tau}\right]\, , \, \frac{1}{2} \log \left[\frac{\sigma ^2-1}{\tau ^2-1}\right] \right) \, = \,\left(x,y,z\right) \, \in \, \mathbb{R}^3
\end{eqnarray}
such that the pull-back of the flat metric (\ref{flat_space}) on
$\Phi_{sq}$ is just the metric  (\ref{dsM2odd}):
\begin{equation}\label{moncrivello}
  \mathit{m}^\star_{sq} \left(ds^{2}_{\mathbb{E}^3}\right) \, = \, ds^2_{\mathfrak{M}_{2|reg}}
\end{equation}
\paragraph{Geometric model of the metric \eqref{dsM2odd} on $\mathit{m}_{sq}\left(\Phi_{sq}\right)\subset \mathbb{R}^3$:} This is the induced metric on the surface 
\begin{equation}\label{fiordiburro}
z\equiv \log \cosh \left(\frac{x}{\sqrt{2}}\right)-\log\cosh \left(\frac{y}{\sqrt{2}}\right) 
\end{equation} 
sitting inside the flat 3D‑space \eqref{flat_space}. Our surface is a \textbf{translation surface} with the two generating profile curves being the same function of $x$ and $y$, i.e. $\log\cosh \left(\frac{x}{\sqrt{2}}\right)$ and $\log\cosh \left(\frac{y}{\sqrt{2}}\right)$, respectively. 
\par
\begin{figure}[htb]
\begin{center}
\includegraphics[width=8cm]{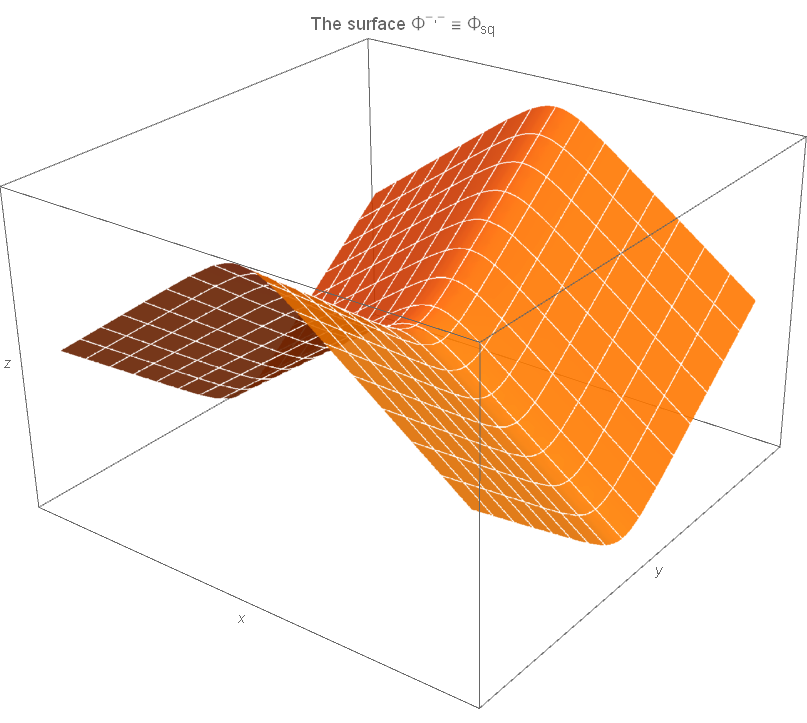}
\includegraphics[width=7cm]{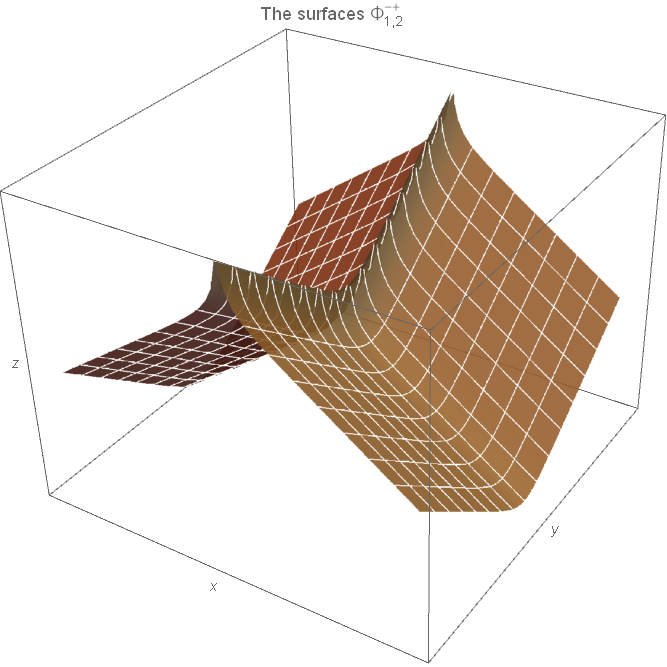}
\includegraphics[width=7cm]{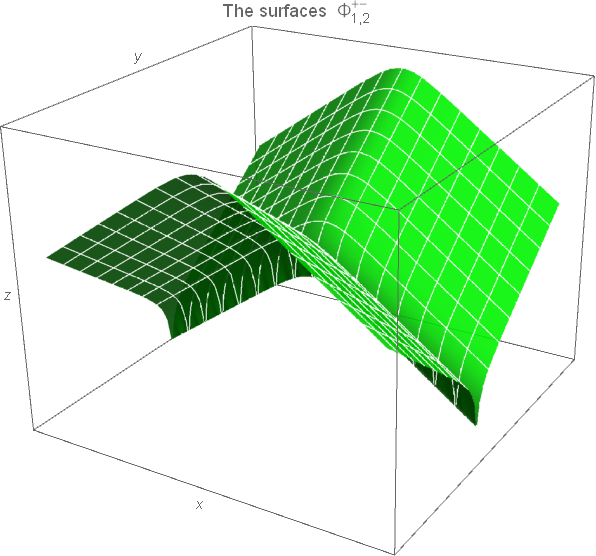}
\includegraphics[width=7cm]{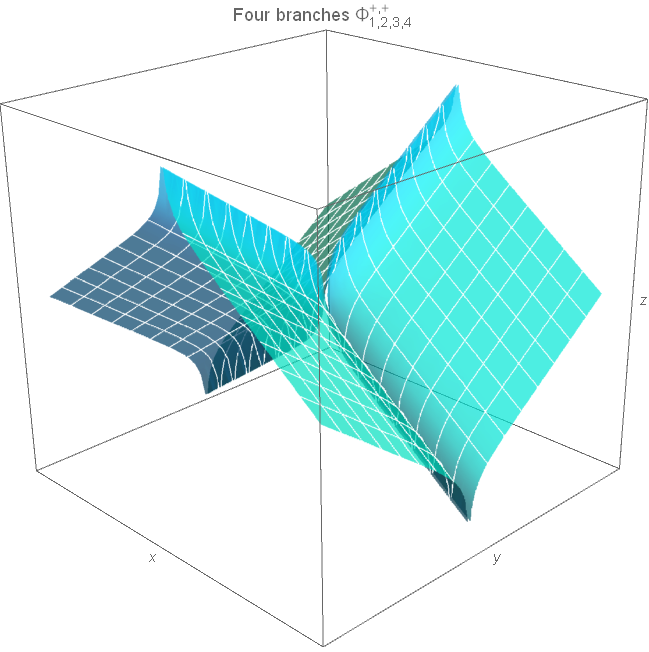}
 \caption{\label{quattrofogli} In this figure we display the images as  surfaces in $\mathbb{R}^3$ of the 9 disconnected manifolds that form the manifold $\mathfrak{M}_{2|reg}$. The disconnected sheets are all translational surfaces where either a hyperbolic tangent or hyperbolic cotangent profile is transported along  a profile made with either one of the same two functions. The further splitting of the four cases is due to the two branches of coth[x]. }
\end{center}
\end{figure}

A convenient way of rewriting the surface equation (\ref{fiordiburro}) is the following one
\begin{equation}\label{surfmenomeno}
  z = \frac{1}{2} \log \left(\frac{1-\tanh ^2\left(\frac{x}{\sqrt{2}}\right)}{1-\tanh
   ^2\left(\frac{y}{\sqrt{2}}\right)}\right)
\end{equation}
and the visualization of  $\mathit{m}_{sq}\left(\Phi_{sq}\right)$ is provided in the first picture of Fig. \ref{quattrofogli}.   
The other three macro region cases correspond to the possible
replacement schemes of $\tanh(w)$ with $\coth(w)$ for $w=x,y$. The additional splitting in 2 or 4 subcases is due to the two branches of the function $\coth(w)$.
\paragraph{Geometric model of the metric \eqref{dsM2odd} on $\mathit{m}_{sq}\left(\Phi_{1,2}^{-,+}\right)\subset \mathbb{R}^3$:}
The image $\mathit{m}_{-,+}\left(\Phi_{1,2}^{-,+}\right)$ of the region $\Phi_{1,2}^{-,+}$ in $\mathbb{R}^3$ is provided by the locus:
\begin{equation}\label{surfmenopiu}
  z = \frac{1}{2} \log \left(\frac{-1+\coth ^2\left(\frac{x}{\sqrt{2}}\right)}{1-\tanh
   ^2\left(\frac{y}{\sqrt{2}}\right)}\right)
\end{equation}
and the embedding map is as follows:
\begin{eqnarray}
\label{m_menopiu}
  \mathit{m}_{-,+}&:& \Phi_{1,2}^{-,+} \, \longrightarrow \, 
  \mathbb{R}^3 \nonumber \\
  \mathit{m}_{-,+}(\sigma_1,\sigma_2) &=& \left\{\frac{\log \left(\frac{\sigma_1 +1}{1-\sigma_1 }\right)}{\sqrt{2}},\frac{\log
   \left(\frac{\sigma_2 +1}{\sigma_2 -1}\right)}{\sqrt{2}},\frac{1}{2} \log
   \left(\frac{1-\sigma_1 ^2}{\sigma_2 ^2-1}\right)\right\} 
\end{eqnarray}
The visualization of the surface (\ref{surfmenopiu}) is provided by the second picture of Fig. \ref{quattrofogli}. The splitting in two manifolds is due to the two disconnected  branches of the $\coth$-function and it clearly visible in the figure.
\paragraph{Geometric model of the metric \eqref{dsM2odd} on $\mathit{m}_{sq}\left(\Phi_{1,2}^{+,-}\right)\subset \mathbb{R}^3$:}
The image $\mathit{m}_{+,-}\left(\Phi_{1,2}^{+,-}\right)$ of the region $\Phi_{1,2}^{+,-}$ in $\mathbb{R}^3$ is provided by the locus:
\begin{equation}\label{surfpiumeno}
  z = \frac{1}{2} \log \left(\frac{1+\tanh ^2\left(\frac{x}{\sqrt{2}}\right)}{-1+\coth
   ^2\left(\frac{y}{\sqrt{2}}\right)}\right)
\end{equation}
and the embedding map is as follows:
\begin{eqnarray}
\label{m_piumeno}
  \mathit{m}_{+,-}&:& \Phi_{1,2}^{+,-} \, \longrightarrow \, 
  \mathbb{R}^3 \nonumber \\
  \mathit{m}_{+,-}(\sigma_1,\sigma_2) &=& \left\{\frac{\log \left(\frac{\sigma_1 +1}{-1+\sigma_1 }\right)}{\sqrt{2}},\frac{\log
   \left(\frac{\sigma_2 +1}{-\sigma_2 +1}\right)}{\sqrt{2}},\frac{1}{2} \log
   \left(\frac{-1+\sigma_1 ^2}{-\sigma_2 ^2+1}\right)\right\} 
\end{eqnarray}
The visualization of the surface (\ref{surfpiumeno}) is provided by the third picture of Fig. \ref{quattrofogli}. The splitting in two manifolds is due to the two branches of the $\coth$-function and it clearly visible in the figure.
\paragraph{Geometric model of the metric \eqref{dsM2odd} on $\mathit{m}_{+,+}\left(\Phi_{1,2}^{+,+}\right)\subset \mathbb{R}^3$:}
The image $\mathit{m}_{+,+}\left(\Phi_{1,2}^{+,+}\right)$ of the region $\Phi_{1,2}^{+,-}$ in $\mathbb{R}^3$ is provided by the locus:
\begin{equation}\label{surfpiumeno}
  z = \frac{1}{2} \log \left(\frac{-1+\coth ^2\left(\frac{x}{\sqrt{2}}\right)}{-1+\coth
   ^2\left(\frac{y}{\sqrt{2}}\right)}\right)
\end{equation}
and the embedding map is as follows:
\begin{eqnarray}
\label{m_piumeno}
  \mathit{m}_{+,-}&:& \Phi_{1,2}^{+,-} \, \longrightarrow \, 
  \mathbb{R}^3 \nonumber \\
  \mathit{m}_{+,-}(\sigma_1,\sigma_2) &=& \left\{\frac{\log \left(\frac{\sigma_1 +1}{-1+\sigma_1 }\right)}{\sqrt{2}},\frac{\log
   \left(\frac{\sigma_2 +1}{\sigma_2 -1}\right)}{\sqrt{2}},\frac{1}{2} \log
   \left(\frac{-1+\sigma_1 ^2}{\sigma_2 ^2-1}\right)\right\} 
\end{eqnarray}
The visualization of the surface (\ref{surfpiumeno}) is provided by the fourth picture of Fig. \ref{quattrofogli}. The splitting in 4 manifolds is due to the two branches of the $\coth$-function and it clearly visible in the figure.
\subsubsection{Perspective meaning of the 9 branches}
\label{perspetbranche} As we stated at the beginning, the \textit{physical range} of the variables is that corresponding to the central square region $\Phi_{sq}$ that constitutes a geodesically complete Cartan-Hadamard manifold. As shown later on, the points on the boundary of $\Phi_{sq}$ are all at infinite distance from any interior point, as it happens in the hyperbolic plane. As also illustrated below, in section \ref{geodm2reg}, no geodesic in each of the 9 regions reaches any interior point of any other of the 8 remaining regions. Hence one might be tempted to exclude them at all from the game. Yet a question that remains open is whether the analytic continuation of the convergent partition functions to values out of the convergence range can be interpreted as the partition functions of the underlying microscopic system in different phases. In such a case, recalling that the metric under study is obtained by freezing to zero a specific magnetic field, it is clear that reactivating externally such a field we reconnect the separated regions and geodesic connecting their internal points become possible. The implications of these simple considerations are all to be studied in detail in future publications.
\subsubsection{Geodesics of the $\mathfrak{M}_{2|reg}$ model.}
\label{geodm2reg}
From the explicit form of the line element given in eq.(\ref{dsM2odd}), by replacing the differentials with the time derivatives $\dot{\sigma},\dot{\tau}$ one obtains the geodesic lagrangian\footnote{The overall normalization of the Lagrangian is irrelevant in the derivation of the Euler-Lagrange equations, which is the only use of the lagrangian in this context: namely a quick trick to obtain the geodesic equations that avoids the cumbersome calculation of the Christoffel symbols  (see \cite{pietrobook} volume 1). For this reason we do not care about the constant mentioned in equation  (\ref{dsM2oddlag}) that is actually $-1$.}
\begin{equation}\label{dsM2oddlag}
 \mathcal{L}(\dot{\sigma_1},\dot{\sigma_2},\sigma_1,\sigma_2) \, = \,  \text{const} \times \left( \frac{\left(\sigma_1
   ^2+2\right) \dot{\sigma_1}^2}{\left(\sigma_1
   ^2-1\right)^2}+\frac{\left(\sigma_2 ^2+2\right) \dot{\sigma_2}^2}{\left(\sigma_2 ^2-1\right)^2}\,-\,\frac{2 \, \sigma_1 \, \sigma_2 \, \dot{\sigma_1} \, \dot{\sigma_2}}{\left(\sigma_1
   ^2-1\right) \left(\sigma_2 ^2-1\right)} \right)
\end{equation}
whose Euler-Lagrange equations are the equations for the geodesics. They are highly non-linear PDE.s and the only possibility is to study them by numerical integration. We show
some examples both plotted over the curvature density plot and mapped to the corresponding geometrical realization of the manifold as a translational surface.
\begin{figure}[htb]
\begin{center}
\includegraphics[width=5cm]{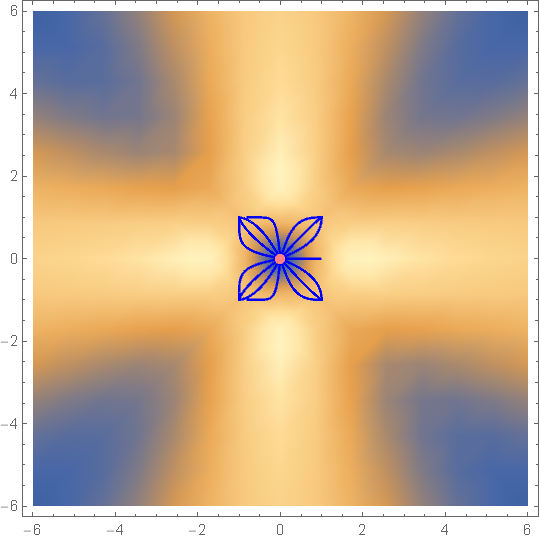}
\includegraphics[width=7cm]{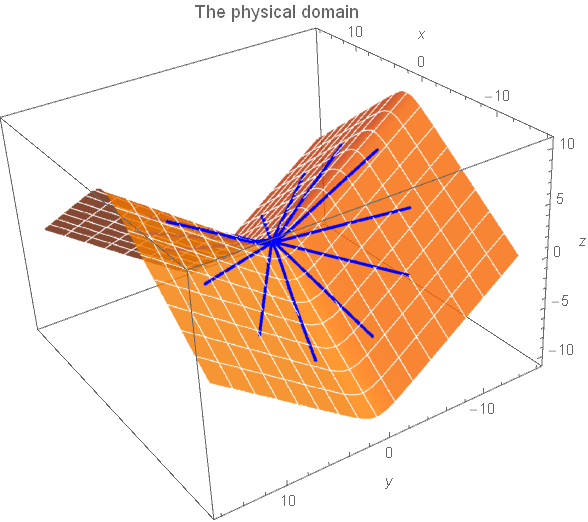}\\
\includegraphics[width=5cm]{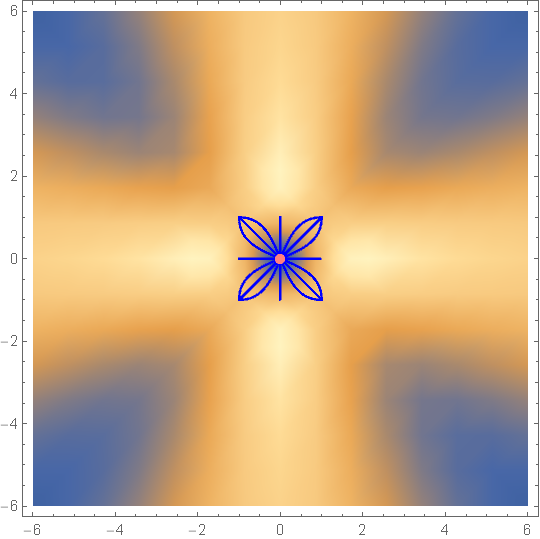}
\includegraphics[width=7cm]{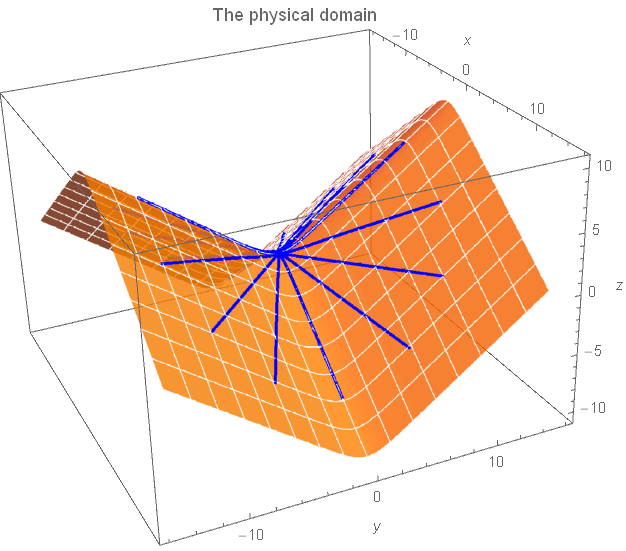}
 \caption{\label{geodes2oddA} In this figure we display geodesics departing from the origin in the central square "physical domain". Comments on their behavior is provided in the main body text.}
\end{center}
\end{figure}
\par 
In Fig. \ref{geodes2oddA} we present examples where the starting points of the considered geodesics is inside the physical region $\Phi_{sq}$. In particular in Fig. \ref{geodes2oddA} we use as starting point the very center, location of the curvature minimum. The difference in the two pictures are simply the initial directions. In the first example we have divided the $2\pi$ angle by 25 and the considered initial angle is $\phi_0 \, = \, \frac{2\pi}{25}\times n=0,1,\dots, 24$. In the second example we have done the same thing dividing $2\pi$ in 16 rather than in 25 parts. The main difference is that when we divide $2\pi$ by a number that is a multiple of 8, we always include in the set of calculated geodesics those that are aligned with the four coordinate axes directions, namely $\phi_0=0,\pm \frac{\pi}{2}, \pm \pi$ and also those that are aligned with the four pincipal diagonals namely depart at $\phi_0=0,\pm \frac{\pi}{4}$. In the square plot, these geodesics are also straight segments  that join the center with the four vertices of the square and with the four middle points of its sides. Instead, in  the three-dimensional geometrical model, the entire perimeter of the square is at infinity, so that all geodesics appear similar and just divergent to infinity. 
\begin{figure}[htb]
\begin{center}
\includegraphics[width=5cm]{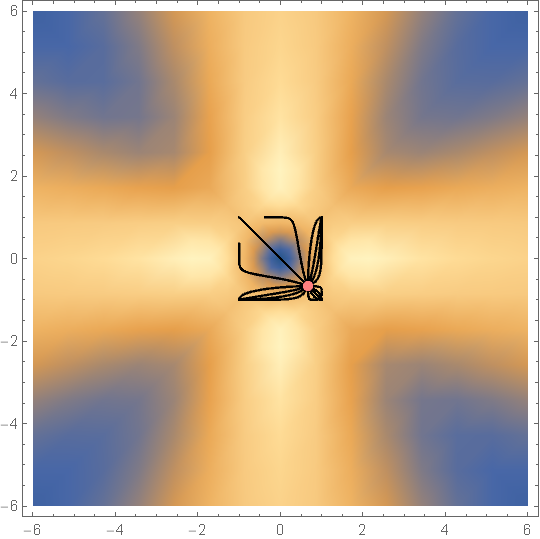}
\includegraphics[width=7cm]{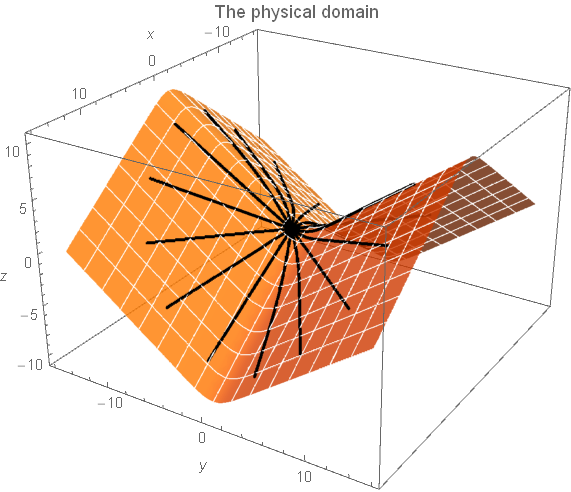}\\
\includegraphics[width=5cm]{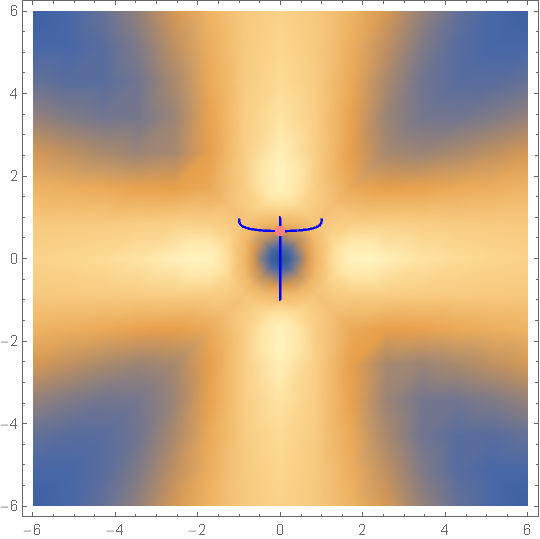}
\includegraphics[width=7cm]{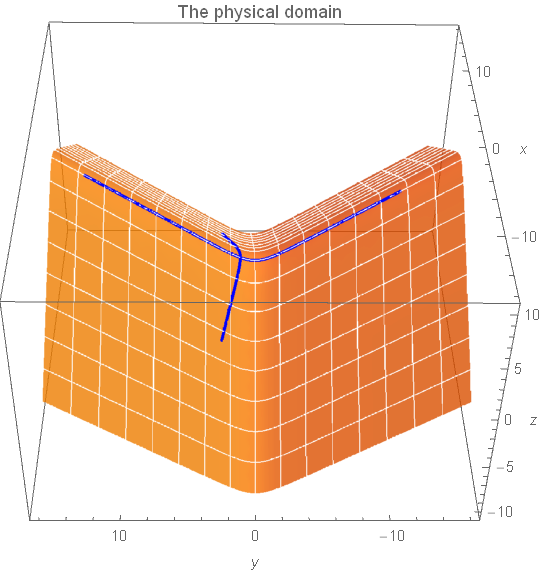}
 \caption{\label{geodescenteccentric} In this figure we present geodesics with an initial point interior to the region $\Phi_{sq}$ different from the origin $(0,0)$. The phenomenon of the existence of six ending points of geodesics on the boundary of $\Phi_{sq}$ is visible also in these example.  } 
\end{center}
\end{figure}
Yet, notwithstanding the Euclidean signature of the metric, the inspection of the plots displayed in Fig. \ref{geodes2oddA} reveals a notable phenomenon that is reminiscent of the causal structure of the boundary of Lorentz space-times in General Relativity. Indeed it is as if the original model in $(\sigma_1,\sigma_2)$-coordinate were the conformal model of space-time, whose role, instead,  is here  played by the translation surface  embedded in $\mathbb{R}^3$. From the point of view of the translation surface, infinity is just infinity, yet as it is the image of the square perimeter it has a structure. The structure visible in the plane plot appears to be the following: the four vertices of the square are the analog of spatial and time infinity in the Penrose diagram, namely the ending points of all geodesics of a certain type and the type seems to be decided by the initial orientation angle $\phi_0$ independently from the position of the starting point. This property is confirmed also by the examples presented in 
Fig. \ref{geodescenteccentric} that have an initial starting point different from the origin $(0,0)$ but always interior to the square $\Phi_{sq}$.
\par 
Such a property concerning the asymptotic terminal points of geodesics seems deeply related with the general property of the Riemann tensor the four images on the embedded surface being the two lines descending or climbing the axes of  steepest descent meeting at the saddle point in the origin. 
\par 
\begin{figure}[htb]
\begin{center}
\includegraphics[width=5cm]{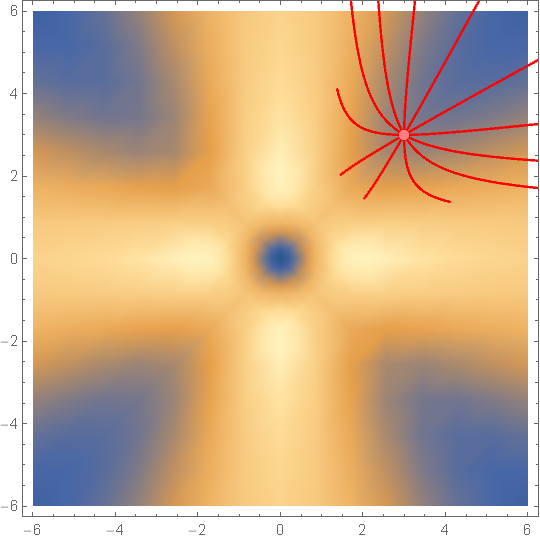}
\includegraphics[width=7cm]{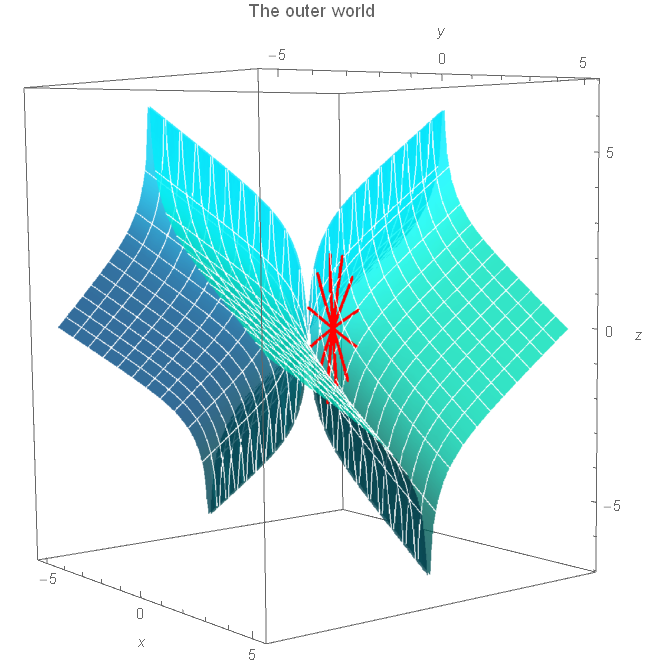}\\
\includegraphics[width=5cm]{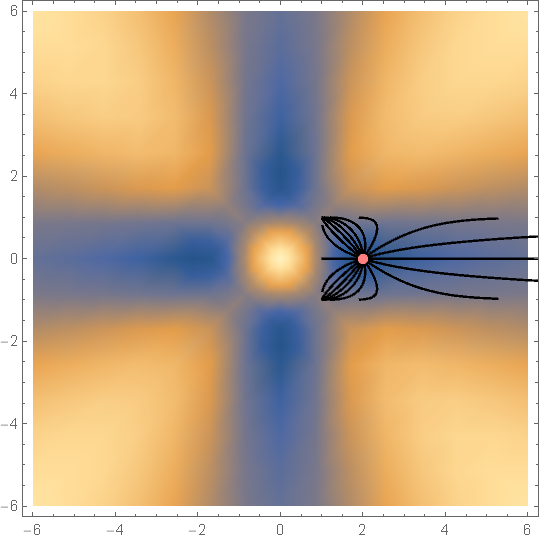}
\includegraphics[width=7cm]{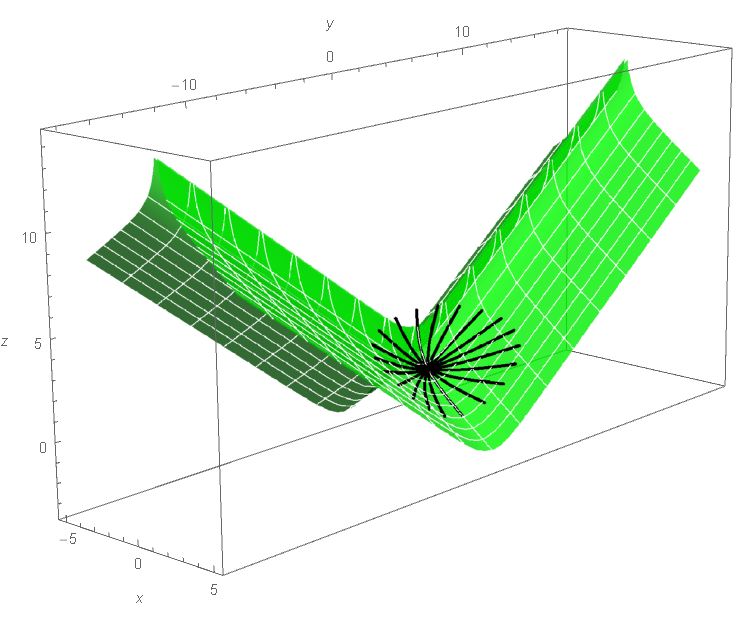}
 \caption{\label{outerworldgeo} In this figure we display images of geodesics in the outer regions different from the physical square $\Phi_{sq}$. Comments on the figures are in the main text.  } 
\end{center}
\end{figure}
For completeness in Fig. \ref{outerworldgeo} we present a couple of examples of geodesics belonging to the outer world different from $\Phi_{sq}$. In the first line of that picture we display a set of geodesic whose starting point is in the region $\Phi_2^{+,+}$: as one sees from the picture the entire development of each of the chosen geodesics remains in such a region. Correspondingly in the geometric model of the same region we see the geodesics as lines traced on one of the four folds of the translation surface. The same happens for the set of geodesics displayed in the second line that start in the region $\Phi^{+,-}$ and their remain both in bidimensional plane
$\sigma_1,\sigma_2$ and in its image as translation surface.
What once again is clearly shown in the plane-plots and obscured in their image on the corresponding translational surface is the limiting points whereto the geodesic converge at infinity.  
\section{Vanishing of $n-1$ contiguous magnetic fields and the universal $n$-dimensional space $\mathfrak{M}_{n|reg}$}
\label{uninmodel}
As we already remarked above, the magnetic fields are not interchangeable: they fall in a precise hierarchical order that follows from the sequential ordering of the principal subalgebras, whose Casimir square-roots are conjugate one-by-one  to a definite $h_i$. This being clarified it follows that when $q\gg 1$ is large the magnetic fields can be switched on and off according with complicated combinatorics, each time constraining the dynamical system spanned by the generalized temperatures $\boldsymbol{\beta}$ and the sequential magnetic coordinates $\mu_i$ (or equivalently in even $q$ case $\hat{\mu}_{i-1}$) to evolve on curved submanifolds. As we have seen, when we switch off just one magnetic field, we generate a non-trivial two-dimensional manifold that has a universal character. 
\par
\begin{figure}[htb]
\begin{center}
\includegraphics[width=13cm]{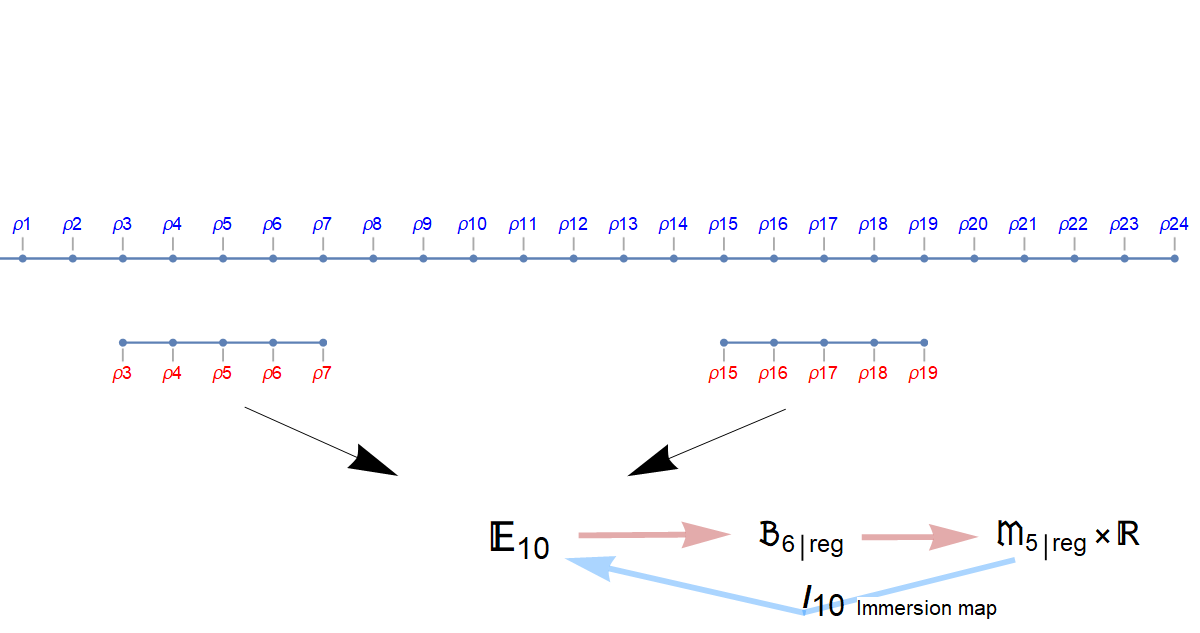}
 \caption{\label{belinfante} Using the thermo-case $\nu=11$ and choosing a sequence of $(n-1)\,=\, 4$ vanishing magnetic fields starting at $j=3$, in this figure we provide a graphical conceptual summary of the generation process of the curved manifold $\mathfrak{M}_{n|reg}$ by means of the immersion map $\iota_{2n}$.} 
\end{center}
\end{figure}

As we anticipated in section \ref{unovanni}, if we let an uninterrupted sequence of $(n-1)$ magnetic fields, starting at $h_j$, go to zero as shown in eq.(\ref{cornicione}), then we single out a flat submanifold $\mathbb{E}_{2n} \subset \mho_{2\nu+3}$ that, on its turn, contracts to a submanifold $\mathfrak{B}_{n+1|reg}$, as shown in eq.(\ref{cuntrattafoglia}),
through the $h$-vanishing equations. At the same time, the latter equations uniquely define the immersion-map $\iota_{2n}$ of the abstract manifold $\mathfrak{B}_{n+1|reg}$ 
into the flat manifold $\mathbb{E}_{2n}$, so that the metric on
$\mathfrak{B}_{n+1|reg}$ is precisely defined by the pull-back of the $\mathbb{E}_{2n}$ flat-metric through $\iota_{2n}^\star$, as stated in eq.(\ref{iota2nstar}). The whole process is conceptually summarized in Fig. \ref{belinfante}.
\par 
We anticipated in eq.(\ref{strippato}) that $\mathfrak{M}_{n+1|reg}$ always turns out to be diffeomorphic and isometric to $\mathbb{E}_1 \simeq \mathbb{R}$ times a non trivial Riemannian space $\mathfrak{M}_{n|reg}$ and have shown that this does indeed happen in the case $n=2$. To this effect we utilized the vielbein formalism and we relayed on the structure of the anholonomic Ricci tensor: in appendix \ref{tregambone} we do the same thing for the $n=3$ case. However it is obvious that the reason for this general fact must be encoded in the very structure of the $\iota_{2n}$ immersion-map. In this section we precisely study the $\iota_{2n}$ map and we uncover its simple and elegant structure which encodes all properties of the manifolds $\mathfrak{B}_{n+1|reg}$ and of their respective non trivial submanifolds  $\mathfrak{M}_{n|reg}$.
\subsection{The $\iota_{2n}$ immersion map}
According to the discussion initiated in section \ref{unovanni}
and relying on the fundamental embedding equation (\ref{roncoscrivia}), after fixing the initial point $j$ of the $(n-1)$ sequence, we identify the relevant $2n$ coordinates $\rho_a$ (see Fig.  \ref{belinfante} for orientation) and we rename them $\xi_1,\dots,\,\xi_{2n}$:
\begin{equation}\label{parlabuono}
 \left \{\rho_j, \,\rho_{j+1},\, \dots,\, \rho_{j+n}, \,\rho_{j+\nu+1},\,\rho_{j+\nu+2}\,\dots,\rho_{j+n+\nu+1} \right \} \, \longrightarrow \,\left\{\xi_1, \, \dots,\,  \xi_{2n} \right\} 
\end{equation}
Next we rename the involved $\beta_i$:
\begin{equation}\label{betaciangio}
  \beta_{j} \, = \, s_1, \,\,  \beta_{j+1} \, = \, s_2 , \,\, \dots, \,\, \beta_{j+n} \, = \, s_n
\end{equation}
and the involved $\mu_i$:
\begin{equation}\label{identificatia}
  \mu_{j} \, = \, w_1, \,\,  \mu_{j+1} \, = \, w_2 , \,\, \dots, \,\, \mu_{j+n} \, = \, w_n
\end{equation}
In this way we have rewritten in a universal format all the instances of relevant thermo-coordinates corresponding to any uninterrupted sequence of $(n-1)$ vanishing magnetic fields.
\par
The relevant point is that the relation (\ref{roncoscrivia}) can be uniquely inverted on the three subsets (\ref{parlabuono}),(\ref{betaciangio})  and (\ref{identificatia}).
Indeed we find:
\begin{equation}
\label{liagroppello}
\varphi\, =\,\left\{\begin{array}{cclcr}
  \xi_i &=& \log\left[\frac{s_i \,+\,w_i}{2}\right] &;& i=1\,\dots\, n \\
  \xi_{i+n} &=& \log\left[\frac{-s_i \,+\,w_i}{2}\right] &;& i=1\,\dots\, n\\ 
  \end{array}\right.
\end{equation}
The map $\varphi$ is a morphism:
\begin{equation}\label{gabriello}
  \varphi \quad : \quad \mathbb{R}^{2n} \,\longrightarrow \, \mathbb{R}^{2n}
\end{equation}
and the pull-back through $\varphi^\star$ of the flat metric
(\ref{crocchetta2n}) is just a segment of the full thermo-dynamic metric (\ref{distaquadodd}):
\begin{eqnarray}\label{vermiciattolo}
  \varphi^\star\left[ds^2_{\mathbb{E}_{2n}}\right] & = & \, \sum_{i=1}^{2n}\,\frac{1}{\left(s_i^2-w^2_i\right)^2}\,
\left[2\,\left(s_i^2+w^2_i\right)\,(ds_i^2 + dw_i^2)\, - \, 8 s_i \, w_i \,ds_i \,dw_i\right] 
\end{eqnarray}
which obviously is also flat as one can verify by calculating the Riemann tensor with standard textbook definitions.
\par
The vanishing equations of the $(n-1)$ magnetic fields are extremely simple in terms of the $w_i$-variables: it suffices to set:
\begin{equation}\label{grunaldo}
  w_1 \, = \, w_2 \, = \, \dots \, \, = \, w_n \, = \, w
\end{equation}
If we insert eq.(\ref{grunaldo}) into the $\varphi$-morphism (\ref{liagroppello}) we obtain the explicit form of the $\iota_{2n}$ embedding map advocated in eq.(\ref{immersione2n}):
\begin{eqnarray}
\label{calindri}
\iota_{2n} &:& \quad \mathfrak{B}_{n+1|reg} \,\longrightarrow \, \mathbb{E}_{2n} \nonumber\\
\iota_{2n}& =&\left\{\begin{array}{cclcr}
  \xi_i &=& \log\left[\frac{s_i \,+\,w}{2}\right] &;& i=1\,\dots\, n \\
  \xi_{i+n} &=& \log\left[\frac{-s_i \,+\,w}{2}\right] &;& i=1\,\dots\, n\\ 
  \end{array}\right.
\end{eqnarray}
If we insert eq.(\ref{grunaldo}) into the metric expression
(\ref{vermiciattolo}) we get the metric of the manifold $\mathfrak{B}_{n+1|reg}$, namely:
\begin{eqnarray}
\label{olinno}
  ds^2_{\mathfrak{B}_{n+1|reg}}(\mathbf{s},w) &=& 2 \, \sum_{k=1}^{n} \frac{s_k^2\,+ \, w^2}{\left(s_k^2\, - \, w^2\right)^2}\,\mathrm{d}s_k^2\, - \, 8 \,\sum_{k=1}^n  \frac{s_k \, w}{\left(s_k^2\, - \, w^2\right)^2}\,\mathrm{d}s_k \,\mathrm{d}w \, + \, 2 \, \left(\sum_{k=1}^n \frac{s_k^2\,+ \, w^2}{\left(s_k^2\, - \, w^2\right)^2}\right)\,\mathrm{d}w^2
\end{eqnarray}
and we obviously have:
\begin{equation}\label{panissa}
  \iota_{2n}^\star\left[ds^2_{\mathbb{E}_{2n}}\right] \, = \, ds^2_{\mathfrak{B}_{n+1|reg}}
\end{equation}
which is just a tautology, by construction.
\par
Next we want to show that as a Riemannian manifold, $\mathfrak{B}_{n+1|reg}$ splits as follows:
\begin{equation}\label{fraschetta}
  \mathfrak{B}_{n+1|reg} \, \simeq \, \mathbb{E}_1 \times \mathfrak{M}_{n|reg}
\end{equation} 
To verify this it suffices to make a coordinate change of the following form:
\begin{eqnarray}
\label{fargnocco}
  s_i &=& \sigma_i \, w \quad\quad \quad\quad\quad\quad\quad ; \quad i=1,\dots, n\nonumber\\
  w &=& \sqrt{\frac{e^{\varrho }}{\prod _{i=1}^n \left(\sigma
   _i^2-1\right){}^{\frac{1}{n}}}}  
\end{eqnarray}
where $\left\{\varrho,\sigma_i\right\}$ are the new $(n+1)$ coordinates.
\par
Upon such coordinate change the metric in eq.(\ref{olinno}) becomes:
\begin{eqnarray}\label{bracciodiferro}
  ds^2_{\mathfrak{B}_{n+1|reg}}&=& \,\frac{n}{2} \, \mathrm{d}\varrho^2  \, + \, ds^2_{\mathfrak{M}_{n|reg}} \nonumber\\
  ds^2_{\mathfrak{M}_{n|reg}}&=& 
   g^{[n]}_{ij}(\boldsymbol{\sigma})\, \mathrm{d}\sigma_i \times \mathrm{d}\sigma_j
\end{eqnarray}
where the metric tensor $g^{[n]}_{ij}(\boldsymbol{\sigma})$ has the following very simple, yet non trivial form:
\begin{eqnarray}\label{caponata}
  g^{[n]}_{i,i}(\boldsymbol{\sigma}) &=& \,  \frac{2}{n} \, \frac{n+(n-1)\sigma_i^{2}}{(-1+\sigma_i^2)^2} \quad\quad\quad\quad\, ; \quad i=1,\dots, n\nonumber\\
  g^{[n]}_{i,j}(\boldsymbol{\sigma}) &=& - \, \frac{2}{n}\, \frac{\sigma_i \, \sigma_j}{(-1+\sigma_i^2)(-1+\sigma_j^2)} \, \quad ; \quad i\neq j, \quad  i,j=1,\dots, n
\end{eqnarray}
From its structure it is suggested that the metric (\ref{caponata}) might display singularities when any of the $\sigma_i$ coordinates touches the boundary $|\sigma_i|\, =\,1$. Yet in order to verify whether the coordinate singularities of the metric are  true geometrical singularities it is better to calculate the Riemann tensor.  Indeed the singularities of the Riemann tensor, if any, are true geometrical singularities of the Riemannian manifold. If the apparent singularities of the metric coefficients are regular points of the Riemann tensor, then there is no true singularity.
\subsection{Peculiar properties of the Riemann tensor of $\mathfrak{M}_{n|reg}$}
By direct systematic calculation within the standard tensor calculus formulation of differential geometry, we discovered a general property of the Riemann tensor of the metrics (\ref{caponata}), that is instead more hidden in the vielbein formulation. The reason of this anomalous evenience is that, at variance with the general trend, the vielbein formulation is less handy in this case, than the ordinary tensor formulation, since it hides the discrete symmetries that are instead patent in the explicit expression of the metric (\ref{caponata}). Whatever might be the reason of what we just said, here is the experimental result that we have been able to derive by explicit calculation of the Riemann tensor for an extended number of instances of the dimension $n$.
\par
Let:
\begin{equation}\label{rimagno}
  \text{Rie}^{\lambda}_{\mu\nu,\sigma} \, = \, \partial_\mu \Gamma^\lambda_{\nu\sigma}\, -\, \partial_\nu \Gamma^\lambda_{\mu\sigma} \, + \, \text{more}
\end{equation}
be the standard Riemann tensor calculated in terms of the Christoffel symbols. As it is well known, if we lower the unique upper index with the metric:
\begin{equation}\label{simello}
 \text{Rie}_{\mu\nu|\sigma\tau} \, \equiv \, \text{Rie}^{\lambda}_{\mu\nu,\sigma} \, g_{\lambda\tau}
\end{equation}
we obtain a tensor $\text{Rie}_{\mu\nu|\sigma\tau}$ that has two pairs of antisymmetric indices $\text{Rie}_{\mu\nu|\sigma\tau}\, = \,- \, \text{Rie}_{\mu\nu|\tau\sigma} $ and is symmetric in the exchange of the pairs:
\begin{equation}\label{kragelund}
  \text{Rie}_{\mu\nu|\tau\sigma} \, = \, \text{Rie}_{\tau\sigma|\mu\nu} 
\end{equation}
Eq.(\ref{kragelund}) is true for any Riemannian metric; for all the metrics of the form (\ref{caponata}) we have a stronger result. Seen as a matrix in the $\frac{n(n-1)}{2}$ space, whose basis elements are the antisymmetric pairs of covariant vectors, the Riemann tensor is not only symmetric, it is diagonal, namely we have:
\begin{equation}\label{ciromalotti}
  \text{Rie}_{\mu\nu|\tau\sigma}\, = \, \Lambda_{\mu\nu} (\boldsymbol{\sigma}) \, \delta_{\mu\nu|\tau\sigma}
\end{equation}
where the $\frac{n(n-1)}{2}$ eigenvalue functions $\Lambda_{\mu\nu} (\boldsymbol{\sigma})$ are the only non vanishing components of the whole Riemann tensor.  For instance for $n=4$, instead of the 20 independent components displayed by the Riemann tensor of a generic metric, the Riemann tensor of the metric (\ref{caponata}) has only 6 of them.
\par
We have not yet made a systematic study of the general structure of this peculiar Riemann tensor, so that we just limit our conveyed information to be a flash of the cases $n=3$ (discussed in detail within the vielbein formalism in appendix \ref{tregambone}) and $n=4$, in order to give the reader  \textit{a taste} of the result. For $n=3$ we have
 \begin{eqnarray}\label{casen3}
   \text{Rie}_{12|12}&=& -\frac{2 \left(\sigma _3^2+1\right)}{\left(\sigma _1^2-1\right) \left(\sigma
   _2^2-1\right) \left(\left(\sigma _2^2+\sigma _3^2+2\right) \sigma _1^2+2 \sigma
   _3^2+\sigma _2^2 \left(\sigma _3^2+2\right)+3\right)} \nonumber\\
   \text{Rie}_{13|13}&=& -\frac{2 \left(\sigma _2^2+1\right)}{\left(\sigma _1^2-1\right) \left(\sigma
   _3^2-1\right) \left(\left(\sigma _2^2+\sigma _3^2+2\right) \sigma _1^2+2 \sigma
   _3^2+\sigma _2^2 \left(\sigma _3^2+2\right)+3\right)} \nonumber\\
   \text{Rie}_{23|23}&=&-\frac{2 \left(\sigma _1^2+1\right)}{\left(\sigma _2^2-1\right) \left(\sigma
   _3^2-1\right) \left(\left(\sigma _2^2+\sigma _3^2+2\right) \sigma _1^2+2 \sigma
   _3^2+\sigma _2^2 \left(\sigma _3^2+2\right)+3\right)} \nonumber\\
 \end{eqnarray}
 For $n=4$ we have:
 {\scriptsize
 \begin{eqnarray}\label{casen3}
   \text{Rie}_{12|12}&=& - \, \frac{2 \left(\sigma _3^2+1\right) \left(\sigma _4^2+1\right)}{\left(\sigma
   _1^2-1\right) \left(\sigma _2^2-1\right) \left(\left(\left(\sigma _3^2+\sigma
   _4^2+2\right) \sigma _2^2+2 \sigma _4^2+\sigma _3^2 \left(\sigma
   _4^2+2\right)+3\right) \sigma _1^2+3 \sigma _3^2+2 \sigma _3^2 \sigma _4^2+3
   \sigma _4^2+\sigma _2^2 \left(\left(\sigma _4^2+2\right) \sigma _3^2+2 \sigma
   _4^2+3\right)+4\right)} \nonumber\\
  \text{Rie}_{13|13}&=& - \,\frac{2 \left(\sigma _2^2+1\right) \left(\sigma _4^2+1\right)}{\left(\sigma
   _1^2-1\right) \left(\sigma _3^2-1\right) \left(\left(\left(\sigma _3^2+\sigma
   _4^2+2\right) \sigma _2^2+2 \sigma _4^2+\sigma _3^2 \left(\sigma
   _4^2+2\right)+3\right) \sigma _1^2+3 \sigma _3^2+2 \sigma _3^2 \sigma _4^2+3
   \sigma _4^2+\sigma _2^2 \left(\left(\sigma _4^2+2\right) \sigma _3^2+2 \sigma
   _4^2+3\right)+4\right)} \nonumber\\
   \text{Rie}_{14|14}&=& - \,\frac{2 \left(\sigma _2^2+1\right) \left(\sigma _3^2+1\right)}{\left(\sigma
   _1^2-1\right) \left(\sigma _4^2-1\right) \left(\left(\left(\sigma _3^2+\sigma
   _4^2+2\right) \sigma _2^2+2 \sigma _4^2+\sigma _3^2 \left(\sigma
   _4^2+2\right)+3\right) \sigma _1^2+3 \sigma _3^2+2 \sigma _3^2 \sigma _4^2+3
   \sigma _4^2+\sigma _2^2 \left(\left(\sigma _4^2+2\right) \sigma _3^2+2 \sigma
   _4^2+3\right)+4\right)} \nonumber\\
    \text{Rie}_{23|23}&=& - \,\frac{2 \left(\sigma _1^2+1\right) \left(\sigma _4^2+1\right)}{\left(\sigma
   _2^2-1\right) \left(\sigma _3^2-1\right) \left(\left(\left(\sigma _3^2+\sigma
   _4^2+2\right) \sigma _2^2+2 \sigma _4^2+\sigma _3^2 \left(\sigma
   _4^2+2\right)+3\right) \sigma _1^2+3 \sigma _3^2+2 \sigma _3^2 \sigma _4^2+3
   \sigma _4^2+\sigma _2^2 \left(\left(\sigma _4^2+2\right) \sigma _3^2+2 \sigma
   _4^2+3\right)+4\right)} \nonumber\\
   \text{Rie}_{24|24}&=& - \, \frac{2 \left(\sigma _1^2+1\right) \left(\sigma _3^2+1\right)}{\left(\sigma
   _2^2-1\right) \left(\sigma _4^2-1\right) \left(\left(\left(\sigma _3^2+\sigma
   _4^2+2\right) \sigma _2^2+2 \sigma _4^2+\sigma _3^2 \left(\sigma
   _4^2+2\right)+3\right) \sigma _1^2+3 \sigma _3^2+2 \sigma _3^2 \sigma _4^2+3
   \sigma _4^2+\sigma _2^2 \left(\left(\sigma _4^2+2\right) \sigma _3^2+2 \sigma
   _4^2+3\right)+4\right)} \nonumber\\
   \text{Rie}_{34|34}&=& -\,\frac{2 \left(\sigma _1^2+1\right) \left(\sigma _2^2+1\right)}{\left(\sigma
   _3^2-1\right) \left(\sigma _4^2-1\right) \left(\left(\left(\sigma _3^2+\sigma
   _4^2+2\right) \sigma _2^2+2 \sigma _4^2+\sigma _3^2 \left(\sigma
   _4^2+2\right)+3\right) \sigma _1^2+3 \sigma _3^2+2 \sigma _3^2 \sigma _4^2+3
   \sigma _4^2+\sigma _2^2 \left(\left(\sigma _4^2+2\right) \sigma _3^2+2 \sigma
   _4^2+3\right)+4\right)} \nonumber\\
   &&\text{all other components vanish}
 \end{eqnarray}
 }
 It appears from the above formulae that there is a general rule behind the form of the Riemann tensor that we plan to decode and study within the framework of a next publication, yet the most important task is to unveil the relation of this peculiar structure of the Riemann tensor with the other general result
 that we mention next, namely the embedding of the manifold $\mathfrak{M}_{n|reg}$ equipped with metric (\ref{caponata}) as 
a codimension $n-1$ hypersurface in a flat Euclidean space $\mathbb{R}^{2n-1}$ which is what we can derive from inspection
of the structure of the immersion map \ref{calindri}.
\subsection{The embedding of $\mathfrak{M}_{n|reg}$ into flat Euclidean space $\mathbb{R}^{2n-1}$ following from the immersion map} 
Let us consider the immersion map (\ref{calindri}) and let us insert into it the change of coordinates (\ref{fargnocco}); introducing the basis of $2n$ elementary functions
\begin{eqnarray}\label{comescoglio}
  \mathit{f}_i(\boldsymbol{\sigma}) &=& \log[1+\sigma_i] \quad:\quad i=1,\dots,n \nonumber\\
  \mathit{f}_{n+i}(\boldsymbol{\sigma}) &=& \log[1-\sigma_i]\quad:\quad i=1,\dots,n 
\end{eqnarray}
we have that the immersion map reduces to:
\begin{equation}\label{alettino}
\iota_{2n} \quad : \quad  \xi_{i} \, = \, \left(\frac{\varrho}{2} -\log[2]\right)\, + \, \sum_{j=1}^{2n}\mathcal{E}^{[2n]}_{ij} \, \mathit{f}_j(\boldsymbol{\sigma}) \quad ; \quad i=1,\dots, 2n
\end{equation}
where $\mathcal{E}^{[m=2n]}$ is a constant square matrix with the very structure mentioned below :
\begin{equation}\label{E2nmatrix}
 \forall m\in \mathbb{N} \quad : \quad \mathcal{E}^{[m]}\, \equiv \, \mathrm{Id}_{m}\, - \, \frac{1}{m} \, J^{[m]}
\end{equation}
where we have introduced the following definition:
\begin{equation}\label{Jmdefi}
\forall\ell \, \in \, \mathbb{N} \quad:\quad  J^{[\ell]}_{ik}:=1 \ \text{ for every pair } (i,k)\, \quad ; \quad i,\,k \, =\, 1,2,\dots,\ell
\end{equation}
For instance for $n=5$ we have:
\begin{equation}\label{E2x5}
  \mathcal{E}^{[10]}\, = \, \left(
\begin{array}{cccccccccc}
 \frac{9}{10} & -\frac{1}{10} & -\frac{1}{10} & -\frac{1}{10} & -\frac{1}{10} &
   -\frac{1}{10} & -\frac{1}{10} & -\frac{1}{10} & -\frac{1}{10} & -\frac{1}{10} \\
 -\frac{1}{10} & \frac{9}{10} & -\frac{1}{10} & -\frac{1}{10} & -\frac{1}{10} &
   -\frac{1}{10} & -\frac{1}{10} & -\frac{1}{10} & -\frac{1}{10} & -\frac{1}{10} \\
 -\frac{1}{10} & -\frac{1}{10} & \frac{9}{10} & -\frac{1}{10} & -\frac{1}{10} &
   -\frac{1}{10} & -\frac{1}{10} & -\frac{1}{10} & -\frac{1}{10} & -\frac{1}{10} \\
 -\frac{1}{10} & -\frac{1}{10} & -\frac{1}{10} & \frac{9}{10} & -\frac{1}{10} &
   -\frac{1}{10} & -\frac{1}{10} & -\frac{1}{10} & -\frac{1}{10} & -\frac{1}{10} \\
 -\frac{1}{10} & -\frac{1}{10} & -\frac{1}{10} & -\frac{1}{10} & \frac{9}{10} &
   -\frac{1}{10} & -\frac{1}{10} & -\frac{1}{10} & -\frac{1}{10} & -\frac{1}{10} \\
 -\frac{1}{10} & -\frac{1}{10} & -\frac{1}{10} & -\frac{1}{10} & -\frac{1}{10} &
   \frac{9}{10} & -\frac{1}{10} & -\frac{1}{10} & -\frac{1}{10} & -\frac{1}{10} \\
 -\frac{1}{10} & -\frac{1}{10} & -\frac{1}{10} & -\frac{1}{10} & -\frac{1}{10} &
   -\frac{1}{10} & \frac{9}{10} & -\frac{1}{10} & -\frac{1}{10} & -\frac{1}{10} \\
 -\frac{1}{10} & -\frac{1}{10} & -\frac{1}{10} & -\frac{1}{10} & -\frac{1}{10} &
   -\frac{1}{10} & -\frac{1}{10} & \frac{9}{10} & -\frac{1}{10} & -\frac{1}{10} \\
 -\frac{1}{10} & -\frac{1}{10} & -\frac{1}{10} & -\frac{1}{10} & -\frac{1}{10} &
   -\frac{1}{10} & -\frac{1}{10} & -\frac{1}{10} & \frac{9}{10} & -\frac{1}{10} \\
 -\frac{1}{10} & -\frac{1}{10} & -\frac{1}{10} & -\frac{1}{10} & -\frac{1}{10} &
   -\frac{1}{10} & -\frac{1}{10} & -\frac{1}{10} & -\frac{1}{10} & \frac{9}{10} \\
\end{array}
\right)
\end{equation}
The matrix $\mathcal{E}^{[10]}$ has always rank $(2n-1)$ and its Null-space is the line:
\begin{equation}\label{lineale}
  \boldsymbol{\mathcal{L}}_1 \, \equiv \, r \, \times \, \boldsymbol{v[1]} \quad ; \quad \boldsymbol{v[1]}\, \equiv\, \left\{\underbrace{1,1,\dots,1,1}_{2n}\right\} \quad ; \quad r\in \mathbb{R}
\end{equation}
Hence introducing the orthogonal decomposition of the standard 
$\mathbb{E}_{2n}$ Euclidean space into the line $\boldsymbol{\mathcal{L}}_1\subset\mathbb{E}_{2n} $ plus its orthogonal hyperplane $\mathbb{E}_{2n-1}^{\boldsymbol{\bot}}$  :
\begin{equation}\label{hyperplanoL1}
  \mathbb{E}_{2n} \, = \, \boldsymbol{\mathcal{L}}_1 \oplus \mathbb{E}_{2n-1}^{\boldsymbol{\bot}}
\end{equation}
we arrive at the intrinsic algebraic interpretation of the matrix $\mathcal{E}^{[2n]}$ entering the explicit expression
(\ref{alettino}) of the immersion map $\iota_{2n}$. Indeed $\mathcal{E}^{[2n]}$ is just the projector of $\mathbb{E}_{2n}$ onto the Hyperplane orthogonal to the line (\ref{lineale}):
\begin{equation}\label{swampproj}
  \mathcal{E}^{[2n]}\, : \, \mathbb{E}_{2n} \, \longrightarrow \, \mathbb{E}_{2n-1}^{\boldsymbol{\bot}}
\end{equation}
That $\mathcal{E}^{[2n]}$ is a true projector is verified by the calculation of its eigenvalues that are:
\begin{equation}\label{autovalliE2n}
  \text{Eigenvalues of $\mathcal{E}^{[2n]}$} \, = \, \left\{\underbrace{1,1,\dots,1,1}_{2n-1},0\right\}
\end{equation}
Hence restricted to the orthogonal Hyperplane $\mathbb{E}_{2n-1}^{\boldsymbol{\bot}}$ the map $\mathcal{E}^{[2n]}$ is the identity map, while restricted to the line  $\boldsymbol{\mathcal{L}}_1$ it is zero. Then we can give an abstract interpretation of the immersion map (\ref{alettino}). Define the embedding of $\mathfrak{M}_{n|reg}$ into $\mathbb{E}_{2n}$ :
\begin{eqnarray}\label{bordighera}
\boldsymbol{H} & : & \mathfrak{M}_{n|reg} \, \longrightarrow \,\mathbb{E}_{2n}\nonumber\\
\forall\boldsymbol{\sigma} \in \mathfrak{M}_{n|reg} \quad :\quad   \boldsymbol{H}(\boldsymbol{\sigma})& \equiv & \left\{\mathit{f}_1(\boldsymbol{\sigma}),\mathit{f}_2(\boldsymbol{\sigma}),
  \dots,\mathit{f}_{2n}(\boldsymbol{\sigma})\right\} \, \in \, \mathbb{E}_{2n}
\end{eqnarray}
we can write:
\begin{equation}\label{beloigrib}
\mathbb{E}_{2n} \, \ni \,\boldsymbol{\xi}(\varrho,\boldsymbol{\sigma})\,=\,\underbrace{\left(\frac{\varrho}{2} \, - \, \log[2]\right ) \, \boldsymbol{v[1]}}_{\in \boldsymbol{\mathcal{L}}_1} \, \oplus \, \underbrace{\mathcal{E}^{[2n]} \left[\boldsymbol{H}(\boldsymbol{\sigma})\right]}_{\in  \mathbb{E}_{2n-1}^{\boldsymbol{\bot}}}
\end{equation}
Hence in order to make the immersion map:
\begin{eqnarray}\label{frugarolo}
  \hat{\boldsymbol{H}} & : & \mathfrak{M}_{n|reg} \, \longrightarrow \,\mathbb{E}_{2n-1}^{\boldsymbol{\bot}}\nonumber\\
\forall\boldsymbol{\sigma} \in \mathfrak{M}_{n|reg} & :&   \hat{\boldsymbol{H}}(\boldsymbol{\sigma})\, \equiv \, \mathcal{E}^{[2n]} \left[\boldsymbol{H}(\boldsymbol{\sigma})\right] \, \in \,\mathbb{E}_{2n-1}^{\boldsymbol{\bot}}
\end{eqnarray}
explicit, we just have to define a complete basis of vectors 
of the vector space $\mathbb{E}_{2n-1}^{\boldsymbol{\bot}}$, defined as the hyperplane orthogonal to $\boldsymbol{v[1]}$ in  
$\mathbb{E}_{2n}$. Such a problem is not a new one in algebra: indeed it is the standard problem one meets in the construction of the root space of the simple Lie algebra $\mathfrak{a}_{\ell=2n-1}$ (see for instance the relevant chapter in \cite{fre2023book}). The looked for basis is provided by the $2n-1$ simple roots:
\begin{eqnarray}
\label{radicisemplici}
  \boldsymbol{\alpha}_1 &=& \{1,-1,0,0,0,0,\dots, 0\} \nonumber\\
  \boldsymbol{\alpha}_2 &=& \{0,1,-1,0,0,0,\dots, 0\}  \nonumber\\
  \boldsymbol{\alpha}_3 &=& \{0,0,1,-1,0,0,\dots, 0\}  \nonumber\\
  \ldots &=& \ldots \nonumber\\
  \boldsymbol{\alpha}_{2n-2} &=& \{0,0,0,\dots,1,-1, 0\}  \nonumber\\
  \boldsymbol{\alpha}_{2n-1} &=& \{0,0,0,\dots,0,1, -1\} 
\end{eqnarray}
The matrix of scalar products of the simple roots is the Cartan matrix $\mathcal{C}_{ij}$, namely:
\begin{equation}\label{Cmatra}
  \langle \boldsymbol{\alpha}_i\, , \, \boldsymbol{\alpha}_j\rangle \, \equiv \,  \mathcal{C}_{[2n]ij} \, = \, \left(
                                   \begin{array}{cccccccc}
                                     2 & -1 & 0 & 0 & 0 & \cdots & 0 & 0 \\
                                     -1 & 2 & -1 & 0 & 0 & \cdots & 0 & 0 \\
                                     0 &-1 & 2 & 1 & 0 & \cdots & 0 & 0 \\
                                     \vdots & \cdots & \cdots & \cdots & \cdots & \cdots & \cdots& \vdots \\
                                      \vdots & \cdots & \cdots & \cdots & \cdots & \cdots & \cdots& \vdots \\
                                     0 & 0&\cdots & 0 & -1 & 2 & -1 & 0  \\
                                     0 & 0 & \cdots & 0 & 0 & 1 & 2 & -1 \\
                                     0 & 0 &\cdots & 0 & 0 & 0 & -1 & 2 \\
                                   \end{array}
                                 \right)  
\end{equation}
The basis of vectors dual to the simple roots is that of the simple weights $w^j$ ($j=1,\dots,2n-1$) such that:
\begin{equation}\label{grumotallo}
 \langle \boldsymbol{\alpha}_i\, , \, \boldsymbol{w}^j\rangle \, = \, \delta^j_i
\end{equation}
The simple weights are easily constructed in terms of the inverse of the Cartan matrix that for convenience we name as follows:
\begin{equation}\label{ICmatra}
  \mathcal{IC}_{[2n]} \, \equiv \, \mathcal{C}^{-1}_{2n}
\end{equation}
indeed we have:
\begin{equation}\label{ICmatra}
  w^j \, \equiv \,  \mathcal{IC}_{[2n]}^{ji} \, \alpha_i
\end{equation}
Given any vector $\boldsymbol{v}\in \mathbb{E}_{2n-1}^{\boldsymbol{\bot}}$ we can expand it into the basis of the simple roots:
\begin{eqnarray}\label{forthechiaro}
  \boldsymbol{v} & = & v^i \, \boldsymbol{\alpha}_i \nonumber\\
  v^i & = & \langle \boldsymbol{v},\boldsymbol{w}^i \rangle
\end{eqnarray}
 In particular this can be done for the vector $\mathcal{E}^{[2n]} \left[\boldsymbol{H}(\boldsymbol{\sigma})\right]$ appearing in eq.(\ref{frugarolo}) and we have:
\begin{eqnarray}\label{barlicco}
  \mathcal{E}^{[2n]} \left[\boldsymbol{H}(\boldsymbol{\sigma})\right]&=&\sum_{j=1}^{2n}\mathcal{E}^{[2n]}_{i,j} \,\mathit{f}_j(\boldsymbol{\sigma}) \,=\, \sum_{k=1}^{2n-1} T^{k}(\boldsymbol{\sigma}) \, \boldsymbol{\alpha}_k \nonumber \\
  T^{k}(\boldsymbol{\sigma}) &=& \langle \boldsymbol{w}^k \, ,\, \mathcal{E}^{[2n]} \left[\boldsymbol{H}(\boldsymbol{\sigma})\right] \rangle \, = \, \langle   \boldsymbol{w}^k \, , \, \mathcal{E}^{[2n]} \cdot \boldsymbol{H}(\boldsymbol{\sigma}) \rangle
\end{eqnarray}
In equation (\ref{barlicco}) the only dependence on the coordinate $\boldsymbol{\sigma}$ is in the vector $\boldsymbol{H}(\boldsymbol{\sigma}$ whose components are the $2n$ independent functions (\ref{comescoglio}) the other ingredients are just numerical constants that define the rectangular $(2n-1) \times (2n)$ matrix :
\begin{equation}\label{prugerro}
  \Pi^{k}_{\phantom{k} i}:\,= \,\sum_{j}^{2n} \, \boldsymbol{w}^k_j \, \mathcal{E}^{[2n]}_{ji}
\end{equation}
such that:
\begin{equation}\label{frosinone}
   T^{k}(\boldsymbol{\sigma}) \, = \, \sum_{i=1}^{2n}\Pi^{k}_{\phantom{k} i} \mathit{f}_i(\boldsymbol{\sigma})
\end{equation}
Consider now a flat Euclidean space $\hat{\mathbb{E}}_{2n-1} \,\simeq \,\mathbb{R}^{2n-1}$ where naming $\{y^k,\dots,y^{2n-1}\}$ the coordinates, the metric is:
\begin{equation}\label{kakkolino}
  ds^2_{\hat{\mathbb{E}}_{2n-1}} \, \stackrel{\text{def}}{=} \, \sum_{k,\ell=1}^{2n-1} \mathrm{d}y^i \, \mathrm{d}y^\ell \,\mathcal{C}_{[2n]k\ell}
\end{equation}
$\mathcal{C}_{[2n]k\ell}$ being the Cartan matrix (\ref{Cmatra}). The functions $T^{k}(\boldsymbol{\sigma})$ in eq.(\ref{frosinone}), derived from eq.(\ref{barlicco}) and therefore completely determined by the original form the thermo--metric upon  the vanishing of an uninterrupted sequence of (n-1) magnetic fields, constitute the embedding of the $\mathfrak{M}_{n|reg}$ manifold into the flat Euclidean space 
$\hat{\mathbb{E}}_{2n-1}$:
\begin{equation}\label{Tmappa}
 T \, = \, \mathfrak{M}_{n|reg} \, \longrightarrow \, \hat{\mathbb{E}}_{2n-1}
\end{equation}
 Indeed one can easily verify that:
\begin{equation}\label{granularis}
 T^\star\left[ ds^2_{\hat{\mathbb{E}}_{2n-1}}\right] \, = \, \sum_{k,\ell=1}^{2n-1}\,\mathrm{d}[T^{k}(\boldsymbol{\sigma})]\, \mathrm{d}[T^{\ell}(\boldsymbol{\sigma})] \,\mathcal{C}_{[2n]k\ell} \, = \, ds^2_{\mathfrak{M}_{n|reg}}
\end{equation}
the metric $ds^2_{\mathfrak{M}_{n|reg}}$ being defined in eq.(\ref{bracciodiferro}).
\par
The calculation of the rectangular map (\ref{prugerro}) is completely algorithmic and most easily automatized in Wolfram MATHEMATICA once the simple roots (\ref{radicisemplici}) are given.  For completeness we quote a quadruplet of examples starting from the case $n=5$ we utilized above:
\begin{equation}\label{pin5}
  \Pi^{[5]} \, = \,\left(
\begin{array}{cccccccccc}
 \frac{9}{10} & -\frac{1}{10} & -\frac{1}{10} & -\frac{1}{10} & -\frac{1}{10} &
   -\frac{1}{10} & -\frac{1}{10} & -\frac{1}{10} & -\frac{1}{10} & -\frac{1}{10}
   \\
 \frac{4}{5} & \frac{4}{5} & -\frac{1}{5} & -\frac{1}{5} & -\frac{1}{5} &
   -\frac{1}{5} & -\frac{1}{5} & -\frac{1}{5} & -\frac{1}{5} & -\frac{1}{5} \\
 \frac{7}{10} & \frac{7}{10} & \frac{7}{10} & -\frac{3}{10} & -\frac{3}{10} &
   -\frac{3}{10} & -\frac{3}{10} & -\frac{3}{10} & -\frac{3}{10} & -\frac{3}{10}
   \\
 \frac{3}{5} & \frac{3}{5} & \frac{3}{5} & \frac{3}{5} & -\frac{2}{5} &
   -\frac{2}{5} & -\frac{2}{5} & -\frac{2}{5} & -\frac{2}{5} & -\frac{2}{5} \\
 \frac{1}{2} & \frac{1}{2} & \frac{1}{2} & \frac{1}{2} & \frac{1}{2} &
   -\frac{1}{2} & -\frac{1}{2} & -\frac{1}{2} & -\frac{1}{2} & -\frac{1}{2} \\
 \frac{2}{5} & \frac{2}{5} & \frac{2}{5} & \frac{2}{5} & \frac{2}{5} & \frac{2}{5}
   & -\frac{3}{5} & -\frac{3}{5} & -\frac{3}{5} & -\frac{3}{5} \\
 \frac{3}{10} & \frac{3}{10} & \frac{3}{10} & \frac{3}{10} & \frac{3}{10} &
   \frac{3}{10} & \frac{3}{10} & -\frac{7}{10} & -\frac{7}{10} & -\frac{7}{10} \\
 \frac{1}{5} & \frac{1}{5} & \frac{1}{5} & \frac{1}{5} & \frac{1}{5} & \frac{1}{5}
   & \frac{1}{5} & \frac{1}{5} & -\frac{4}{5} & -\frac{4}{5} \\
 \frac{1}{10} & \frac{1}{10} & \frac{1}{10} & \frac{1}{10} & \frac{1}{10} &
   \frac{1}{10} & \frac{1}{10} & \frac{1}{10} & \frac{1}{10} & -\frac{9}{10} \\
\end{array}
\right)
\end{equation}
for $n=4$, we have instead:
\begin{equation}\label{pin4}
  \Pi^{[4]} \, = \,\left(
\begin{array}{cccccccc}
 \frac{7}{8} & -\frac{1}{8} & -\frac{1}{8} & -\frac{1}{8} & -\frac{1}{8} &
   -\frac{1}{8} & -\frac{1}{8} & -\frac{1}{8} \\
 \frac{3}{4} & \frac{3}{4} & -\frac{1}{4} & -\frac{1}{4} & -\frac{1}{4} &
   -\frac{1}{4} & -\frac{1}{4} & -\frac{1}{4} \\
 \frac{5}{8} & \frac{5}{8} & \frac{5}{8} & -\frac{3}{8} & -\frac{3}{8} &
   -\frac{3}{8} & -\frac{3}{8} & -\frac{3}{8} \\
 \frac{1}{2} & \frac{1}{2} & \frac{1}{2} & \frac{1}{2} & -\frac{1}{2} &
   -\frac{1}{2} & -\frac{1}{2} & -\frac{1}{2} \\
 \frac{3}{8} & \frac{3}{8} & \frac{3}{8} & \frac{3}{8} & \frac{3}{8} &
   -\frac{5}{8} & -\frac{5}{8} & -\frac{5}{8} \\
 \frac{1}{4} & \frac{1}{4} & \frac{1}{4} & \frac{1}{4} & \frac{1}{4} & \frac{1}{4}
   & -\frac{3}{4} & -\frac{3}{4} \\
 \frac{1}{8} & \frac{1}{8} & \frac{1}{8} & \frac{1}{8} & \frac{1}{8} & \frac{1}{8}
   & \frac{1}{8} & -\frac{7}{8} \\
\end{array}
\right)
\end{equation}  
while for $n=3$ we have:
\begin{equation}\label{pin3}
  \Pi^{[3]} \, = \,\left(
\begin{array}{cccccc}
 \frac{5}{6} & -\frac{1}{6} & -\frac{1}{6} & -\frac{1}{6} & -\frac{1}{6} &
   -\frac{1}{6} \\
 \frac{2}{3} & \frac{2}{3} & -\frac{1}{3} & -\frac{1}{3} & -\frac{1}{3} &
   -\frac{1}{3} \\
 \frac{1}{2} & \frac{1}{2} & \frac{1}{2} & -\frac{1}{2} & -\frac{1}{2} &
   -\frac{1}{2} \\
 \frac{1}{3} & \frac{1}{3} & \frac{1}{3} & \frac{1}{3} & -\frac{2}{3} &
   -\frac{2}{3} \\
 \frac{1}{6} & \frac{1}{6} & \frac{1}{6} & \frac{1}{6} & \frac{1}{6} &
   -\frac{5}{6} \\
\end{array}
\right)
\end{equation}
and finally for $n=2$ we have:
\begin{equation}\label{pin3}
  \Pi^{[3]} \, = \,\left(
\begin{array}{cccc}
 \frac{3}{4} & -\frac{1}{4} & -\frac{1}{4} & -\frac{1}{4} \\
 \frac{1}{2} & \frac{1}{2} & -\frac{1}{2} & -\frac{1}{2} \\
 \frac{1}{4} & \frac{1}{4} & \frac{1}{4} & -\frac{3}{4} \\
\end{array}
\right)
\end{equation}
\subsection{Another formulation of the embedding of $\mathfrak{M}_{n|reg}$ into flat Euclidean space $\mathbb{R}^{2n-1}$ with standard metric.}
\label{sashaembed}
Obviously the constant, symmetric, positive-definite Cartan matrix (\ref{Cmatra}) can be diagonalized, for each value of $n$, and then rescaled in such a way as to become the Kronecker delta, yet concetually this is not necessary since $\hat{\mathbb{E}}_{2n-1}$ is diffeomorphic to, and it has the same group of isometries as, ${\mathbb{E}}_{2n-1}$. Yet, when it happens to be useful for the visualization purpose of the embedded hypersurface, we mention that another form of the embedding of $\mathfrak{M}_{n|reg}$ into $\mathbb{E}_{2n-1}$ with standard Kronecker delta metric is possible and can be described in general terms for all values of $n$.
\par
Indeed the $n$-dimensional hyper surface  with the Riemann metric (\ref{caponata}) can be embedded  into the ($2n-1$)-dimensional space with flat Euclidean metric
\begin{eqnarray}
ds^2_{\mathfrak{M}_{n|reg}}=\sum_{i=1}^n dx_i^2+\sum_{k=1}^{n-1}dz_k^2,
\label{flat_space_n}
\end{eqnarray}
as an $n$-dimensional graph
\begin{eqnarray}
(x_1,\dots,x_n)\ \longmapsto\ \big(x_1,\dots,x_n,\ z_1(x),\dots, z_{n-1}(x)\big)\ \in\ \mathbb R^{2n-1}\,,
\end{eqnarray}
where 
\begin{eqnarray}
\label{grafuccio}
\left\{x_i=\frac{1}{\sqrt{2}}\log\left(\frac{1+\sigma_i}{1-\sigma_i}\right)\,,\quad z_k(x)=\sqrt2\sum_{i=1}^n \boldsymbol{r}^{(k)}_i\,\log\cosh\!\left(\frac{x_i}{\sqrt2}\right),\quad k=1,\dots,n-1\right\}\,.   
\end{eqnarray}
using the following special vectors
\begin{eqnarray}
\label{roots}
&\boldsymbol{r}^{(k)}=\frac{1}{\sqrt{k(k+1)}}\big(\underbrace{1,\dots,1}_{k},-k,0,\dots,0\big)\,, \quad k=1,\dots,n-1\,,&
\\&
\sum_{a=1}^{n-1} \boldsymbol{r}^{(a)}_i\,\boldsymbol{r}^{(a)}_j \;=\; \mathcal{E}^{n}_{ij}&
\end{eqnarray}
Obviously the factorization of the matrix $\mathcal{E}^{2n}_{ij}$ in terms of the $(2n-1)$-vectors $\boldsymbol{r}^{(k)}$ might have been used  just after eq.(\ref{alettino}) setting;
\begin{eqnarray}\label{bradipo}
  \mathcal{E}^{2n}_{ij} & = &\sum_{a=1}^{2n-1} \boldsymbol{r}^{(a)}_i\,\boldsymbol{r}^{(a)}_j \nonumber\\ 
  \boldsymbol{r}^{(k)}&=&\frac{1}{\sqrt{k(k+1)}}\big(\underbrace{1,\dots,1}_{k},-k,0,\dots,0\big)\,, \quad k=1,\dots,2n-1\,
\end{eqnarray}
yet this would have obscured the nature of the decomposition   
(\ref{hyperplanoL1}) which is the heart of the immersion map
(\ref{alettino}). 
\subsubsection{Geometric interpretation of the embedded surface}
Via the above described immersion map, the manifold $\mathfrak{M}_{n|reg}$ is realized as a generalized translation-type hypersurface: each of its $n-1$ independent height functions  $\{z_k(x)\}$ is a fixed linear combination of the same profile function $\log\cosh$ applied separately to each $x_i$, projected onto the traceless ($S_n$-invariant complement of the diagonal) subspace. This is the natural $n$-dimensional generalization of the $n=2$ translation surface $z=\log\cosh(x/\sqrt2)-\log\cosh(y/\sqrt2)$, and it manifestly realizes the $B_n$ symmetry group geometrically: permuting the $\{x_i\}$ permutes the $\{z_k\}$ correspondingly, and $x_i\to -x_i$  leaves  every height function fixed, since $\log\cosh(x/\sqrt2)$ is an even function  — exactly reproducing the hyperoctahedral isometries as restrictions of ambient permutations/reflections of the $x_i$-axes in $\mathbb R^{2n-1}$. 
\section{About  geodesics of the universal $\mathfrak{M}_{3|reg}$ manifold.}
In appendix \ref{tregambone} we provide a detailed analysis of 
the anholonomic curvature tensor for the $\mathfrak{M}_{3|reg}$ manifold utilizing the vielbein formalism. The goal of the appendix is that of illustrating the behavior of the multicomponent curvature  that generalizes the features observed in two dimensions. As in $n=2$ the physical domain is a square with sides of length 2, in the same way in $n=3$ the physical domain is a cube with edges of  length 2 and one easily 
extrapolates the result to a generic $n$, where the physical domain is a \textbf{hypercube} with edges also of lenght 2. In the $n=2$ case we observed that the origin $\{0,0\}$ is an absolute minimum of the curvature scalar. Similarly from the results of appendix \ref{tregambone} we see that in the origin $\{0,0,0\}$ the curvature $2$-form is diagonal in the sense that:
\begin{equation}\label{curvataorigen}
  \mathfrak{R}^{ij}|_{\{0,0,0\}} \, = \, - \, \frac{1}{6} \, V^i \wedge V^j
\end{equation}
and it has the same form as the curvature $2$-form of a symmetric space $\mathrm{SO(1,n)}/\mathrm{SO(n)}$. We \textbf{conjecture} that such a statement is true for all manifolds $\mathfrak{M}_{n|reg}$. Along the coordinate axes the curvature $2$-form is also diagonal but it reaches the boundary with a different destiny of its components, namely at the crossing point of the coordinate axis with the boundary the curvature $2$-form degenerates into that of a direct product $\mathbb{R}\times \frac{\mathrm{SO(1,n-1)}}{\mathrm{SO(n-1)}}$. This which is certainly true for $n=3$ we \textbf{conjecture} should be true in general. As we show in appendix \ref{tregambone} and we illustrate in Fig. \ref{densimetro} the behavior of the curvature along the six lines stretching from the origin  to the 6 cube faces provide the analogue of the four mountain ridges displayed in Fig. \ref{curva2odd} for the $n=2$ case. Finally the behavior of the curvature $2$-form components along the 8 diagonals that join the origin to the eight vertices of the cube (see eq.s(\ref{angolari})) constitute the analogue of the $4$ valleys shown in Fig. \ref{curva2odd} for the $n=2$ case.
\begin{figure}[ht]
\begin{center}
\includegraphics[width=9cm]{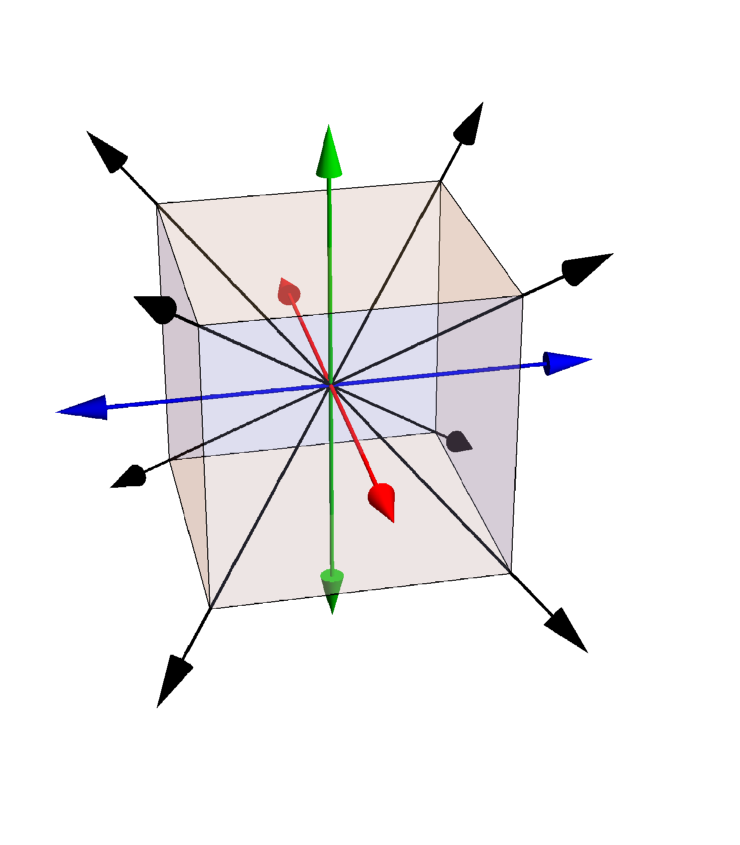}
 \caption{\label{centralina} In this figure we illustrate the 14 special directions along which the limit to the boundary  of the curvature 2-form $\mathfrak{R}^{ij}$ produces different results.  Going to infinity along the red, blue or green directions, namely the 6 in correspondence with the 3 pairs of opposite faces of the cube, one obtains the result of eq.(\ref{fragitoso}). Going to the boundary (namely to the surface of the cube) along any of the 8 black directions that are in one-to-one correspondence with the eight vertices of the unit cube around the origin, one obtains the result of eq.(\ref{angolari}). }
\end{center}
\end{figure} 
\par The behavior of the curvature $2$-form, recalled above and summarized in Fig. \ref{centralina},  is obviously the responsible structure for the peculiar behavior of the geodesics that have a feature observed in the $n=2$ case and, as we are going show in the present section, confirmed in the case $n=3$.
The mentioned feature is that on the boundary at infinity, namely on the boundary of the cube, all geodesic have as terminal point:
\begin{enumerate}
  \item either one of the 8 vertices,
  \item or the central point of  one of the 6  faces,
  \item or the midpoint of one of the 12 edges.  
\end{enumerate}
We illustrate this behavior with numerically calculated examples and we \textbf{conjecture} that such a feature must generalize \textit{mutatis mutandis} to the geodesics of any $\mathfrak{M}_{3|reg}$ manifold. Such a property is what we described before as a sort of \textbf{causal structure} of the border at infinity $\partial_{\infty}\mathfrak{M}_{3|reg}$. Indeed what seems to decide the fate of a geodesic is not its initial point, rather its initial direction. Postponing to further studies and future publication this extremely interesting feature probably strategic for Machine Learning Applications, we focus on the examples that display the claimed property. 
\par
\begin{figure}[htb]
\begin{center}
\includegraphics[width=7cm]{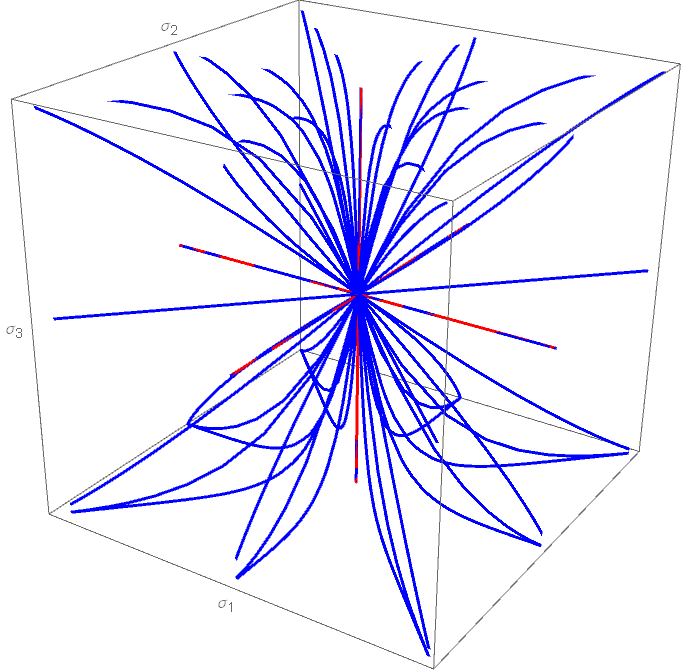}
\includegraphics[width=7cm]{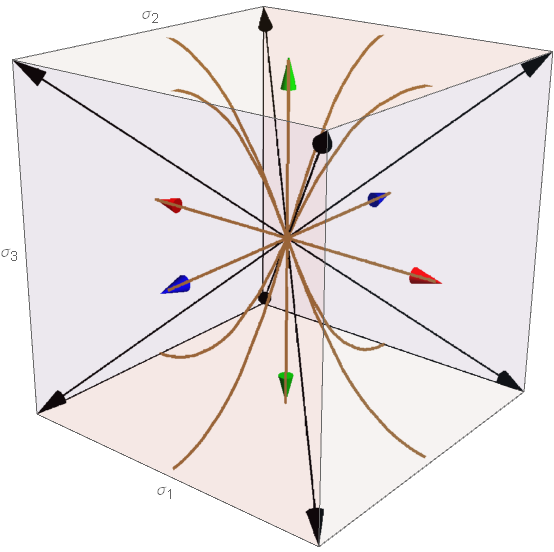}
 \caption{\label{destiny3d} In this picture we show examples of geodesic grids starting from the origin $\{0,0,0\}$. In the picture on the left by choice of the angles we capture geodesics that tend to the vertices of the cube. In the picture on the right by choice of the angles we select geodesics that tend to the side-midpoints.} 
\end{center}
\end{figure}

\begin{figure}[htb]
\begin{center}
\includegraphics[width=7cm]{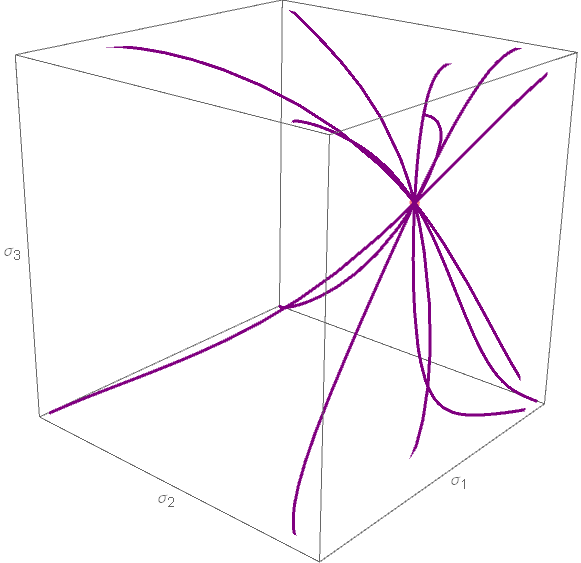}
\includegraphics[width=7cm]{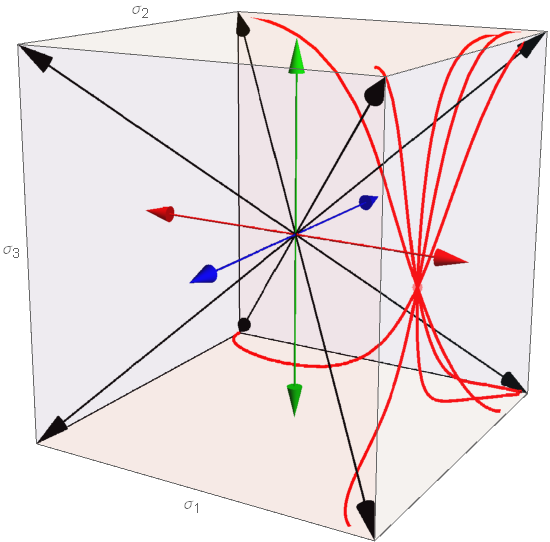}
 \caption{\label{destiny3d1} In this picture we show examples of geodesic grids starting from points inside the cube, different from the origin $\{0,0,0\}$. The destiny of every geodesic to some of the mentioned possible endpoints on the boundary is clear from the pictures.} 
\end{center}
\end{figure}

\begin{figure}[htb]
\begin{center}
\includegraphics[width=6cm]{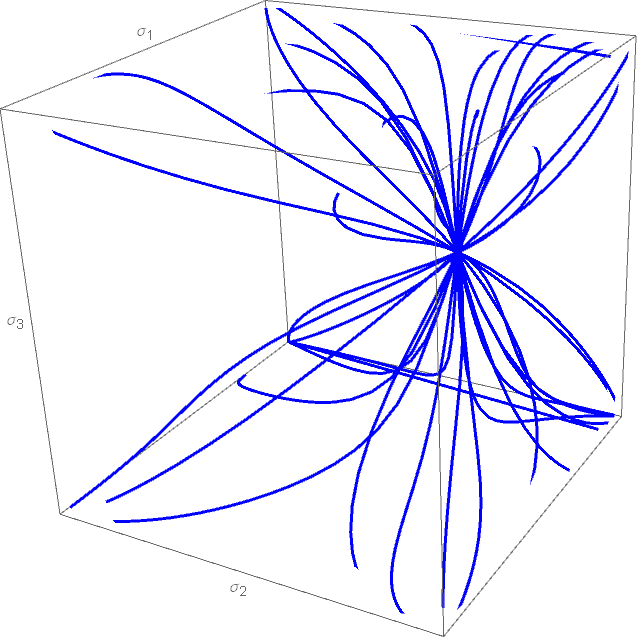}
\includegraphics[width=6cm]{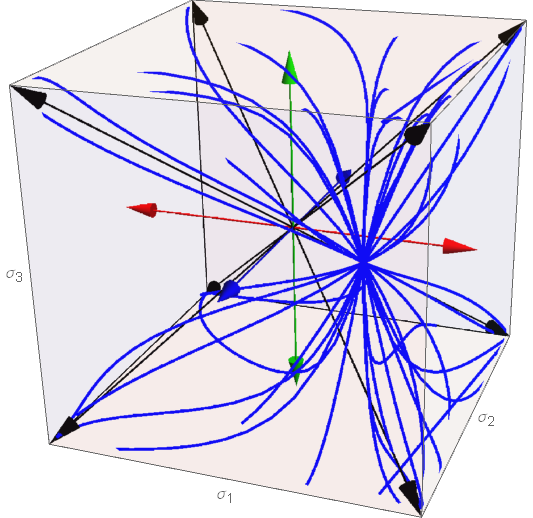}
 \caption{\label{destiny3d2} In this picture we show a denser grid of geodesics starting from an interior point of the cube, different from the origin. In the picture on the left the grid is displayed without the special directions. In the picture on the right the same grid is displayed on the background of the marked special axes. } 
\end{center}
\end{figure}

 
As we did before for $n=2$, we present here  for 
$n=3$, examples of grids of geodesics starting at different points of the manifold. The method is the same and it is general. From eq.s(\ref{bracciodiferro}-\ref{caponata}) we extract the geodesic lagrangian:
\begin{equation}\label{kurdopiatto}
 \mathcal{L}_{n}(\boldsymbol{\dot{\sigma}}\, , \,\boldsymbol{\sigma}) \, = \, \text{const} \, \times \,\left(  \frac{2}{n}\sum_{i=1}^{n} \, \frac{n+(n-1)\sigma_i^2}{(-1+\sigma_i^2)^2} \,\dot{\sigma}_i^2\, - \,\frac{2}{n} \, \sum_{i\neq j}\, \frac{\sigma_i \, \sigma_j}{(-1+\sigma_i^2)(-1+\sigma_j^2)} \, \dot{\sigma}_i\,\dot{\sigma}_j\right)
\end{equation}
and from $\mathcal{L}_{n}(\boldsymbol{\dot{\sigma}}\, , \,\boldsymbol{\sigma})$ we  can derive the corresponding highly non linear Euler-Lagrange equations for $\sigma_i(t)$. We solve the latter numerically with a dedicated MATHEMATICA code  by fixing, as initial conditions, the starting point $\sigma(0)$ and the initial velocities $\dot{\sigma}_i(0)$. The code allows also the specification of the maximal time $T$ for the solution.
The examples are presented in Fig. s\ref{destiny3d},\ref{destiny3d1},\ref{destiny3d2}
\newpage 
\section{The CV Thermo--Metric in the $q=\text{even}$ case}
\label{evencrat}
We finally come to the analysis of the Thermo-Metric in the $q=\text{even}$ case. As we anticipated before, the even case differs from the odd one because of a bunch of three directions that give rise to a characteristic curved 2-dimensional submanifold displaying also a curvature singularity. Instead, all the other directions arrange themselves into a flat manifold, completely isomorphic to the analogue submanifold of the odd case. Furthermore the vanishing of magnetic fields proceeds in the same way as in the odd case and generates exactly the same curved manifolds we have already analyzed in the previous sections. For these reasons we just concentrate on the additional distinguishing phenomenon associated with the first three temperatures and the first $\hat{\mu}$ parameter.
\par
As in the previous $q=\text{odd}$ case, the Thermo-Metric is easily obtained by calculating the Hessian matrix of the stochastic hamiltonian:
\begin{eqnarray}\label{cardinaleven}
ds^2_{s} & \equiv & \mathrm{d}\lambda^i \, \mathrm{d}\lambda^j \, \frac{\partial^2}{\partial\lambda^i \, \partial\lambda^j}\,\mathcal{H}^{sto}_{s}(\boldsymbol{\lambda}) \quad ; \quad i,j=1,\dots, 2s +2 \nonumber\\
 \mathcal{H}^{sto}_{s}(\boldsymbol{\lambda})& \equiv & - \,\log\left[Z_{s}^{\boldsymbol{\mathfrak{d}}}(\boldsymbol{\lambda})\right]\nonumber\\
 \boldsymbol{\lambda}& = & \left\{\beta_0,\,\beta_1,\, \dots , \beta_{s+1}, \, h^1,\, \dots, h^{s} \right\}
\end{eqnarray}
Setting
\begin{equation}\label{carciofoide}
  \beta_1\, = \, u \quad ; \quad \beta_2 \, = \, v \quad ; \quad \hat{\mu}_1 \, = \, w
\end{equation}
and utilizing the original names $\beta_i$ and $\hat{\mu}_k$ for all the other coordinates we obtain the following form of the metric
\begin{eqnarray}
ds^2_{s} &=& \,\frac{\mathrm{d}\beta_0^2}{\beta_0^2} \, + \, \sum_{i=3}^{s+1}\,\frac{1}{\left(\beta_i^2-\hat{\mu}^2_{i-1}\right)^2}\,
\left[2\,\left(\beta_i^2+\hat{\mu}^2_{i-1}\right)\,(d\beta_i^2 + d\hat{\mu}_{i-1}^2)\, - \, 8 \beta_i \, \hat{\mu}_{i-1} \,d\beta_i \,d\hat{\mu}_{i-1}\right] \nonumber\\
&&+ds^2_{even3}  \label{distaquadeven}
\end{eqnarray}
where
\begin{eqnarray}\label{krillo}
  ds^2_{even3} & = & \mathrm{d}w^2
   \left(\frac{1}{(u+w)^2}+\frac{1}{(u-w)^2}+\frac{1}{(v-w)
   ^2}+\frac{1}{(v+w)^2}+\frac{1}{w^2}\right)\nonumber\\
   &&+\mathrm{d}u^2
   \left(\frac{1}{(u+w)^2}+\frac{1}{(u-w)^2}\right)
   -\frac{8\, u\, w\, \mathrm{d}u \, \mathrm{d}w}{\left(u^2-w^2\right)^2}\nonumber\\
   &&+\mathrm{d}v^2
   \left(-\frac{1}{(v+w)^2}-\frac{1}{(v-w)^2}\right)+\frac{8\, v\, w\,
   \mathrm{d}v\, \mathrm{d}w}{\left(v^2-w^2\right)^2}
\end{eqnarray}
As one sees, the part of the thermo-metric in the q=even case that is residual after the subtraction of the three-dimensional metric (\ref{krillo}),  is identical in form to the metric of the q=odd case (see eq.(\ref{cardinalodd}) upon changing $\hat{\mu}_{i-1}\to {\mu}_{i-1}$ and starting the summation at $i=3$ rather than  at $i=1$. Hence the reduced models that are obtained in the odd case by identifying one or more $\mu_i$.s through the vanishing of one or more magnetic fields are found
identically, \textit{mutatis mutandis} also in the even case.
\par
The three-dimensional metric (\ref{krillo}), instead, is quite similar to the three-dimensional metric (\ref{olunitre}) that we found while switching off one magnetic field in the odd-dimensional case yet it differs from it because of the additional term $\frac{\mathrm{d}w^2}{w^2}$. This difference implies some substantial change in the geometry, as we are going to see. As for the other terms in eq.(\ref{distaquadeven}) they provide the metric of a flat $(2s-1)$-dimensional manifold. To see this it suffices to perform the analogue of the transformation (\ref{roncoscrivia}):  
\begin{equation}\label{stradella}
  \beta_0 \, = \, \exp[\rho_0] \quad ; \quad \beta_i \, = \, \exp[\rho_i] \, - \, \exp[\rho_{s-1+i}]\quad ; \quad \hat{\mu}_{i-1} \, = \, \exp[\rho_i] \, + \, \exp[\rho_{s-1+i}] \quad : \quad i=3,\dots,s+1
\end{equation}
and the metric (\ref{distaquadeven}) becomes:
\begin{equation}\label{prometeo}
  ds^2_{s} \, = \, ds^2_{even3}\, + \, \mathrm{d}\rho_0^2\, +\,
  \sum_{i=3}^{2s} \mathrm{d}\rho_i^2 
\end{equation}
\subsection{Analysis of the $ds^2_{even3}$ metric}
In a way similar to the previous treatment of the $ds^2_{\mathfrak{B}_{3|reg}}$ metric we begin with a set of dreibein as follows:
\begin{eqnarray}\label{dreibein2}
  \boldsymbol{\tilde{e}}^1 & = & \frac{\sqrt{2} \left(\mathrm{d}u \left(u^2+w^2\right)-2\, u\, w\,
   \mathrm{d}w\right)}{\left(w^2-u^2\right) \sqrt{u^2+w^2}}                   \nonumber\\ 
  \boldsymbol{\tilde{e}}^2 & = &  \frac{\sqrt{2} \left(\mathrm{d}v \left(v^2+w^2\right)-2\, v\, w\,
   \mathrm{d}w\right)}{\left(w^2-v^2\right) \sqrt{v^2+w^2}}                 \nonumber\\
  \boldsymbol{\tilde{e}}^3 & = & \mathrm{d}w \sqrt{2
   \left(\frac{1}{u^2+w^2}+\frac{1}{v^2+w^2}\right)-\frac{1}
   {w^2}}         \nonumber\\
\end{eqnarray}
we calculate the Ricci tensor with anholonomic indices and we find that it has one null eigenvalue and two identical non vanishing eigenvalues. We derive the local $\mathrm{O(3)}$ gauge transformation that diagonalises the Ricci tensor.
\par
It is given by the following orthogonal matrix
\begin{equation}\label{newotra}
  \hat{\mathcal{O}} \, = \, \left(
\begin{array}{ccc}
 -\frac{\sqrt{\frac{2}{3}} u}{\sqrt{u^2+w^2}} &
   -\frac{\sqrt{\frac{2}{3}} v}{\sqrt{v^2+w^2}} &
   \frac{1}{\sqrt{3} \sqrt{\frac{\left(u^2+w^2\right)
   \left(v^2+w^2\right)}{w^2 \left(u^2+v^2\right)-u^2 v^2+3
   w^4}}} \\
 \sqrt{\frac{w^2 \left(u^2+v^2\right)-u^2 v^2+3
   w^4}{\left(u^2+w^2\right) \left(v^2+3 w^2\right)}} & 0 &
   \sqrt{2} u \sqrt{\frac{v^2+w^2}{\left(u^2+w^2\right)
   \left(v^2+3 w^2\right)}} \\
 -\frac{2 u v}{\sqrt{3} \sqrt{\left(u^2+w^2\right)
   \left(v^2+3 w^2\right)}} & \frac{1}{\sqrt{\frac{2
   v^2}{v^2+3 w^2}+1}} & \sqrt{\frac{2}{3}} v w
   \sqrt{\frac{u^2 \left(\frac{4}{v^2+3
   w^2}-\frac{1}{w^2}\right)+1}{\left(u^2+w^2\right)
   \left(v^2+w^2\right)}} \\
\end{array}
\right)
\end{equation} 
\begin{figure}[ht]
\begin{center}
\includegraphics[width=7cm]{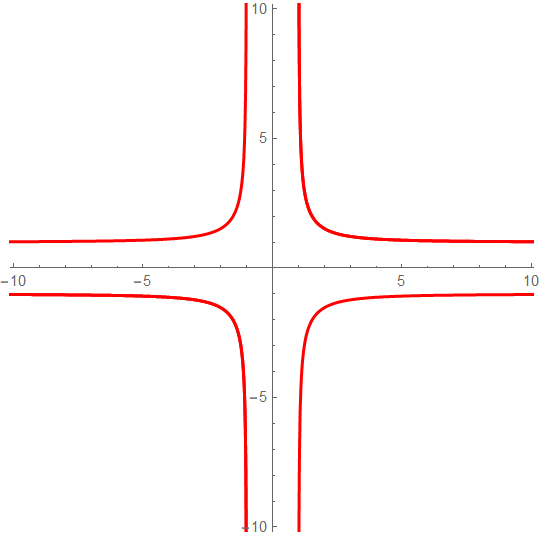}
\includegraphics[width=7cm]{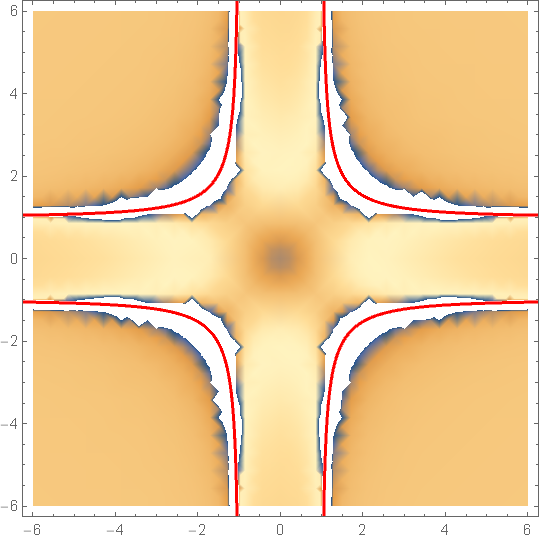}
 \caption{\label{curvaturepari} In this figure we show the singular locus of the curvature in eq.(\ref{crologino}). In the picture on the left the four red curves correspond to the singularities. In the figure on the right the same curves are superimposed on the density plot of the curvature.  }
\end{center}
\end{figure}
Defining the new dreibein as follows:
\begin{equation}\label{kematra}
  \boldsymbol{\bar{e}}^i \, = \, \hat{\mathcal{O}}^{ij}\,\boldsymbol{\tilde{e}}^j 
\end{equation}
we obtain:
\begin{eqnarray}\label{mortadella}
\bar{\boldsymbol{e}}^1 & = & \frac{-\mathrm{d}w \left(w^2 \left(u^2+v^2\right)+u^2 v^2-3
   w^4\right)+2 u\, w\, \mathrm{d}u\, (v-w) (v+w)+2 \,v\, w \, \mathrm{d}v (u-w)
   (u+w)}{\sqrt{3} w \left(w^2-u^2\right)
   \left(w^2-v^2\right)}               \nonumber\\ 
  \bar{\boldsymbol{e}}^2 & = & \frac{(w\, \mathrm{d}u-u \, \mathrm{d}w) \sqrt{u^2 \left(\frac{8}{v^2+3
   w^2}-\frac{2}{w^2}\right)+2}}{w^2-u^2}  \nonumber\\ 
  \bar{\boldsymbol{e}}^3 & = &  \frac{\sqrt{\frac{2}{3}} \left(-v\, \mathrm{d}w \left(u^2 \left(v^2-5
   w^2\right)+w^2 \left(v^2+3 w^2\right)\right)+w\, \mathrm{d}v (w-u)
   (u+w) \left(v^2+3 w^2\right)+2 \,u\, v\, w\, \mathrm{d}u (v-w)
   (v+w)\right)}{w\, \left(w^2-u^2\right) \left(w^2-v^2\right)
   \sqrt{v^2+3 w^2}}              \nonumber\\ 
\end{eqnarray}
and we verify that the first dreibein is an exact form:
\begin{equation}\label{exactamente}
  \bar{\boldsymbol{e}}^1 \, = \, d\boldsymbol{\mathit{f}} \quad ; \quad \boldsymbol{\mathit{f}} \, = \, \frac{\log \left(\frac{\left(w^2-u^2\right)
   \left(w^2-v^2\right)}{w}\right)}{\sqrt{3}}
\end{equation}

\begin{figure}[htb]
\begin{center}
\includegraphics[width=7cm]{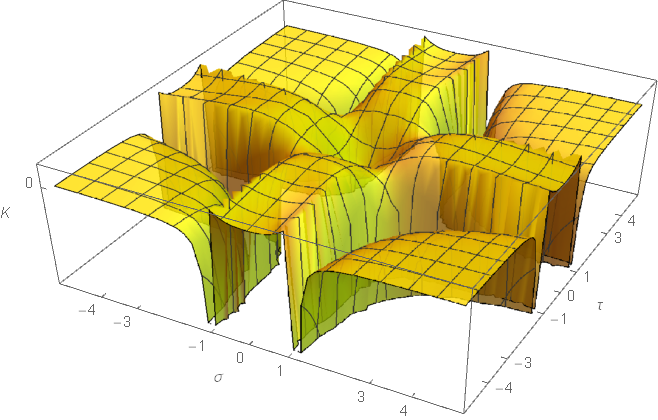}
\includegraphics[width=10cm]{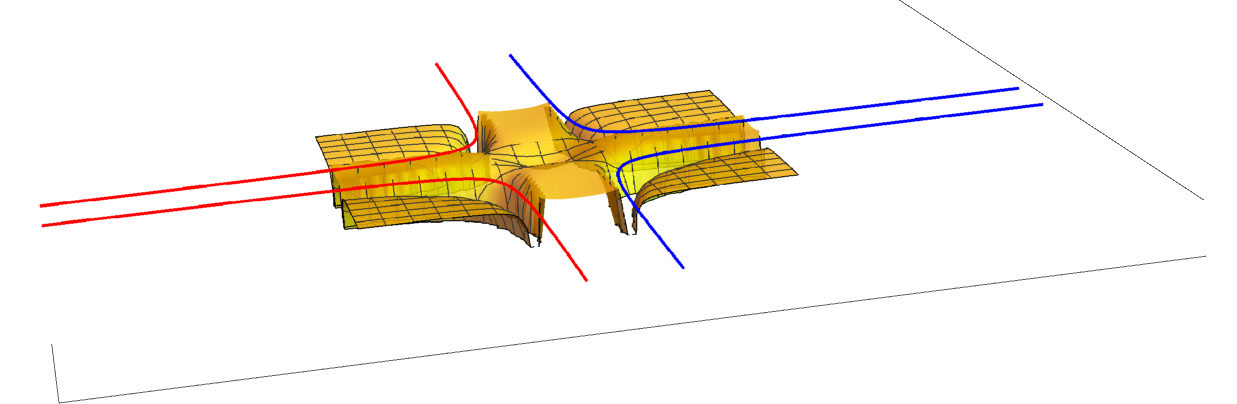}
 \caption{\label{tremendo} In this figure we show the $3$-dimensional plot of the curvature in eq.(\ref{crologino}). In the picture on the left we just display the plot. In the picture on the right we  superimpose the bidimensional plot of the singular curves that precisely run at the bases of the singular walls.  }
\end{center}
\end{figure}
\begin{figure}[htb]
\begin{center}
\includegraphics[width=4cm]{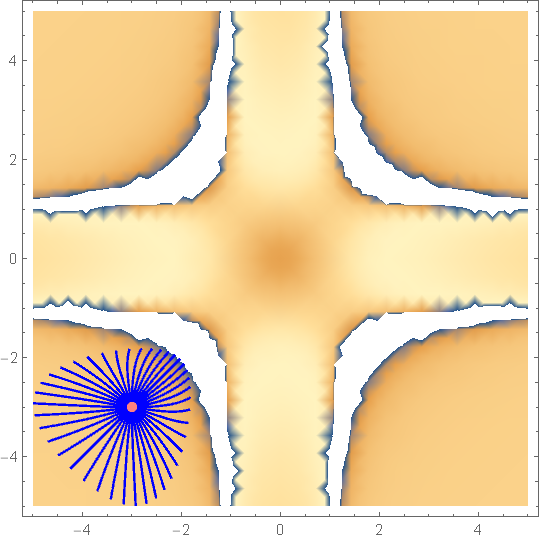}
\includegraphics[width=4cm]{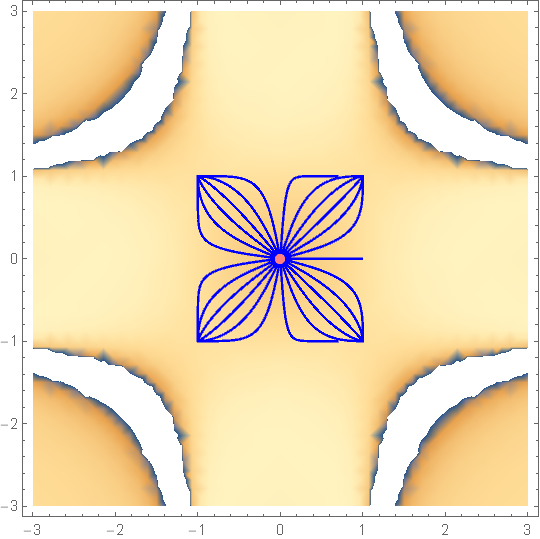}
\includegraphics[width=4cm]{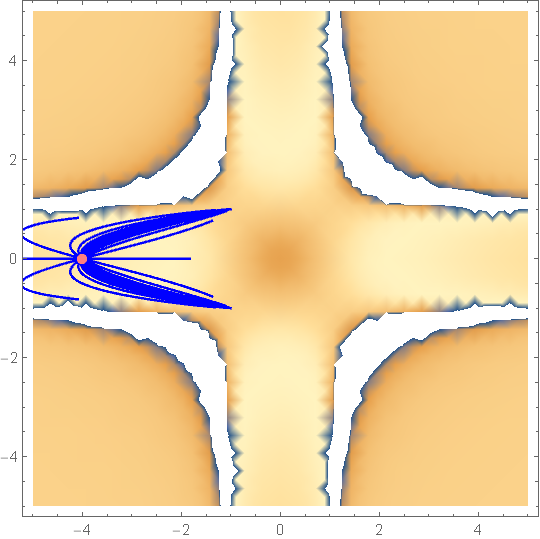}
\includegraphics[width=4cm]{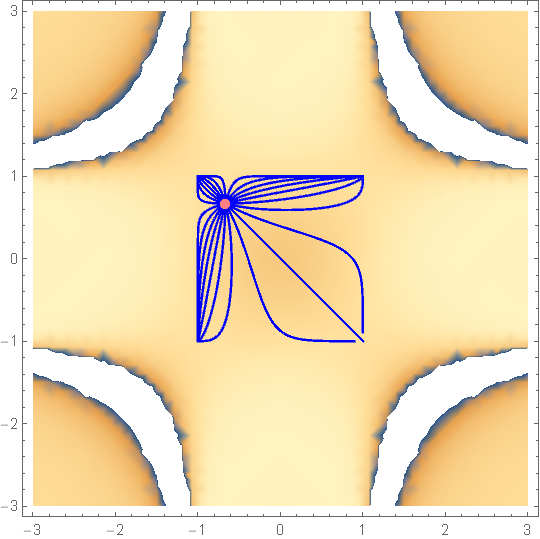}
\includegraphics[width=4cm]{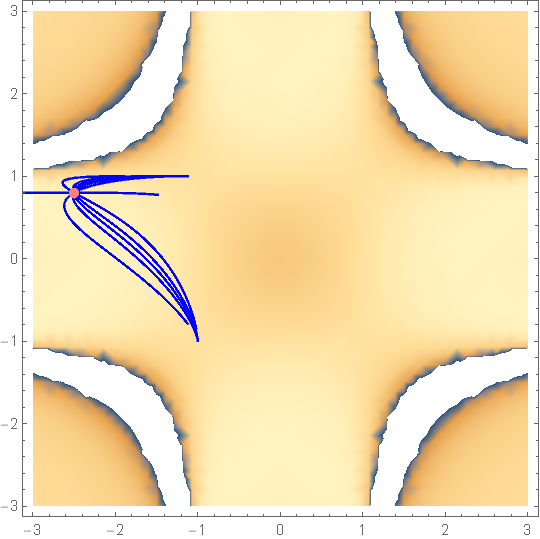}
 \caption{\label{poffarba} In this picture we show a collection of geodesics grids starting at different points of the permitted regions separated by the curvature singulartiy walls,}
\end{center}
\end{figure}
Hence changing variables in a way similar to eq.(\ref{ciulatoprimo})
\begin{equation}\label{ciulatone}
  u\,= \, \sigma \,w \quad ; \quad v \,= \, \tau \,w \quad ; \quad \varrho \, = \, \frac{1}{\sqrt{3}}\log \left(w^3 \sqrt{\left(\sigma ^2-1\right) \left(\tau
   ^2-1\right)}\right)
\end{equation}
in the new coordinates $\sigma,\tau,\varrho$ we find:
\begin{eqnarray}\label{lardone}
\bar{\boldsymbol{e}}^1 & = &  \mathrm{d}\varrho           \nonumber\\ 
  \bar{\boldsymbol{e}}^2 & = & -\frac{\mathrm{d}\sigma \sqrt{\sigma ^2 \left(\frac{8}{\tau
   ^2+3}-2\right)+2}}{\sigma ^2-1} \nonumber\\ 
  \bar{\boldsymbol{e}}^3 & = & \frac{\sqrt{\frac{2}{3}} \left(\frac{2 \sigma  \tau 
   \mathrm{d}\sigma}{\sigma ^2-1}-\frac{\left(\tau ^2+3\right)
   \mathrm{d}\tau}{\tau ^2-1}\right)}{\sqrt{\tau ^2+3}}   \nonumber\\ 
\end{eqnarray}
Eq.(\ref{lardone}) has to be compared with eq.(\ref{mortadella}).
Just as in the previous case the coordinate $\varrho$ spans an additional flat direction while $\bar{\boldsymbol{e}}^2=V^1$ and $\bar{\boldsymbol{e}}^3 = V^2$ can be taken as the zweibein of a two dimensional space $\mathfrak{M}_{2|sing}$. 
\par
Calculating its spin-connection and its curvature $2$-form we find:
\begin{equation}\label{curvateven}
  \mathfrak{R}^{12} \, = \, \mathcal{K}(\sigma,\tau)\, V^1\wedge V^2 \quad; \quad \mathcal{K}(\sigma,\tau) \, = \,-\,\frac{3 \left(\sigma ^2-1\right) \left(\tau
   ^2-1\right)}{\left(\sigma ^2 \left(-\left(\tau
   ^2-1\right)\right)+\tau ^2+3\right)^2}
\end{equation}
Differently from the previous case, the curvature has now polar singularities. Indeed utilizing polar coordinates in the plane $\sigma,\tau$ we get:
\begin{equation}\label{crologino}
  \mathcal{K}(r\cos[\phi],r\sin[\phi])\,=\, - \,\frac{192 \left(r^2 \sin ^2(\phi )-1\right) \left(r^2 \cos
   ^2(\phi )-1\right)}{\left(r^4 \cos (4 \phi )-r^4+8
   r^2+24\right)^2}
\end{equation}
and we see that the curvature is singular along the curves defined by the quartic equation:
\begin{equation}\label{quartic}
  r^4 \cos (4 \phi )-r^4+8
   r^2+24 \, = \, 0
\end{equation}
In Fig. \ref{curvaturepari} we show the four branches of such a singular locus.
In Fig. \ref{tremendo} we provide a three-dimensional view of the curvature plot.
\par
In Fig. \ref{poffarba} we display grids of geodesics starting at various different points of the plane in different regions separated by curvature walls.
\section{Conclusions}
\label{zakliuchenie}
In the conclusions to the second episode of the present thermodynamic tale, namely paper \cite{secondtemperature}, we listed the investigation directions for our further work and the first points a) and b) in such check list were the following ones:
\par
\textit{
\begin{description}
\item[a)]
Careful investigation of the macroscopic K\"ahler geometry whose K\"ahler potential is the Legendre transform, according with eq.(\ref{legendroK}), of the real symplectic potential, identified with the 
    stochastic hamiltonian $G(\boldsymbol{\lambda}) \, = \, -\log[Z(\boldsymbol{\lambda})]$. In relation with this we
    note the very inspiring fact that such K\"ahler potential has a structure very similar to that of special geometries since it is obtained from a symplectic potential with the following structure:  
    \begin{eqnarray}\label{gramellino}
      G(\boldsymbol{\lambda})& = & \beta_0 + \log[\mathbb{P}^{1}(\boldsymbol{\lambda})]+
      \log[\mathbb{P}^{2\nu+2}(\boldsymbol{\lambda})] \nonumber\\
       \mathbb{P}^{2\nu+2}(\lambda)&=& \text{homogeneous polynomial of degree $2\nu+2$ in $\boldsymbol{\lambda}$}\nonumber\\
       \mathbb{P}^{1}(\lambda)&=& \text{linear polynomial of degree $1$ in $\boldsymbol{\lambda}$}
       \end{eqnarray}
       \item[b)] Careful investigation of the symmetry breaking patterns related with the \textit{magnetic fields} $\mathbf{h}$ associated with Casimir functions of sequence subalgebras.
           \end{description}
           }
\par
In the present paper we have precisely performed the above mentioned careful investigations in the case where the symplectic potential of eq.(\ref{gramellino}) is given   by the logarithm of the partition functions corresponding to the \textbf{abelian structures} of the \textbf{principal subalgebra sequence} for the $\boldsymbol{\mathfrak{b}}$ and $\boldsymbol{\mathfrak{d}}$ series of CV manifolds. Before summarizing what we have discovered in this case, we ought to stress that, for the same CV manifolds  there are also other partition functions, corresponding to different abelian structures and, therefore, to different generalization of Souriau thermodynamics. What these different partition functions imply for the thermo-metric is something we are going to study in a forthcoming new paper.  
\par
For the \textit{principal subalgebra sequence} the result we have so far obtained is the following:
\begin{description}
  \item[A)] There is a fundamental distinction between the $q=\text{odd}$ case ($\boldsymbol{\mathfrak{b}}$-series)
      and the $q=\text{even}$ case ($\boldsymbol{\mathfrak{d}}$-series). The $\boldsymbol{\mathfrak{b}}$-symplectic potential with all the available magnetic fields included, produces a completely flat metric. Instead $\boldsymbol{\mathfrak{d}}$-symplectic potential with all the available magnetic fields included, produces a metric that is flat in all directions except two that span a universal non compact submanifold $\mathfrak{M}_{2|sing}$, whose curvature is singular along four infinite curves. This produces the  partition of the two-dimensional space in five separate regions not allowed to communicate with each other by means of geodesics.  
  \item[B)] Whenever some of the magnetic fields are externally frozen to zero, the thermodynamic space is reduced to a submanifold on which the pull-back of the ambient thermo-metric is no longer flat and it develops non trivial curvature.
  \item[C)] When the microscopic $\Omega = CV$ manifolds are of large dimension $\text{dim}_{\mathbb{R}}\Omega \, = \, 2\, m\gg 2$, the macroscopic thermodynamic space $\text{dim}_{\mathbb{R}}\mho[\Omega,\boldsymbol{\mathfrak{p}}^a]=2m$ is also of large dimension, the symplectic potential depending on exactly $m$ variables. Then the vanishing patterns of the several magnetic fields introduce a complex combinatorial scheme. Indeed the generalized magnetic fields, just as the generalized temperatures, are organized into an ordered hierarchy and what matters is 
      whether a set of vanishing magnetic fields constitute a continuous chain $h_{i},h_{i+1},\dots$ or not.   
  \item[D)] Every isolated vanishing magnetic field introduces a universal $\mathfrak{M}_{2|reg}$ 2-dimensional Riemannian manifold $\mathfrak{M}_{2|reg}$ whose curvature is regular. Hence, if we set to zero a certain number $r$ of isolated magnetic fields, we obtain a thermo-manifold that is the tensor product of a flat manifold with $r$ copies of the same universal space $\mathfrak{M}_{2|reg}$. In the $\boldsymbol{\mathfrak{d}}$-series case one has to add to the tensor product also the universal $\mathfrak{M}_{2|sing}$-factor which has the above mentioned singularities.
  \item[E)] Every isolated chain of (n-1)-vanishing magnetic fields $h_{i},h_{i+1},\dots,h_{i+n-1}$ produces a universal $n$-dimensional Riemannian manifold $\mathfrak{M}_{n|reg}$, whose curvature components are all regular. This latter  constitutes a natural generalization of $\mathfrak{M}_{2|reg}$.  We investigated in some depth the structure of the $n$=3 case and we inferred the general pattern of the $\mathfrak{M}_{n|reg}$ geometry for all values of $n$. In particular, we investigated the embedding of the  $\mathfrak{M}_{n|reg}$ manifolds into flat Euclidean $(2n-1)$-space following just from the vanishing magnetic field equations. This embedding reveals the geometrical interpretation of the manifolds $\mathfrak{M}_{n|reg}$ as generalized translational surface with equal profiles in all translation directions. Furthemore the natural Euclidean space in which the $\mathfrak{M}_{n|reg}$ are embedded has a flat metric provided by the Cartan matrix of the algebra $\mathfrak{a}_{2n-1}$.
  \item[F)] The properties of the Riemannian spaces $\mathfrak{M}_{n|reg}$, that form the tiles of the combinatorial magnetization schemes and hence are strategic for any  conceivable application to \textbf{unsupervised learning with reinforcement} are very peculiar and quite challenging. The metric has no continuous isometry, yet it has a rich group of discrete isometries that includes the Symmetric Permutation Group $S_n$ extended with reflections.  The behavior of the Riemann tensor of  $\mathfrak{M}_{n|reg}$ is very special and it is responsible for partitioning the entire manifold in subregions qualifiable as valleys or ridges with respect to the curvature components. The physical domain of the manifold corresponding to the polytope where the partition function integral converges has the structure of an $n$-dimensional Hypercube whose boundary is composed of faces each of which is a hypercube of dimension $n-1$. This boundary that is always at infinite distance from the interior points has \textit{de-facto} a combinatorial structure, made of faces, edges and  vertices. The fascinating point is that the geodesics tend to the boundary not indiscriminately rather to focal points made of vertices 
      central points of the faces and mid point of the edges. 
      Such property, that is reminescent of the causal structure of Lorentzian space-times and Penrose diagrams
      (see for instance volume 2 of \cite{pietrobook}) was discovered by us experimentally, doing numerical integration of the geodesic, yet is quite probable that it might be proved a priori and it deserves a careful theoretical investigation. It superfluous to emphasize that such property might play a decisive role in any application to unsupervised learning with reinforcement. 
\end{description}
As one sees the geometrical landscape of the Thermo--metric is already quite complex for the choice of  Souriau generalization
corresponding to the \textit{principal subalgebra sequence} and such landscape can be further enriched via the inclusion of the Souriau extensions corresponding to different abelian structures. What we  still have to do in order to complete our study of symmetry breaking patterns is summarized in the following four points:
\begin{enumerate}
  \item Extend to all values of $n$ the asymptotic analysis of the  $\mathfrak{M}_{n|reg}$ geometry generated by every isolated (n-1)-chain of vanishing contiguous magnetic fields.  
  \item Find convenient formulations of the curvature walls that partition the parameter space in every combinatorial scheme of magnetic field freezing. 
  \item Consider also the case where the magnetic fields are frozen to constant values different from zero.
  \item Include the thermo geometry associated with other abelian structures.    
\end{enumerate}
The above issues constitute one of the first priorities in our road map of further development. The other equally urgent priorities were already listed in the conclusions of \cite{secondtemperature} and we find it convenient to recall them here:
\begin{description}
\item[c)] Investigation of the relation between $\mathcal{SK}_{3+q}$ Special K\"ahler symmetric spaces 
    and the Special K\"ahler Homogeneous but not symmetric spaces $L(-1,q)$ (see \cite{deWit:1995tf,toineugenio,SKGaggio3,SKGaggio2,SKGaggio1,specHomgeoA2,specHomgeoA1})
    that appear to be provided by a deformed metric with a smaller group of isometries on the same solvable Lie group 
    metrically equivalent to the CV manifolds (work in progress \cite{toinepietromario}). The pattern of symmetry breaking that leads to such homogeneous non-symmetric metric might be related with the symmetry breaking introduced by the "\textbf{sponatenous magnetization}"
    of Casimir functions.
\item[d)] Explicit algorithmic construction of Cartan Neural Networks based on CV-manifolds and investigation of the statistical distributions on their layers by means of the new here developed thermodynamics \cite{r2paperone}
\end{description}

\subsection{Perspectives}
As we noted in the conclusion to \cite{secondtemperature} the use of extended Souriau thermodynamics in Machine Learning can occur at two levels that are strictly correlated:
\begin{enumerate}
  \item In a posteriori analysis of the distributions of information data on the various layers of a \textbf{supervised and trained Cartan Neural Network} in order to study the geometry of categorical representations in the \textit{categorical perception} \cite{groundwork,neurocoding} à la Bonnasse-Gahot and Nadal.
  \item In \textbf{unsupervised learning with reinforcement}, namely generative AI. In that capacity (still to be accurately studied) the variation of magnetic fields might be the mathematical realization of the \textit{agents} that either reinforce or inhibit the spontaneous evolution of the dynamical system (geodesics in thermodynamic space) 
\end{enumerate}
\newpage
\appendix
\section{Detailed study of  $\mathfrak{M}_{3|reg}$ in vielbein formalism} 
\label{tregambone}
For the case $n=3$, the metric (\ref{caponata}) can be realized by means of the following dreibein:
\begin{eqnarray}
\label{dreibagno}
  V^1 &=&\frac{\sqrt{2} \text{d$\sigma $}_1}{\sqrt{-\frac{\left(\sigma
   _1^2-1\right){}^2 \left(\left(\sigma _3^2+2\right) \sigma
   _2^2+2 \sigma _3^2+3\right)}{\left(\sigma _2^2+\sigma
   _3^2+2\right) \sigma _1^2+2 \sigma _3^2+\sigma _2^2
   \left(\sigma _3^2+2\right)+3}}}\nonumber \\
  V^2 &=& \frac{\text{d$\sigma $}_2 \left(\sigma _1^2-1\right)
   \left(\left(\sigma _3^2+2\right) \sigma _2^2+2 \sigma
   _3^2+3\right)-\text{d$\sigma $}_1 \sigma _1 \sigma _2
   \left(\sigma _2^2-1\right) \left(\sigma
   _3^2+1\right)}{\left(\sigma _1^2-1\right)
   \left(\left(\sigma _3^2+2\right) \sigma _2^2+2 \sigma
   _3^2+3\right) \sqrt{-\frac{\left(\sigma _2^2-1\right){}^2
   \left(2 \sigma _3^2+3\right)}{2 \left(\sigma _3^2+2\right)
   \sigma _2^2+4 \sigma _3^2+6}}} \\
  V^3 &=& \frac{\text{d$\sigma $}_3 \left(\sigma _1^2-1\right)
   \left(\sigma _2^2-1\right) \left(2 \sigma
   _3^2+3\right)-\sigma _3 \left(\sigma _3^2-1\right)
   \left(\text{d$\sigma $}_2 \left(\sigma _1^2-1\right)
   \sigma _2+\text{d$\sigma $}_1 \sigma _1 \left(\sigma
   _2^2-1\right)\right)}{\sqrt{3} \left(\sigma _1^2-1\right)
   \left(\sigma _2^2-1\right) \left(2 \sigma _3^2+3\right)
   \sqrt{-\frac{\left(\sigma _3^2-1\right){}^2}{4 \sigma
   _3^2+6}}} 
\end{eqnarray}
Indeed we have:
\begin{equation}\label{taziagiulia}
   ds^2_{\mathfrak{M}_{3|reg}}\, = \, \sum_{i=1}^3 \, V^i\times V^i
\end{equation}
Calculating the spin connection defined by the vanishing torsion constraint:
\begin{equation}\label{Torzero}
  dV^i \, + \, \omega^{ij} \wedge V^j \, = \,0
\end{equation}
and the curvature 2-form:
\begin{equation}\label{curvista}
  \mathfrak{R}^{ij} \,\equiv \,  d\omega^{ij} \, + \, \omega^{ik}\wedge \omega^{kj}
\end{equation}
we have 3 independent components of the 2-form curvature:
\begin{equation}\label{corleone}
  \mathfrak{R}^A \, = \, \left\{\mathfrak{R}^{12},\,\mathfrak{R}^{13},\,\mathfrak{R}^{23}\right\}
\end{equation}
and similarly the anholonomic vielbein basis for $2$-forms is made of three elements:
\begin{equation}\label{enoelroc}
  VV^A \, = \, \left\{V^1\wedge V^2,\,V^1\wedge V^3,\,V^2\wedge V^3\right\}
\end{equation}
Therefore the complete form of the curvature $2$-form, namely the anholonomic Riemann tensor of the metric $ds^2_{\mathfrak{M}_{3|reg}}$ can be encoded into a $3\times 3$ symmetric matrix $K^A_{\phantom{A}B}(\boldsymbol{\sigma})$ such that
\begin{equation}\label{peronostica}
 \mathfrak{R}^A \, = \, K^A_{\phantom{A}B}(\boldsymbol{\sigma}) \,VV^B
\end{equation}
Hence, all intrinsic information about the geometry of the Riemannian space $\mathfrak{M}_{3|reg}$ is encoded in the 6-independent entries of the curvature matrix 
$K^A_{\phantom{A}B}(\boldsymbol{\sigma})$, that have the following form:
{\scriptsize
\begin{eqnarray}
\label{Kdiag}
 K^1_{\phantom{1}1}(\boldsymbol{\sigma})&=& \frac{N_{11}}{D_{11}} \nonumber\\
 N_{11}&=&-4 \left(\sigma _1^2-1\right) \left(\sigma _2^2-1\right) \sigma
   _3^6+\left(-\left(\sigma _2^2-1\right) \sigma _1^4-\left(\sigma _2^4+18 \sigma
   _2^2-17\right) \sigma _1^2+\sigma _2^4+17 \sigma _2^2-16\right) \sigma
   _3^4\nonumber\\
  &&+\left(\left(\sigma _2^2-1\right) \sigma _1^4+\left(\sigma _2^4-19 \sigma
   _2^2+20\right) \sigma _1^2-\sigma _2^4+20 \sigma _2^2-21\right) \sigma _3^2-9
   \left(\sigma _1^2-1\right) \left(\sigma _2^2-1\right) \nonumber\\
  D_{11} &=& 2 \left(2 \sigma _3^2+3\right) \left(\left(\sigma _2^2+\sigma _3^2+2\right) \sigma
   _1^2+2 \sigma _3^2+\sigma _2^2 \left(\sigma _3^2+2\right)+3\right){}^2 \nonumber\\
    K^2_{\phantom{2}2}(\boldsymbol{\sigma})&=& \frac{N_{22}}{D_{22}}\nonumber\\
 N_{22}&=& -3 \left(\sigma _3^2-1\right) \left(\sigma _2^2 \left(\sigma _2^2-1\right)
   \left(\sigma _3^2+1\right){}^2 \sigma _1^4+\left(\left(\sigma _3^2+2\right){}^2
   \sigma _2^6+\left(6 \sigma _3^4+20 \sigma _3^2+17\right)\sigma_2^4\right.\right.\nonumber\\
   &&\left.\left.+\left(\sigma _3^2+2\right) \left(7 \sigma _3^2+10\right) \sigma
   _2^2+\left(2 \sigma _3^2+3\right){}^2\right) \sigma _1^2-\left(\sigma
   _2^2+1\right) \left(\left(\sigma _3^2+2\right) \sigma _2^2+2 \sigma_3^2+3\right){}^2\right)\nonumber\\
 D_{22}&=& 2 \left(2 \sigma _3^2+3\right) \left(\left(\sigma _3^2+2\right) \sigma _2^2+2
   \sigma _3^2+3\right) \left(\left(\sigma _2^2+\sigma _3^2+2\right) \sigma _1^2+2
   \sigma _3^2+\sigma _2^2 \left(\sigma _3^2+2\right)+3\right){}^2\nonumber\\
 K^3_{\phantom{1}3}(\boldsymbol{\sigma})&=& -\frac{3 \left(\sigma _1^2+1\right) \left(\sigma _2^2-1\right) \left(\sigma
   _3^2-1\right)}{2 \left(\left(\sigma _3^2+2\right) \sigma _2^2+2 \sigma
   _3^2+3\right) \left(\left(\sigma _2^2+\sigma _3^2+2\right) \sigma _1^2+2 \sigma
   _3^2+\sigma _2^2 \left(\sigma _3^2+2\right)+3\right)} \nonumber\\    
\end{eqnarray}
}
for the three diagonal components and the following one for the 
three off-diagonal ones
{\scriptsize
\begin{eqnarray}
\label{Koffdiag}
K^1_{\phantom{1}2}(\boldsymbol{\sigma})&=& \frac{\sqrt{3} \sigma _2 \sigma _3 \left(1-\sigma _3^2\right) \left(-\left(\sigma
   _2^2-1\right) \left(\sigma _3^2+1\right) \sigma _1^4+\left(\left(\sigma
   _2^2+2\right) \left(\sigma _3^2+2\right) \sigma _2^2+3 \sigma _3^2+4\right)
   \sigma _1^2-\left(\sigma _2^2+1\right) \left(\left(\sigma _3^2+2\right) \sigma
   _2^2+2 \sigma _3^2+3\right)\right)}{2 \left(2 \sigma _3^2+3\right)
   \sqrt{\left(\sigma _3^2+2\right) \sigma _2^2+2 \sigma _3^2+3}
   \left(\left(\sigma _2^2+\sigma _3^2+2\right) \sigma _1^2+2 \sigma _3^2+\sigma
   _2^2 \left(\sigma _3^2+2\right)+3\right){}^2}\nonumber\\
   K^1_{\phantom{1}3}(\boldsymbol{\sigma})&=&-\frac{\sqrt{3} \sigma _1 \left(\sigma _1^2+1\right) \left(\sigma _2^2-1\right)
   \sigma _3 \left(1-\sigma _3^2\right)}{2 \sqrt{\left(2 \sigma _3^2+3\right)
   \left(\left(\sigma _3^2+2\right) \sigma _2^2+2 \sigma _3^2+3\right)}
   \left(\left(\sigma _2^2+\sigma _3^2+2\right) \sigma _1^2+2 \sigma _3^2+\sigma
   _2^2 \left(\sigma _3^2+2\right)+3\right){}^{3/2}}\nonumber\\
  K^2_{\phantom{1}3}  &=&-\frac{3 \sigma _1 \left(\sigma _1^2+1\right) \sigma _2 \left(\sigma _2^2-1\right)
   \left(\sigma _3^4-1\right)}{2 \sqrt{2 \sigma _3^2+3} \left(\left(\sigma
   _3^2+2\right) \sigma _2^2+2 \sigma _3^2+3\right) \left(\left(\sigma _2^2+\sigma
   _3^2+2\right) \sigma _1^2+2 \sigma _3^2+\sigma _2^2 \left(\sigma
   _3^2+2\right)+3\right){}^{3/2}}
\end{eqnarray}
}
The visualization of a three dimensional curvature that is not
constant in any direction, as the present one happens to be, is always rather problematic. We provide a method below. 
First, however, let us observe some relevant general properties of the matrix $K^A_{\phantom{A}B}(\boldsymbol{\sigma})$ that can be verified analytically. We begin by observing that $K^A_{\phantom{A}B}(\boldsymbol{\sigma})$ is diagonal in the origin and proportional to the identity matrix:
\begin{equation}\label{formalina}
  K^A_{\phantom{A}B}(0,0,0)\, = \, -\frac{1}{6} \times 
  \left(
    \begin{array}{ccc}
      1 & 0 & 0 \\
      0 & 1 & 0 \\
      0 & 0 & 1 \\
    \end{array}
  \right)
\end{equation}
If eq.(\ref{formalina}) were true everywhere, our manifold would be an  Einstein symmetric space, namely $\mathrm{SO(1,3)/SO(3)}$. This means that in the vicinity of the origin our manifold behaves as a slightly deformed hyperboloid. As we move away from the origin the curvature tensor develops all of its components and the manifold behaves in a substantially different way from a symmetric space. However there are directions along which the curvature matrix remains diagonal: namely the coordinate axes. 
Indeed we find:
\begin{eqnarray}\label{carmelitano}
K(x,0,0)& = & \left(
\begin{array}{ccc}
 \frac{3 \left(x^2-1\right)}{2 \left(2 x^2+3\right)^2} & 0 & 0 \\
 0 & \frac{9 x^2-9}{6 \left(2 x^2+3\right)^2} & 0 \\
 0 & 0 & -\frac{x^2+1}{2 \left(2 x^2+3\right)} \\
\end{array}
\right) \nonumber\\
K(0,y,0) & = &\left(
\begin{array}{ccc}
 \frac{3 \left(y^2-1\right)}{2 \left(2 y^2+3\right)^2} & 0 & 0 \\
 0 & -\frac{y^2+1}{2 \left(2 y^2+3\right)} & 0 \\
 0 & 0 & \frac{3 \left(y^2-1\right)}{2 \left(2 y^2+3\right)^2} \\
\end{array}
\right)\nonumber\\
K(0,0,z)& = &\left(
\begin{array}{ccc}
 \frac{-4 z^6-16 z^4-21 z^2-9}{2 \left(2 z^2+3\right)^3} & 0 & 0 \\
 0 & \frac{3 \left(z^2-1\right)}{2 \left(2 z^2+3\right)^2} & 0 \\
 0 & 0 & \frac{3 \left(z^2-1\right)}{2 \left(2 z^2+3\right)^2} \\
\end{array}
\right)
\end{eqnarray}
\par
This means that when we move along the coordinate axes we always have:
\begin{equation}\label{propina}
  \mathfrak{R}^{12} \, \propto \,V^1 \wedge V^2 \quad;\quad  \mathfrak{R}^{13}  \, \propto \, V^1\wedge V^3   \quad;\quad  \mathfrak{R}^{23}  \, \propto \, V^2\wedge V^3  
\end{equation}
but, depending on the axis we choose, the three components evolve differently. At the origin they are all equal among themselves and equal to $-1/6$, but approaching the boundary $x\to\pm 1$ or $y\to \pm 1$ or $z\to \pm 1$, two of the components go to zero and only one survives:
\begin{eqnarray}
\label{fragitoso}
\lim_{x\to \pm 1} K(x,0,0) &=& \left(
\begin{array}{ccc}
 0 & 0 & 0 \\
 0 & 0 & 0 \\
 0 & 0 & -\frac{1}{5} \\
\end{array}
\right) \quad \stackrel{\text{on the boundary}}{\Rightarrow} \mathfrak{R}^{23} \, = \,- \,\ft 15 V^2\wedge V^3, \, \, \mathfrak{R}^{12}\, = \,\mathfrak{R}^{13} \, = \, 0 \nonumber\\
\lim_{y\to \pm 1} K(0,y,0) &=& \left(
\begin{array}{ccc}
 0 & 0 & 0 \\
 0 & -\frac{1}{5} & 0 \\
 0 & 0 & 0 \\
\end{array}
\right) \quad \stackrel{\text{on the boundary}}{\Rightarrow} \mathfrak{R}^{13} \, = \,-\,\ft 15 V^1\wedge V^3, \, \, \mathfrak{R}^{12}\, = \,\mathfrak{R}^{23} \, = \, 0 \nonumber\\ 
\lim_{z\to \pm 1} K(0,0,z) &=& \left(
\begin{array}{ccc}
 -\frac{1}{5} & 0 & 0 \\
 0 & 0 & 0 \\
 0 & 0 & 0 \\
\end{array}
\right) \quad \stackrel{\text{on the boundary}}{\Rightarrow} \mathfrak{R}^{12} \, = \,\ft 15 V^1\wedge V^2, \, \, \mathfrak{R}^{13}\, = \,\mathfrak{R}^{23} \, = \, 0 \nonumber\\ 
\end{eqnarray}
The interpretation of the above limits is simple. While in the vicinity of the origin $\{0,0,0\}$ the manifold $\mathfrak{M}_{3|reg}$ is approximately a $3$-dimensional hyperboloid $\frac{\mathrm{SO(1,3)}}{\mathrm{SO(3)}}$, in the vicinity of the center of each of the $6$ cube faces, $\mathfrak{M}_{3|reg}$ is approximately  $\mathbb{R}\times \frac{\mathrm{SO(1,2)}}{\mathrm{SO(2)}}$, the flat direction being the axis connecting the face center to the origin and the two curved directions being instead those of the plane orthogonal to such axis. 
\begin{figure}[ht]
\begin{center}
\includegraphics[width=5cm]{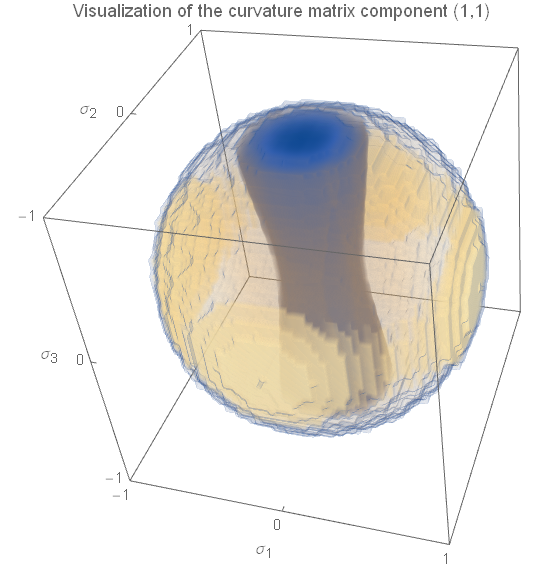}
\includegraphics[width=5cm]{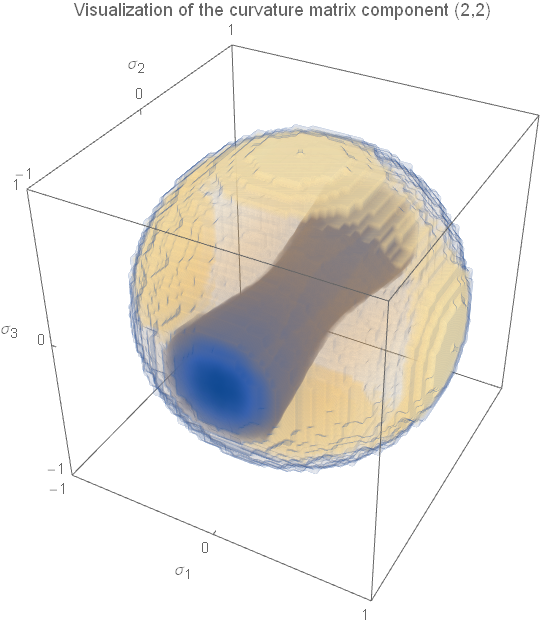}
\includegraphics[width=5cm]{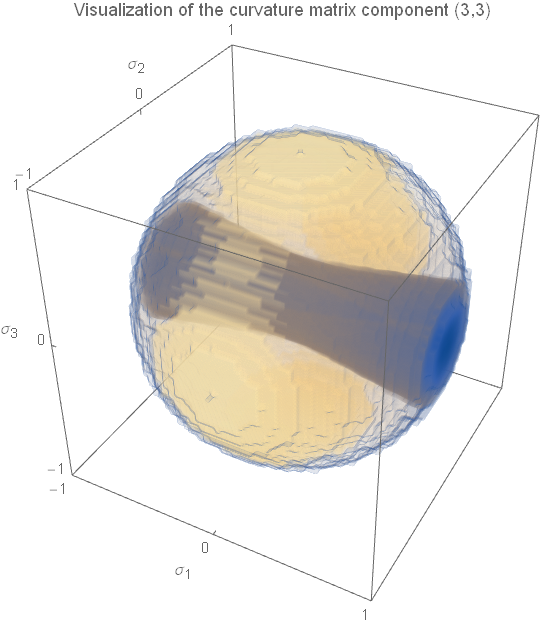}
\includegraphics[width=5cm]{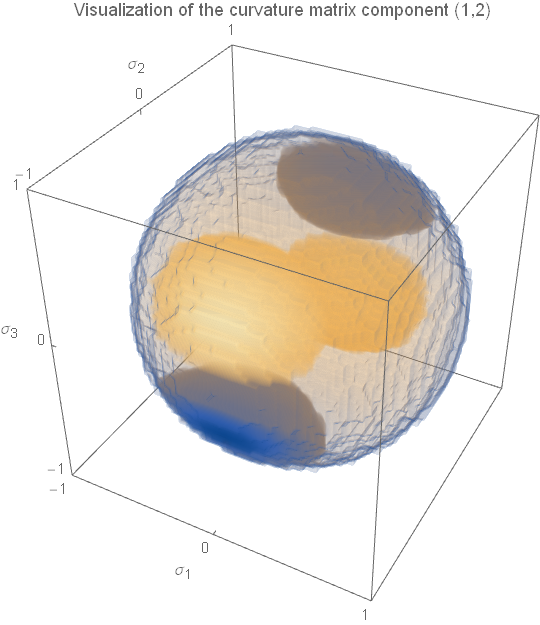}
\includegraphics[width=5cm]{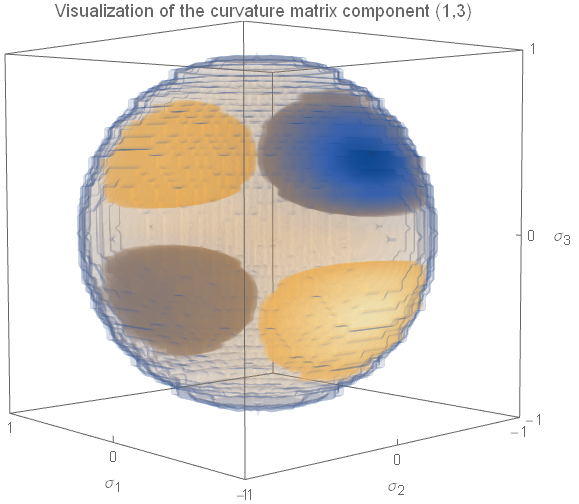}
\includegraphics[width=5cm]{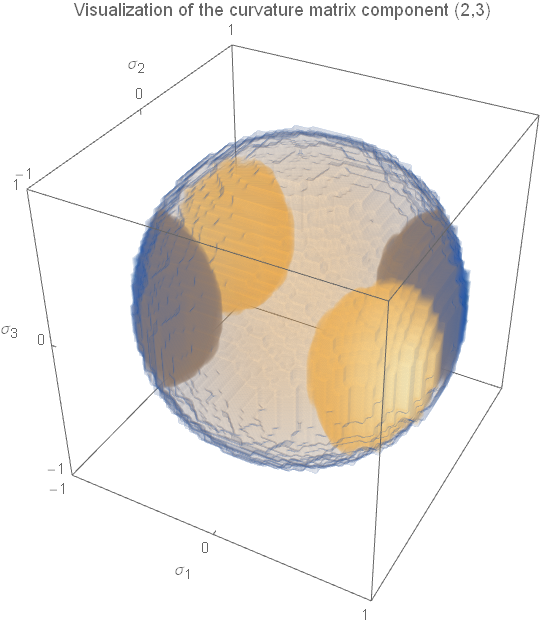}
 \caption{\label{densimetro}In this figure we present the density plots of the 6 components of the curvature matrix}
\end{center}
\end{figure}
There are still eight important directions along which the 
approach to the boundary limit is special. They are the lines that join  the origin and the eight vertices of the unit cube (see Fig. \ref{centralina}). Along these lines the curvature matrix is not diagonal, yet in the limit $x\to \pm 1$ it vanishes completely:
\begin{equation}\label{angolari}
  \lim_{x\to \pm 1}K(\pm x,\pm x,\pm x) \, = \, 
\left(  \begin{array}{ccc}
 0 & 0 & 0 \\
  0 & 0 & 0 \\
   0 & 0 & 0 \\
\end{array}
\right)
\end{equation}
The summary of our results about the complicated asymptotic behavior of the curvature  is provided in Fig. \ref{centralina}.
In conclusion on the border at infinity (namely on the boundary of the cube) the  curvature of the manifold $\mathfrak{M}_{3|reg}$ degenerates in such a way that it vanishes completely at the vertices of the cube and instead it has just one component equal to $-\ft 1 5$ in the center of each of the six faces. 
\par
A possible way to visualize the behavior of the intrinsic curvature matrix (\ref{Kdiag}),(\ref{Koffdiag}) is provided by the 3D density Plot of its six components. This is shown in Fig. 
\ref{densimetro}.
Using the color code to distinguish \textit{larger values} = \textit{lighter yellowish color} from \textit{smaller values} =
\textit{darker bluish color}, the density plots provide a global vision of the functions behavior. Looking at the first three figures corresponding to the diagonal entries of the curvature matrix, the three blue tubes enveloping each of the coordinatee axes are a graphical representation of the approach to asymptotic limits (\ref{fragitoso}). Indeed 
the asymptotic value $-\ft 15$ is negative and corresponds to dark blue. The light yellow is $0$ and one clearly sees on the boundary  of the unit ball the four light islands signaling the vanishing of the curvature component along the other two axes. 
It is harder to interpret  the behavior of the off-diagonal components  as depicted in the density plots presented in the second line of Fig. (\ref{densimetro}).
\par

\end{document}